\documentclass[aps,preprint,amsmath,amssymb,amsfonts]{revtex4}
\usepackage{epsfig}
\usepackage{graphicx}
\usepackage{subfigure}
\usepackage{dcolumn}
\usepackage{bm}
\usepackage{amsthm}
\usepackage{enumitem}
\usepackage{slashed}

\newcommand{\del}{\partial}

\newcommand{\tr}{\operatorname{tr}}
\newcommand{\str}{\operatorname{str}}

\newcommand{\bbR}{\mathbb{R}}
\newcommand{\bbC}{\mathbb{C}}

\newcommand{\bbP}{\mathbb{P}}

\newcommand{\calA}{\mathcal{A}}
\newcommand{\calB}{\mathcal{B}}
\newcommand{\calC}{\mathcal{C}}
\newcommand{\calD}{\mathcal{D}}
\newcommand{\calE}{\mathcal{E}}
\newcommand{\calP}{\mathcal{P}}

\newcommand{\dSCFT}[2]{dS\textsubscript{#1}/CFT\textsubscript{#2}}
\newcommand{\AdSCFT}[2]{AdS\textsubscript{#1}/CFT\textsubscript{#2}}

\begin{document}

\title{The holographic dual of the Penrose transform}

\author{Yasha Neiman}
\email{yashula@gmail.com}

\affiliation{Okinawa Institute of Science and Technology, 1919-1 Tancha, Onna-son, Okinawa 904-0495, Japan}

\date{\today}

\begin{abstract}
We consider the holographic duality between type-A higher-spin gravity in $AdS_4$ and the free $U(N)$ vector model. In the bulk, linearized solutions can be translated into twistor functions via the Penrose transform. We propose a holographic dual to this transform, which translates between twistor functions and CFT sources and operators. We present a twistorial expression for the partition function, which makes global higher-spin symmetry manifest, and appears to automatically include all necessary contact terms. In this picture, twistor space provides a fully nonlocal, gauge-invariant description underlying both bulk and boundary spacetime pictures. While the bulk theory is handled at the linear level, our formula for the partition function includes the effects of bulk interactions. Thus, the CFT is used to solve the bulk, with twistors as a language common to both. A key ingredient in our result is the study of ordinary spacetime symmetries within the fundamental representation of higher-spin algebra. The object that makes these ``square root'' spacetime symmetries manifest becomes the kernel of our boundary/twistor transform, while the original Penrose transform is identified as a ``square root'' of CPT.
\end{abstract}

\maketitle
\tableofcontents
\newpage

\section{Introduction}

Quantum spacetime is infamously difficult to address directly. This leads one to search for alternative geometric frameworks, which may survive the breakdown of locality at the Planck scale. The most productive approach to date is AdS/CFT \cite{Maldacena:1997re,Witten:1998qj,Aharony:1999ti} -- a retreat from the bulk spacetime onto its asymptotic boundary. There, one can operate with a fixed classical geometry, since the Planck length effectively vanishes due to an infinite warp factor. AdS/CFT relates two spacetime pictures, with two different notions of locality: an approximate locality in the higher-dimensional bulk, and a precise locality on the lower-dimensional boundary. The duality itself is of necessity non-local. Furthermore, the bulk and boundary pictures each contain a different set of gauge redundancies -- the well-known price of locality -- which are absent in the dual picture. A question then suggests itself: is there some third geometric framework, completely divorced from spacetime locality, underlying both the bulk and boundary descriptions? To find such a framework, one must focus on non-local, gauge-invariant objects in both bulk and boundary. Such is arguably the strategy behind the study of Ryu-Takayanagi surfaces \cite{Ryu:2006bv}, kinematic space in the context of MERA \cite{Czech:2015qta}, and other such relations between quantum-informational quantities and bulk geometry. 

At the same time, there exists a much older proposal for a geometric framework to replace spacetime: Penrose's twistor theory \cite{Penrose:1986ca,Ward:1990vs}. There, we effectively trade locality for causality as the fundamental principle, replacing points with twistors -- the ``maximally lightlike'' extended shapes in spacetime. Originally conceived as a framework for quantum General Relativity, twistor theory has now become a workhorse for scattering amplitude calculations in maximally supersymmetric Yang-Mills \cite{Adamo:2011pv} and supergravity \cite{Skinner:2013xp}. Might it be possible, then, to use twistor space as a basis for the non-local description underlying both bulk and boundary in AdS/CFT?

In the present paper, we answer this question in the affirmative, in the context of one simple model -- the duality \cite{Klebanov:2002ja} between type-A higher-spin gravity in $AdS_4$ and a free $U(N)$ vector model on its 3d boundary. Higher-spin gravity \cite{Vasiliev:1995dn,Vasiliev:1999ba} is an interacting theory of infinitely many massless fields, in this case one for each integer spin. On the boundary, these fields are dual to an infinite tower of conserved currents in the free CFT. The simplicity of this holographic model stems from its infinite-dimensional higher-spin symmetry -- similar in some ways to supersymmetry, but stronger. We must note that, in this simple version, higher-spin gravity is highly unrealistic: while it does contain a massless spin-2 ``graviton'', its interactions are nothing like those of GR, and in fact appear to be non-local at the cosmological scale. In this sense, we are dealing with a toy model. On the other hand, higher-spin gravity has the virtue of being formulated in four bulk dimensions, and is easily compatible with a positive cosmological constant.

A crucial simplifying feature of our higher-spin model is that it allows us to deal exclusively with free theories. In the bulk, we consider the linearized version of higher-spin gravity, i.e. free massless fields of all spins, which can be mapped into twistor space via the Penrose transform. On the boundary, we have the free CFT, which we map into twistor space using a novel ``holographic dual'' of the Penrose transform. This boundary version of the transform is more powerful than its bulk counterpart, since the correlators of the free CFT encode not only the linearized bulk theory, but also the bulk interactions. Thus, we're essentially using the boundary CFT to solve the bulk theory, using twistor space as a common language between the two. 

Note that twistor theory is a dimension-specific tool: it was originally constructed for massless 4d theories, subject either to conformal 4d symmetry (e.g. Yang-Mills) or to 4d isometries (e.g. GR or higher-spin gravity). On the other hand, AdS/CFT exploits the relation between conformal symmetry in $d$ dimensions and isometries in $d+1$ dimensions. Thus, the intersection between twistor theory and holography will naturally take place in either \AdSCFT{5}{4} or \AdSCFT{4}{3}. The \AdSCFT{5}{4} case was discussed in \cite{Adamo:2016rtr}, and has the promise of general applicability: since the 4d boundary theory is conformal, one can always think of it as ``massless''. In contrast, in the \AdSCFT{4}{3} case considered in this paper, we expect that twistor methods will be relevant only in the special setup of higher-spin theory, since it's only there that the 4d bulk fields are all massless.

The rest of the paper is structured as follows. In section \ref{sec:summary}, we summarize the main results, with only a cursory explanation of the notations. Section \ref{sec:geometry} is a geometric introduction to twistor space and its relation to bulk and boundary spinor spaces. Our geometry is carried out in 5d flat spacetime, within which both bulk and boundary are embedded. In section \ref{sec:algebra}, we introduce the higher-spin algebra, including structures that arise when focusing on a bulk or boundary point. In section \ref{sec:linear_HS}, we formulate the linearized bulk theory and the Penrose transform. In section \ref{sec:spacetime_subgroup}, we resume our discussion of higher-spin algebra, focusing on the representation of ordinary spacetime symmetries within the higher-spin adjoint and fundamental. This will lead us to a geometric viewpoint on the Penrose transform, which in turn will suggest its boundary dual. In section \ref{sec:CFT}, we discuss the boundary CFT in a bilocal language, and present the holographic dual of the Penrose transform. In section \ref{sec:holography}, we establish the holographic relationship between the bulk and boundary pictures, by calculating expectation values of local boundary currents. An analogous matching for the local field strengths of boundary sources is left for later work. Section \ref{sec:discuss} is devoted to discussion and outlook.

Throughout the paper, we consider for simplicity Euclidean spacetime, i.e. the bulk is Euclidean Anti de Sitter space ($EAdS_4$). However, as discussed in section \ref{sec:discuss}, we envision an eventual application to Lorentzian de Sitter ($dS_4$). 

\section{Summary of results} \label{sec:summary}

\subsection{Penrose transform}

In some ways, higher-spin gravity is the most natural application of twistor theory, more so than Yang-Mills or General Relativity. In Yang-Mills and GR, twistors serve ``merely'' as the spinor representation of isometries or conformal transformations in 4d spacetime. In higher-spin theory, we utilize a greater power of these objects, using them to generate an infinite-dimensional extension of spacetime symmetries -- the higher-spin (HS) group. The role of twistors in higher-spin algebra is identical to the role of vectors in Clifford algebra: 
\begin{align}
 \text{Clifford algebra:}\quad \left\{\gamma_\mu,\gamma_\nu \right\} = -2\eta_{\mu\nu} \ ; \quad 
 \text{Higher-spin algebra:}\quad \left[Y_a,Y_b \right]_\star = 2iI_{ab} \ , \label{eq:Clifford_HS}
\end{align}
where $Y_a$ are twistor coordinates, and $I_{ab}$ is the twistor metric. In both cases \eqref{eq:Clifford_HS}, the ordinary action of spacetime symmetries (realized as rotations in a higher-dimensional flat space) is implemented by the algebra's adjoint representation, i.e. by multiplication on both sides. This should raise a curiosity about the \emph{fundamental} representation: what if we multiply by the group element on one side only? In the case of Clifford algebra, this leads one to discover spinors. In the case of higher-spin algebra, it leads to the Penrose transform! Specifically, \emph{the Penrose transform is a CPT reflection in the fundamental representation of the higher-spin group}. In other words, \emph{the Penrose transform is a square root of CPT}:
\begin{align}
  \delta_x(Y)\star F(Y)\star\delta_x(Y) &= F(\text{CPT}\text{ of }Y\text{ around origin }x) \ ; \label{eq:CPT} \\ 
  \pm F(Y)\star i\delta_x(Y) &\equiv C(x;Y) = \text{Penrose transform of }F(Y)\text{ at the point }x \ . \label{eq:sqrt_CPT}
\end{align}
Here, $F(Y)$ is a spacetime-independent twistor function, $x$ is a bulk point, $C(x;Y)$ is a master field encoding a solution to the free massless field equations, and $\delta_x(Y)$ is a certain $x$-dependent delta function in twistor space. One may think of $\delta_x(Y)$ as a ``twistor-bulk propagator''. The factor of $\pm i$ in \eqref{eq:sqrt_CPT} is for later convenience.

We should point out that the statement \eqref{eq:sqrt_CPT} is both old and new. On one hand, it was always clear that the twistor formalism of higher-spin theory is closely related to the Penrose transform (for a relatively recent treatment, see \cite{Gelfond:2008td}). Also, right-multiplication by a delta-function as in \eqref{eq:sqrt_CPT} has long been recognized \cite{Didenko:2009td,Didenko:2012tv,Iazeolla:2011cb,Iazeolla:2012nf,Iazeolla:2017vng} as an important operation, relating the adjoint and ``twisted adjoint'' representations of higher-spin algebra, and allowing the construction of higher-spin invariants, as well as some explicit solutions to the Vasiliev equations. However, to our knowledge, it was never quite spelled out that this operation literally \emph{is} the Penrose transform, i.e. that it relates free massless fields to \emph{spacetime-independent} twistor functions. The reason for this is that the standard formulation of higher-spin theory works with ``twistors'' made up of spinors within a local orthonormal frame on a featureless base manifold. In such a framework, spacetime-independent twistor functions simply don't arise as a natural object. 

In contrast, in this paper, we work with global, spacetime-independent, Penrose-style twistors, associated with a background AdS\textsubscript{4} spacetime. Specifically, our approach to higher-spin theory is a linearized version of the reformulation \cite{Neiman:2015wma} of the \emph{full non-linear} Vasiliev equations on a fixed AdS\textsubscript{4} background. At the linearized level, the existence and utility of such a formulation is not surprising. At the non-linear level, the reformulation \cite{Neiman:2015wma} is a less trivial matter, as it manages to avoid complicating the field equations, and retains the full higher-spin gauge symmetry. This is possible in higher-spin theory (as opposed to GR), because spacetime translations are contained in the local gauge group along with rotations, \emph{independently from diffeomorphisms}. This in turn is related to the unfolded language of higher-spin theory, which bundles the fields' spacetime derivatives together with the fields themselves. 

\subsection{Holographic dual of the Penrose transform}

Coming back to eqs. \eqref{eq:CPT}-\eqref{eq:sqrt_CPT}, the next question is: can we find a context in which the $\sqrt{\text{CPT}}$ nature of the Penrose transform becomes manifest? It turns out that the answer is yes, and that it is intimately related to another ``square root'' relation -- the fact that fundamental higher-spin fields in the bulk are dual to \emph{quadratic} operators in the boundary CFT. In fact, the free vector model on the boundary is best expressed in a \emph{bilocal} language, in which the relatively complicated local operators $\phi(\ell)\overset{\leftrightarrow}{\nabla}\dots\overset{\leftrightarrow}{\nabla}\bar\phi(\ell)$ are replaced by the simple product $\phi(\ell)\bar\phi(\ell')$, where $\ell,\ell'$ are boundary points. Consider, then, a boundary-bilocal object in the higher-spin algebra -- a ``twistor-boundary-boundary propagator'': 
\begin{align}
 K(\ell,\ell';Y) = \frac{\sqrt{-2\ell\cdot\ell'}}{4\pi}\,\delta_{\ell}(Y)\star\delta_\ell'(Y) \ . \label{eq:K_summary}
\end{align}
On this object, it turns out that the Penrose transform acts explicitly as a ``square root'' of CPT, by applying CPT to \emph{one of the two} boundary points:
\begin{align}
\begin{split}
  i\delta_x(Y)\star K(\ell,\ell';Y) &= \pm K(\text{CPT}\text{ of }\ell\text{ around origin }x\ ,\ \ell'\ ;\ Y) \ ; \\
  K(\ell,\ell';Y)\star i\delta_x(Y) &= \pm K(\ell\ ,\ \text{CPT}\text{ of }\ell'\text{ around origin }x\ ;\ Y) \ .
\end{split} \label{eq:K_sqrt_CPT}
\end{align}
This property applies not only to CPT reflections, but to all of $SO(1,4)$, since the latter can be constructed (in (A)dS, but not in flat spacetime!) by combining CPT reflections around different origins. Thus, while $SO(1,4)$ is manifestly realized on arbitrary functions $f(Y)$ in the \emph{adjoint} representation of the HS algebra, it is also manifestly realized in the \emph{fundamental} representation when acting on $K(\ell,\ell';Y)$, by transforming one of the two boundary points $\ell,\ell'$. In particular, for the infinitesimal $SO(1,4)$ generators $M_{\mu\nu} = (-i/8)Y\gamma_{\mu\nu}Y$, we have:
\begin{align}
 \begin{split}
   M_{\mu\nu}\star K(\ell,\ell';Y) &= \ell_\mu\frac{\del K}{\del\ell^\nu} - \ell_\nu\frac{\del K}{\del\ell^\mu} \ ; \\
   -K(\ell,\ell';Y)\star M_{\mu\nu} &= \ell'_\mu\frac{\del K}{\del\ell'^\nu} - \ell'_\nu\frac{\del K}{\del\ell'^\mu} \ . 
 \end{split} \label{eq:K_sqrt_infinitesimal_summary}
\end{align} 
The $\sqrt{-\ell\cdot\ell'}$ prefactor in \eqref{eq:K_summary} is necessary for eqs. \eqref{eq:K_sqrt_CPT}-\eqref{eq:K_sqrt_infinitesimal_summary} to hold, and it gives $K(\ell,\ell';Y)$ the appropriate conformal weight for a two-point function of massless scalars on the boundary. The numerical factor in \eqref{eq:K_summary} is irrelevant to eqs. \eqref{eq:K_sqrt_CPT}-\eqref{eq:K_sqrt_infinitesimal_summary}, but is necessary for the CFT results below. As we will see, the sign ambiguities in \eqref{eq:K_sqrt_CPT} are inherent to the HS algebra.

Moving on now from geometry to physics, our main result is that while $\delta_x(Y)$ solves the linearized bulk theory, $K(\ell,\ell';Y)$ solves the boundary CFT! Specifically, we begin with the CFT action with $U(N)$ singlet, single-trace sources, written in the spirit of \cite{Das:2003vw} in a bilocal form:
\begin{align}
 S_{\text{CFT}}[\Pi(\ell',\ell)] = -\int d^3\ell\,\bar\phi_I\Box\phi^I - \int d^3\ell' d^3\ell\,\bar\phi_I(\ell')\Pi(\ell',\ell)\phi^I(\ell) \ . \label{eq:S_summary}
\end{align}
We then define a ``holographic dual of the Penrose transform'', which packages the sources $\Pi(\ell,\ell')$ into a twistor function $F(Y)$:
\begin{align}
 F(Y) &= \int d^3\ell\,d^3\ell'K(\ell,\ell';Y)\,\Pi(\ell',\ell) \ . \label{eq:bdry_transform_summary}
\end{align}
This allows us to write the partition function in the manifestly higher-spin-invariant form:
\begin{align}
 Z_{\text{CFT}}[F(Y)] \sim \exp\left(\frac{N}{4}\tr_\star\ln_\star[1+F(Y)]\right) \equiv \left(\textstyle\det_\star[1+F(Y)]\right)^{N/4} \ , \label{eq:Z_summary}
\end{align}
where ``$\tr_\star$'' stands for the HS-invariant trace operation $\tr_\star F(Y) = F(0)$. From \eqref{eq:Z_summary}, we can extract the expectation value of the bilocal operator $\phi^I(\ell)\bar\phi_I(\ell')$ in the presence of sources:
\begin{align}
  \left<\phi^I(\ell)\bar\phi_I(\ell')\right> = \frac{N}{4}\tr_\star\left(K(\ell',\ell;Y)\star[1 - F(Y) + \dots] \right) \ , \label{eq:phi_phi_summary}
\end{align}
where the dots indicate higher orders in the source $F(Y)$. This twistor formulation of the CFT makes global HS symmetry manifest, while doing away with the gauge redundancy of the sources $\Pi(\ell',\ell)$.  

Crucially, we will see that, up to some subtleties involving discrete symmetries, the twistor functions $F(Y)$ in the bulk and boundary pictures can be identified with each other. Specifically, we will show that, away from sources, the asymptotic boundary data of the linearized bulk solution \eqref{eq:sqrt_CPT} reproduces the linearized expectation values \eqref{eq:phi_phi_summary} of the CFT operators, once the latter are translated into local currents. Thus, the 2-point correlators (more precisely, the 2-bilocal correlators) of the partition function \eqref{eq:Z_summary} are directly associated with the linearized bulk solution. The higher-point functions in \eqref{eq:Z_summary} can then be interpreted as encoding the effects of bulk interactions.

We note that the relation between conformal 3d fields $\phi^I(\ell),\bar\phi_I(\ell')$ and the fundamental HS representation \eqref{eq:K_sqrt_infinitesimal_summary} was realized from different points of view in \cite{Shaynkman:2001ip,Iazeolla:2008ix}. The partition function in a form similar to \eqref{eq:Z_summary} was obtained previously in \cite{Gelfond:2013xt}. On the CFT side, the main difference between our approach and that of \cite{Gelfond:2013xt} is that the latter operates directly with the current operators, while we are making contact with the fundamental fields $\phi^I(\ell),\bar\phi_I(\ell')$, i.e. with the underlying local structure of the boundary theory.

\subsection{Avoiding contact terms} \label{sec:summary:contact}

In the HS holography literature, when one calculates the correlation functions $\langle j(\ell_1)\dots j(\ell_n)\rangle$ of local CFT operators, the calculation is usually restricted to separated points, i.e. $\ell_1,\dots,\ell_n$ are all taken to be distinct. By themselves, these are not enough to capture the value of $Z_{CFT}$ for a general finite configuration of sources. Indeed, to calculate such values, we would need integrals of the form $d^3\ell_1\dots d^3\ell_n$, where some of the points $\ell_1,\dots,\ell_n$ may coincide, though only on lower-dimensional submanifolds of the integration domain. Thus, the full partition function at the single-trace level requires also some knowledge of the correlators' behavior at coincident points. This extra requirement is similar to, but weaker than, a knowledge of the \emph{multi-trace} correlators: the latter are equivalent to simply making single-trace insertion points coincide, as opposed to the coincidence appearing as a lower-dimensional possibility in a larger integral. This distinction is a consequence of the simplicity of our particular CFT: if the source-free action contained any multi-trace couplings, we would have no choice but to always take multi-trace insertions into account.

To be more specific, there are two kinds of problems that we can encounter on coincident-point submanifolds. First, the separated-point correlator may not be integrable through these submanifolds. Second, the answer may violate gauge invariance, or, equivalently, current conservation. Fixing such problems requires regularization, as well as adding contact terms both in the action and in the definition of the currents. For example, when a charge current $\mathbf{j} = i\phi\overset{\leftrightarrow}{\mathbf{\nabla}}\bar\phi$ (suppressing $U(N)$ indices) is coupled to a gauge potential $\mathbf{A}$, the current's expectation value is divergent at points where $\mathbf{A}$ is nonzero. Specifically, the relevant 2-point function has a non-integrable $\sim 1/r^4$ short-distance singularity, which becomes $\sim 1/r^{2s+2}$ in the spin-$s$ case. This divergence is directly related to the fact that the true conserved current contains an extra contact term $\mathbf{A}\phi\bar\phi$; in other words, the derivative in the definition of $\mathbf{j}$ must be gauge-covariantized. The $\mathbf{A}\phi\bar\phi$ term has its own short-distance singularity, which cancels the previous one and leaves us with a finite \& conserved current.

Most of these issues are resolved automatically by switching to the bilocal language \eqref{eq:S_summary}. There, we only ever find-short distance singularities of the form $\sim 1/r$ (the fundamental propagator of the $\phi$ fields), which is integrable, and therefore doesn't require regularization. From the local point of view, the bilocal language can be viewed as an extreme form of point-split regularizartion. Conversely, from the bilocal point of view, the local language corresponds to a singular choice of gauge, where the source $\Pi(\ell',\ell)$ is distributional with support on $\ell=\ell'$.

That being said, the bilocal language does not solve everything. In particular, given some values of the bilocal source, we may still wish to know the expectation value of a local current. It turns out that upon naive calculation, the resulting current isn't locally conserved: the two points in the bilocal $\Pi(\ell',\ell)$ act as a source/sink pair (as one can see by examining eq. \eqref{eq:example_j} below). Thus, the need for contact terms arises whenever we're interested in a local expectation value, even if the sources are bilocal.

Finally, we come to the fully nonlocal twistor formulation \eqref{eq:Z_summary} of the partition function. Here, we find that the need for contact terms seems to disappear entirely. This should not be too surprising, if we put two facts together:
\begin{enumerate}
	\item The points at which the (local or bilocal) sources are non-vanishing are gauge-dependent. In particular, at any given point, one can gauge away the value of a spin-$s$ gauge potential and its first $2s-2$ derivatives. 
	\item The twistor language does away with both locality and gauge redundancy.
\end{enumerate}
Specifically, as we'll discuss in section \ref{sec:holography:general_currents}, the currents that can be derived from a twistorial expression of the form \eqref{eq:Z_summary} are always automatically conserved. 

\section{Spacetime and twistor geometry} \label{sec:geometry}

In this section, we present some elements of geometry in the $EAdS_4$ bulk, its 3d boundary, and twistor space. Throughout, we view the bulk and boundary as embedded in a flat 5d spacetime. Similar embedding-space approaches to higher-spin theory and holography may be found e.g. in \cite{Bekaert:2012vt,Didenko:2012vh}, in the context of general dimensions. Those approaches employ a tensor formalism, while our emphasis will be on spinors and twistors. In particular, this section will focus on the embedding of bulk and boundary spinor spaces within the global twistor space. 

\subsection{Spacetime} \label{sec:geometry:spacetime}

\subsubsection{Bulk and boundary}

We define $EAdS_4$ as the hyperboloid of future-pointing unit timelike vectors in flat 5d Minkowski space $\bbR^{1,4}$: 
\begin{align}
 EAdS_4 = \left\{x^\mu\in\bbR^{1,4}\, |\, x_\mu x^\mu = -1, \ x^0 > 0 \right\} \ . \label{eq:EAdS}
\end{align}
The metric $\eta_{\mu\nu}$ of $\bbR^{1,4}$ has signature $(-,+,+,+,+)$. We use indices $(\mu,\nu,\dots)$ for $\bbR^{1,4}$ vectors, which we raise and lower using $\eta_{\mu\nu}$. The isometry group of $EAdS_4$ is just the rotation group $O(1,4)$ in the 5d spacetime (more precisely -- the component $O^\uparrow(1,4)$ that preserves time orientation).

The tangent space at a point $x\in EAdS_4$ consists simply of the vectors $v^\mu$ that satisfy $x\cdot v = 0$. The $EAdS_4$ metric at $x$ can be identified with the projector onto this tangent space:
\begin{align}
 q_{\mu\nu}(x) = \eta_{\mu\nu} + x_\mu x_\nu \ .
\end{align}
The covariant derivative of vectors in $EAdS_4$ can be defined as the flat $\bbR^{1,4}$ derivative, followed by a projection back onto the hyperboloid:
\begin{align}
 \nabla_\mu v_\nu = q_\mu^\rho(x)\, q_\nu^\sigma(x)\, \del_\rho v_\sigma \ .
\end{align}

In addition to the ambient $\bbR^{1,4}$ picture, it is sometimes useful to use an intrinsic coordinate system for $EAdS_4$. Of particular interest are the Poincare coordinates:
\begin{align}
 x^\mu(z,\mathbf{r}) = \frac{1}{z}\left(\frac{1+z^2+r^2}{2}, \mathbf{r}, \frac{1-z^2-r^2}{2} \right) \ , \label{eq:Poincare}
\end{align}
where $\mathbf{r}$ is a flat 3d coordinate, and the metric reads:
\begin{align}
 dx_\mu dx^\mu = \frac{dz^2 + d\mathbf{r}\cdot d\mathbf{r}}{z^2} \ .
\end{align}

The asymptotic boundary of $EAdS_4$ is the conformal 3-sphere of future-pointing null directions in $\bbR^{1,4}$. Thus, we represent boundary points by null vectors $\ell^\mu$, with the equivalence $\ell^\mu\cong\lambda\ell^\mu$. The $O(1,4)$ symmetry group then becomes the conformal symmetry of the boundary. The limit where a bulk point $x$ approaches the boundary can be represented as an extreme boost in $\bbR^{1,4}$, where the unit vector $x^\mu$ approaches a null direction $\ell^\mu$ as:
\begin{align}
 x^\mu\rightarrow \ell^\mu/z \ , \quad z\rightarrow 0 \ . \label{eq:limit}
\end{align}

One can fix the conformal frame on the boundary by choosing a section of the $\bbR^{1,4}$ lightcone. Perhaps the most convenient is the flat section:
\begin{align}
 \ell^\mu(\mathbf{r}) = \left(\frac{1+r^2}{2}, \mathbf{r}, \frac{1-r^2}{2} \right) \ , \label{eq:flat}
\end{align}
which can be viewed as the bulk-to-boundary limit \eqref{eq:limit} of the Poincare coordinates \eqref{eq:Poincare}. The section \eqref{eq:flat} can be defined as the intersection of the lightcone $\ell\cdot\ell = 0$ with the null hyperplane:
\begin{align}
\ell\cdot n = -\frac{1}{2} \ ; \quad n^\mu = \left(\frac{1}{2},\mathbf{0},-\frac{1}{2}\right) \label{eq:flat_hyperplane}
\end{align}

The metric on the flat section \eqref{eq:flat} is simply $d\ell_\mu d\ell^\mu = d\mathbf{r}\cdot d\mathbf{r}$. In particular, the $\bbR^{1,4}$ scalar product $\ell\cdot\ell'$ is directly related to the 3d Euclidean distance in the frame \eqref{eq:flat}:
\begin{align}
 \ell\cdot\ell' = -\frac{1}{2}\left(\mathbf{r} - \mathbf{r'}\right)^2 \ . \label{eq:flat_distance}
\end{align}

\subsubsection{Massless scalars and conserved currents on the boundary} \label{sec:geometry:spacetime:currents}

Boundary quantities with conformal weight $\Delta$ are represented by functions $f(\ell)$ on the lightcone, subject to the homogeneity condition $f(\lambda\ell) = \lambda^{-\Delta}f(\ell)$, or, equivalently:
\begin{align}
 \ell^\mu\frac{\del}{\del\ell^\mu}f(\ell) = -\Delta f(\ell) \ . \label{eq:homogeneity}
\end{align}
In particular, a free massless scalar on the 3d boundary has conformal weight $\Delta=1/2$. Scalars with this weight admit a conformally covariant Laplacian $\Box$, which in the $\bbR^{1,4}$ language is given simply by \cite{Eastwood:2002su}:
\begin{align}
 \Box\phi(\ell) = \frac{\del\phi(\ell)}{\del\ell_\mu\del\ell^\mu} \ . \label{eq:Laplacian}
\end{align}
Here, it's assumed that we've extended the function $\phi(\ell)$ into non-null values of $\ell^\mu$, where it remains subject to the homogeneity condition \eqref{eq:homogeneity}. The Laplacian \eqref{eq:Laplacian} does not otherwise depend on this artificial extension of $\phi(\ell)$ into $\ell\cdot\ell\neq 0$, as it vanishes for any function that is zero at $\ell\cdot\ell = 0$:
\begin{align}
 \Box\left((\ell\cdot\ell) f(\ell)\right) = 0 \ \ \text{at} \ \  \ell\cdot\ell = 0 \ , \ \ \text{for any } f(\ell) \text{ with weight } \Delta = 5/2 \ .
\end{align}
One can verify explicitly that eq. \eqref{eq:Laplacian} defines the usual 3d Laplacian on the flat section \eqref{eq:flat}.

Boundary currents of various spin and their conservation laws are also easy to describe in the $O(1,4)$-covariant framework. A spin-$s$ current is represented by a totally symmetric and traceless tensor $j^{\mu_1\dots\mu_s}$. To bring the tensor's indices from $\bbR^{1,4}$ down to the boundary's 3d tangent space, we impose a constraint and an equivalence relation:
\begin{align}
 \ell_{\mu_1}j^{\mu_1\mu_2\dots\mu_s} &= 0 \ ; \label{eq:current_tensor_constraint} \\
 j^{\mu_1\mu_2\dots\mu_s} &\cong j^{\mu_1\mu_2\dots\mu_s} + \ell^{(\mu_1}\theta^{\mu_2\dots\mu_s)} \ , \label{eq:current_tensor_equivalence}
\end{align}
where $\theta^{\mu_1\dots\mu_{s-1}}$ is a totally symmetric and traceless tensor satisfying $\ell_{\mu_1}\theta^{\mu_1\mu_2\dots\mu_{s-1}} = 0$. The presence of tensor indices makes the notion of conformal weight a bit subtle. In this paper, our tensor indices lie in the $\bbR^{1,4}$ ambient space, and we define the conformal weight $\Delta$ via $j^{\mu_1\dots\mu_s}(\lambda\ell) = \lambda^{-\Delta}j^{\mu_1\dots\mu_s}(\ell)$. For the corresponding tensor with indices in the boundary's tangent or cotangent bundle, this implies a conformal weight of $\Delta+s$ or $\Delta-s$, respectively. For a spin-$s$ tensor $j^{\mu_1\dots\mu_s}$ with the particular weight $\Delta = s+1$, one can define a conformally covariant divergence:
\begin{align}
 (\operatorname{div} j)^{\mu_1\dots\mu_{s-1}} = \frac{\del j^{\mu_1\dots\mu_{s-1}\mu_s}}{\del\ell^{\mu_s}} \ , \label{eq:tensor_div}
\end{align}
where we again extend $j^{\mu_1\dots\mu_s}(\ell)$ away from $\ell\cdot\ell = 0$, while maintaining the constraint \eqref{eq:current_tensor_constraint} and the homogeneity condition $j^{\mu_1\dots\mu_s}(\lambda\ell) = \lambda^{-s-1}j^{\mu_1\dots\mu_s}(\ell)$. To see that the result doesn't otherwise depend on this artificial extension, we note that eq. \eqref{eq:tensor_div} can be rewritten in terms of the derivative $\ell\wedge(\del/\del\ell)$, which only acts tangentially to the $\ell\cdot\ell = 0$ lightcone:
\begin{align}
 \left(\ell_\nu\frac{\del}{\del\ell^{\mu_s}} - \ell_{\mu_s}\frac{\del}{\del\ell^\nu} \right) j^{\mu_1\dots\mu_{s-1}\mu_s} 
   = (\operatorname{div} j)^{\mu_1\dots\mu_{s-1}}\ell_\nu + j^{\mu_1\dots\mu_{s-1}}{}_\nu \ .
\end{align}
It remains to verify that the formula \eqref{eq:tensor_div} is consistent with the equivalence relation \eqref{eq:current_tensor_equivalence}. It is here that the conformal weight $\Delta = s+1$ will be important. One must be careful to extend eq. \eqref{eq:current_tensor_equivalence} away from $\ell\cdot\ell = 0$ in a way that doesn't conflict with the constraint \eqref{eq:current_tensor_constraint}. To do this, we introduce a fixed null vector $n\neq\ell$, and replace $\ell^\mu$ in \eqref{eq:current_tensor_equivalence} with:
\begin{align}
 \tilde\ell^\mu = \ell^\mu - \frac{\ell\cdot\ell}{\ell\cdot n}\,n^\mu \ .
\end{align}
One then finds that the divergence \eqref{eq:tensor_div} is indeed consistent with \eqref{eq:current_tensor_equivalence}, via:
\begin{align}
 \delta j^{\mu_1\dots\mu_s} = \tilde\ell^{(\mu_1}\theta^{\mu_2\dots\mu_s)} \quad \Longrightarrow \quad 
 \delta(\operatorname{div} j)^{\mu_1\dots\mu_{s-1}} = \frac{s-1}{s}\,\ell^{(\mu_1}\frac{\del}{\del\ell^\nu}\theta^{\mu_2\dots\mu_{s-1})\nu} \ .
\end{align}

\subsection{Twistors} \label{section:geometry:twistors}

Here, we introduce spinors and twistors in $EAdS_4$ from the viewpoint described in \cite{Neiman:2013hca}. Our focus here is on algebraic properties; see \cite{Neiman:2013hca} for a more detailed geometric perspective.

The twistors of $EAdS_4$ are just the 4-component Dirac spinors of the isometry group $SO(1,4)$. We use indices $(a,b,\dots)$ for twistors. The twistor space is equipped with a symplectic metric $I_{ab}$, which is used to raise and lower indices via:
\begin{align}
 U_a = I_{ab}U^b \ ; \quad U^a = U_b I^{ba} \ ; \quad I_{ac}I^{bc} = \delta_a^b \ .
\end{align}
Tensor and twistor indices are related through the gamma matrices $(\gamma_\mu)^a{}_b$, which satisfy the Clifford algebra $\{\gamma_\mu,\gamma_\nu\} = -2\eta_{\mu\nu}$. These 4+1d gamma matrices can be realized as the usual 3+1d ones, with the addition of $\gamma_5$ (in our notation, $\gamma_4$) for the fifth direction in $\bbR^{1,4}$. In $2\times 2$ block notation, the matrices $I_{ab}$ and $(\gamma_\mu)^a{}_b$ can be represented e.g. as:
\begin{align}
 \begin{split}
   I_{ab} &= \begin{pmatrix} 0 & -i\sigma_2 \\ -i\sigma_2 & 0 \end{pmatrix} \ ; \\
   (\gamma^0)^a{}_b &= \begin{pmatrix} 0 & 1 \\ 1 & 0 \end{pmatrix} \ ; \quad 
   (\gamma^4)^a{}_b = \begin{pmatrix} 0 & -1 \\ 1 & 0 \end{pmatrix} \ ; \quad 
   (\gamma^k)^a{}_b = \begin{pmatrix} -i\sigma^k & 0 \\ 0 & i\sigma^k \end{pmatrix} \ , \label{eq:gamma_null}
 \end{split}
\end{align}
where $\sigma^k$ with $k=1,2,3$ are the Pauli matrices. The representation \eqref{eq:gamma_null} is geared to simplify the ``null'' matrices $\gamma_0\pm\gamma_4$. An alternative representation, which simplifies the ``timelike'' matrix $\gamma_0$, reads:
\begin{align}
\begin{split}
I_{ab} &= \begin{pmatrix} -i\sigma_2 & 0 \\ 0 & i\sigma_2 \end{pmatrix} \ ; \\
(\gamma^0)^a{}_b &= \begin{pmatrix} 1 & 0 \\ 0 & -1 \end{pmatrix} \ ; \quad 
(\gamma^4)^a{}_b = \begin{pmatrix} 0 & -1 \\ 1 & 0 \end{pmatrix} \ ; \quad 
(\gamma^k)^a{}_b = \begin{pmatrix} 0 & i\sigma^k \\ i\sigma^k & 0 \end{pmatrix} \ . \label{eq:gamma_timelike}
\end{split}
\end{align}
The matrices $\gamma^\mu_{ab}$ are antisymmetric and traceless in their twistor indices. We define the antisymmetric product of gamma matrices as:
\begin{align}
 \gamma^{\mu\nu}_{ab} \equiv \gamma^{[\mu}_{ac}\gamma^{\nu]c}{}_b \ .
\end{align}
The $\gamma^{\mu\nu}_{ab}$ are symmetric in their twistor indices. We use the matrices $\gamma_\mu^{ab}$ to convert between 4+1d vectors and traceless bitwistors as:
\begin{align}
 \xi^{ab} = \gamma_\mu^{ab}\xi^\mu \ ; \quad \xi^\mu = -\frac{1}{4}\gamma^\mu_{ab}\xi^{ab} \ . \label{eq:conversion_5d}
\end{align}
Similarly, $\gamma_{\mu\nu}^{ab}$ can be used to convert between bivectors and symmetric twistor matrices:
\begin{align}
 f^{ab} = \frac{1}{2}\gamma_{\mu\nu}^{ab}f^{\mu\nu} \ ; \quad f^{\mu\nu} = \frac{1}{4}\gamma^{\mu\nu}_{ab} f^{ab} \ . \label{eq:conversion_bivectors}
\end{align}
Useful identities include: 
\begin{align}
 \begin{split}
   &\gamma^\mu_{ab}\gamma_\nu^{ab} = -4\delta^\mu_\nu \ ; \quad \gamma^{\mu\nu}_{ab}\gamma_{\rho\sigma}^{ab} = 8\delta^{[\mu}_{[\rho}\delta^{\nu]}_{\sigma]} \ ; \quad
   \gamma_\mu^{ab}\gamma^\mu_{cd} = I^{ab}I_{cd} - 4\delta^{[a}_{[c} \delta^{b]}_{d]} \ ; \quad \gamma_{\mu\nu}^{ab}\gamma^{\mu\nu}_{cd} = 8\delta^{(a}_{(c} \delta^{b)}_{d)} \ ; \\
   &\epsilon^{abcd} = 3I^{[ab}I^{cd]} \ ; \quad \epsilon^{abcd}I_{cd} = 2I^{ab} \ ; \quad \epsilon^{abcd}\gamma^\mu_{cd} = -2\gamma^{\mu ab} \ ; \quad
   \gamma_\mu^{[ab}\gamma_\nu^{cd]} = \frac{1}{3}\eta_{\mu\nu}\epsilon^{abcd} \ .
 \end{split} \label{eq:twistor_identities}
\end{align}
Here, $\epsilon^{abcd}$ is the totally antisymmetric symbol with inverse $\epsilon_{abcd} = 3I_{[ab}I_{cd]}$, such that $\epsilon_{abcd}\epsilon^{abcd} = 4!$. The metric $I_{ab}$ has unit determinant with respect to $\epsilon^{abcd}$. We use $\epsilon^{abcd}$ to define a measure on twistor space via:
\begin{align}
 d^4U \equiv \frac{\epsilon_{abcd}}{4!(2\pi)^2}\,dU^a dU^b dU^c dU^d \ . \label{eq:twistor_measure}
\end{align}
Here and elsewhere, we include $2\pi$ factors in the measure, in such a way that they will not appear explicitly in our Fourier and Gaussian integrals. Note that our choice for the overall sign of $\epsilon^{abcd}$ is the opposite from that in \cite{Neiman:2013hca}, and indeed, in the basis \eqref{eq:gamma_null}, we get $\epsilon^{1234} = -1$. This choice will end up being more convenient for relations such as \eqref{eq:measure_product_bulk}.

\subsubsection{Index-free notation} \label{sec:geometry:twistors:index_free}

In order to streamline the formulas below, we now introduce some index-free notation for products in $\bbR^{1,4}$ and in twistor space. $x\cdot x$ will represent the scalar product $x_\mu x^\mu$ in $\bbR^{1,4}$. The twistor matrices $\delta_a^b$ and $(\gamma_\mu)^a{}_b$ will be written in index-free notation as $1$ and $\gamma_\mu$. Combined with the index conversion \eqref{eq:conversion_5d}, this means that the matrix $(x^\mu\gamma_\mu)^a{}_b$ for a vector $x^\mu\in\bbR^{1,4}$ will be written simply as $x$ (this is just the Feynman slash convention, without the slash). Products in the index-free notation imply bottom-to-top index contractions. So, e.g. for two twistors $U^a,V^a$ and two vectors $\ell^\mu,x^\mu$, we have:
\begin{align}
 \begin{split}
   &UV \equiv U_a V^a = -I_{ab}U^a V^b \ ; \quad \ell\cdot x \equiv \ell_\mu x^\mu = -\frac{1}{4}\tr(\ell x) \ ; \\
   &(xU)^a \equiv x^a{}_b U^b \ ; \quad U\ell xU \equiv U_a\ell^a{}_b x^b{}_c U^c = -\ell_\mu x_\nu\gamma^{\mu\nu}_{ab} U^a U^b \ .
 \end{split}
\end{align}
A product $U\Gamma_1\dots\Gamma_n V$, where $U$ and $V$ are twistors and the matrices $\Gamma_1,\dots,\Gamma_n$ are either symmetric or antisymmetric, can be reversed as follows:
\begin{align}
V\Gamma_n\dots\Gamma_1 U = (-1)^{n_\text{sym}+1}(U\Gamma_1\dots\Gamma_n V) \ , \label{eq:reverse}
\end{align}
where $n_\text{sym}$ is the number of symmetric matrices among the $\Gamma_1,\dots,\Gamma_n$.

\subsubsection{Twistor integrals} \label{sec:geometry:twistors:integrals}

In calculations below, we will need to evaluate integrals over twistor space, as well as over various spinor subspaces. These integrals are somewhat delicate, because the relevant spaces are complex, and one has to worry about appropriate integration contours. To some extent, this is a result of our choice of signature: in Lorentzian $AdS_4$, the twistors and boundary spinors (but not the bulk spinors) have a natural real structure. However, this real structure doesn't necessarily help, because the natural real contours may not be the ones along which the integrals converge. Luckily, the only integrals we will need explicitly are of delta functions and Gaussians. These can be defined by analytical continuation from appropriate real-line integrals. 

The first integral formula that we'll need is:
\begin{align}
 \int d^4U d^4V f(U)\, e^{iUV} = f(0) \ . \label{eq:delta_integral_raw}
\end{align}
This can equivalently be written as:
\begin{align}
 \int d^4U\,\delta(U) f(U) = f(0) \ , \label{eq:delta_integral}
\end{align}
where the twistor delta function is defined as:
\begin{align}
 \delta(U) = \int d^4V e^{iVU} \ . \label{eq:delta_U}
\end{align}
The second twistor integral that we will use is the Gaussian:
\begin{align}
 \int d^4U\, e^{(UAU)/2} = \frac{\pm 1}{\sqrt{\det A}} \ ; \quad \det A = \frac{1}{8}\left(\tr A^2\right)^2 - \frac{1}{4}\tr A^4 \ , \label{eq:Gaussian}
\end{align}
where $A_{ab}$ is a symmetric twistor matrix, and we use its tracelessness for the last expression in \eqref{eq:Gaussian}. Note that the $2\pi$ factors are already taken care of by the definition \eqref{eq:twistor_measure} of the measure. 

The sign in \eqref{eq:Gaussian} is ambiguous due to the square root, and in general will depend on how exactly we analytically continue from the case of a real contour and real negative-definite $A_{ab}$. In fact, we'll see that in the context of the HS symmetry group, this sign ambiguity is crucial, and cannot be globally fixed. Specifically, within the HS group, the subgroup $SO(1,4)$ of ordinary spacetime symmetries is represented by twistor Gaussians, and its topology is only consistent when the sign ambiguity \eqref{eq:Gaussian} is taken into account.

Finally, we note that the sign ambiguity in Gaussian integrals also reflects on the delta function \eqref{eq:delta_U}. The integral in \eqref{eq:delta_U} can be regularized and evaluated by inserting a broad Gaussian into the integrand. However, the result of this Gaussian integral is only defined up to sign. Therefore, while the integral \eqref{eq:delta_integral} involving $\delta(U)$ is well-defined, $\delta(U)$ itself is defined as a limit of ordinary functions only up to sign. An alternative way to see this is to define $\delta(U)$ as the limit of a series of ever-narrowing Gaussians, which are constrained to have a unit integral. Since these integrals are only defined up to sign, the same is true for the series that limits to $\delta(U)$.

\subsection{Bulk spinors}

When we choose a point $x\in EAdS_4$, the Dirac representation of $SO(1,4)$ becomes identified with the Dirac representation of the rotation group $SO(4)$ at $x$. It then decomposes into left-handed and right-handed Weyl spinor representations, corresponding to $SO(4) = SO(3)_L\times SO(3)_R$. The decomposition is accomplished by a pair of projectors:
\begin{align}
 \begin{split}
   P_L{}^a{}_b(x) &= \frac{1}{2}\left(\delta^a_b - x^\mu\gamma_\mu{}^a{}_b \right) = \frac{1}{2}\left(\delta^a_b - x^a{}_b \right) \ ; \\
   P_R{}^a{}_b(x) &= \frac{1}{2}\left(\delta^a_b + x^\mu\gamma_\mu{}^a{}_b \right) = \frac{1}{2}\left(\delta^a_b + x^a{}_b \right) \ . \label{eq:projectors}
 \end{split}
\end{align}
These serve as an $x$-dependent version of the familiar chiral projectors in $\bbR^4$. We note that $P_L$ and $P_R$ get interchanged under the ``antipodal map'' $x^\mu\rightarrow-x^\mu$. In the Euclidean AdS context, this is a formal operation that takes us away from the hyperboloid \eqref{eq:EAdS} and into its $x^0<0$ counterpart. 

Given a twistor $U^a$, we denote its left-handed and right-handed components at $x$ as $u_{L/R}^a(x) = (P_{L/R}){}^a{}_b(x)U^b$. As in our treatment of tensors, it is possible to use the $(a,b,\dots)$ indices for both $SO(4,1)$ and $SO(4)$ Dirac spinors. The projectors $P^L_{ab}(x)$ and $P^R_{ab}(x)$ serve as the spinor metrics for the left-handed and right-handed Weyl spinor spaces. For a 2d spinor space, a symplectic metric also acts as a measure, i.e. we can define:
\begin{align}
 d^2u_L \equiv \frac{P^L_{ab}(x)}{2(2\pi)}\,dU^a dU^b \ ; \quad d^2u_R \equiv \frac{P^R_{ab}(x)}{2(2\pi)}\,dU^a dU^b \ . \label{eq:measure_chiral}
\end{align}
Alternatively, the measures can be defined as the inverses of $P_L^{ab}$ and $P_R^{ab}$, as in:
\begin{align}
 \frac{du_L^a du_L^b}{2\pi} \equiv P_L^{ab}(x)\,d^2u_L \ ; \quad \frac{du_R^a du_R^b}{2\pi} \equiv P_R^{ab}(x)\,d^2u_R \ . \label{eq:measure_chiral_inverse}
\end{align}
The two chiral spinor measures combine to form the twistor measure \eqref{eq:twistor_measure}, via: 
\begin{align}
 d^4U = d^2u_L d^2u_R \ . \label{eq:measure_product_bulk}
\end{align}

The power of this formalism for describing spinors is that the twistors, i.e. the spinors of $\bbR^{1,4}$, are flat: we can transport them freely from one point in $EAdS_4$ to another. What changes from point to point is the twistor's decomposition into left-handed and right-handed spinors. In particular, the covariant derivative for Weyl spinors in $EAdS_4$ can be constructed by embedding the spinor inside a twistor, taking the flat $\bbR^{1,4}$ derivative, and projecting back into the appropriate spinor space. For e.g. a left-handed spinor field $\psi_L^a(x)$, this can be written as:
\begin{align}
 \nabla_\mu\psi^a_L(x) = q_\mu^\nu(x) P_L{}^a{}_b(x)\,\del_\nu\psi_L^b(x) \ . \label{eq:spinor_covariant}
\end{align}
An important special case is the covariant derivative of the left-handed and right-handed components $y_L(x),y_R(x)$ of a \emph{spacetime-independent} twistor $Y$:
\begin{align}
 \nabla_\mu\, y_L^a = -\frac{1}{2}(\gamma_\mu)^a{}_b\,y_R^b \ ; \quad \nabla_\mu\, y_R^a = \frac{1}{2}(\gamma_\mu)^a{}_b\,y_L^b \ .
\end{align}
This is just Penrose's twistor equation, in the presence of a cosmological constant.

A vector $\xi^\mu\in\bbR^{1,4}$, when evaluated at a point $x\in EAdS_4$, decomposes into an $O(4)$ scalar (the radial component, encoded by the scalar product $\xi\cdot x$) and an $O(4)$ vector (the tangential component, encoded by the vector $\xi^\mu + (\xi\cdot x)x^\mu$ or the bivector $\xi^{[\mu}x^{\nu]}$). For the twistor matrix $\xi = \xi^\mu\gamma_\mu$, this decomposition can be expressed in terms of chiral projections of the twistor indices:
\begin{align}
 O(4)\text{ scalar:} \quad &P_L\xi P_L = (\xi\cdot x)P_L \ ; \quad P_R\xi P_R = -(\xi\cdot x)P_R \ ; \\
 O(4)\text{ vector:} \quad &P_L\xi P_R + P_R\xi P_L = \xi + (\xi\cdot x)x \ ; \quad P_L\xi P_R - P_R\xi P_L = \frac{1}{2}(\xi x - x\xi) \ . \label{eq:vector}
\end{align}
In particlar, displacements $dx^\mu$ along the $EAdS_4$ hyperboloid have only mixed-chirality components, as in \eqref{eq:vector}. 

\subsection{Boundary spinors} \label{sec:geometry:spinors_boundary}

At a boundary point $\ell^\mu$, the decomposition of twistor space is somewhat different. While the 4d bulk has two Weyl spinor spaces at each point, the 3d boundary has a single (2-component) Dirac spinor space. Let us now describe how this spinor space arises from the $\bbR^{1,4}$, twistorial perspective.

In the asymptotic limit \eqref{eq:limit}, both the left-handed and right-handed projectors degenerate into multiples of $\ell^{ab}$:
\begin{align}
 P_L^{ab}(x)\rightarrow -\frac{1}{z}P^{ab}(\ell) \ ; \quad P_R^{ab}(x)\rightarrow \frac{1}{z}P^{ab}(\ell) \ , \label{eq:limit_spinor_spaces}
\end{align}
where we've defined:
\begin{align}
 P^{ab}(\ell) \equiv \frac{1}{2}\ell^{ab} \ . \label{eq:P_ell}
\end{align}
Thus, the two subspaces $P_L(x)$ and $P_R(x)$ degenerate into a single subspace $P(\ell)$, spanned by the bitwistor $P^{ab}(\ell) \sim \ell^{ab}$. Equivalently, $P(\ell)$ is the subspace annihilated by the matrix $\ell^a{}_b$:
\begin{align}
 u^a\in P(\ell) \quad \Longleftrightarrow \quad \ell^{[ab}u^{c]} = 0 \quad \Longleftrightarrow \quad \ell^a{}_b u^b = 0 \ .
\end{align}
The subspace $P(\ell)$ can be identified as the spinor space on the 3d boundary. Though $P(\ell)$ is null under the twistor metric $I_{ab}$, one can use the inverse of the matrix \eqref{eq:P_ell} to define a metric and a measure $d^2u$ on $P(\ell)$, in analogy with the bulk definition \eqref{eq:measure_chiral_inverse}:
\begin{align}
 \frac{du^a du^b}{2\pi} \equiv P^{ab}(\ell)\, d^2u = \frac{1}{2}\ell^{ab} d^2u \ . \label{eq:measure}
\end{align}
The measure $d^2u$ scales inversely with the null vector $\ell^\mu$, i.e. it has conformal weight $\Delta = 1$. We should therefore think of $P(\ell)$ as the space of boundary \emph{co}spinors, i.e. the square roots of boundary covectors. 

The space of \emph{contravariant} boundary spinors, i.e. the square roots of boundary vectors, is the space $P^*(\ell)$  dual to $P(\ell)$ under the twistor metric. It is easy to see that this is the \emph{quotient space} of twistors modulo terms in $P(\ell)$:
\begin{align}
(u^*)^a \cong (u^*)^a + u^a \ , \quad u^a \in P(\ell) \ . \label{eq:dual_space}
\end{align}
$P^*(\ell)$ can be equipped with a metric and measure inversely related to that of \eqref{eq:measure}, i.e. given simply by the matrix \eqref{eq:P_ell}:
\begin{align}
d^2u^* \equiv \frac{P_{ab}(\ell)}{2(2\pi)}\,(du^*)^a (du^*)^b = \frac{\ell_{ab}}{8\pi}\,(du^*)^a (du^*)^b \ , \label{eq:dual_measure}
\end{align}
with conformal weight $\Delta = -1$. Multiplication by the matrix \eqref{eq:P_ell} defines a mapping between $P(\ell)$ and $P^*(\ell)$, via:
\begin{align}
(u^*)^a \in P^*(\ell) \quad \longleftrightarrow \quad P^a{}_b(\ell) (u^*)^b = \frac{1}{2}\ell^a{}_b (u^*)^b \in P(\ell) \ . \label{eq:u_u*}
\end{align}
This mapping is consistent with the measures \eqref{eq:measure},\eqref{eq:dual_measure}. It can be viewed as the map between boundary spinors and cospinors via the spinor metric \eqref{eq:dual_measure}. It should be stressed that the spinor spaces $P(\ell)$ and $P^*(\ell)$ depend only on the \emph{direction} of $\ell^\mu$, which corresponds to the choice of boundary point. However, the measures \eqref{eq:measure},\eqref{eq:dual_measure} and the mapping \eqref{eq:u_u*} depend also on the \emph{scaling} of $\ell^\mu$, which corresponds to a choice of conformal frame.

Note that for \emph{bulk} spinors, there was no need for such subtleties. There, we have no arbitrary rescaling of the spinor metrics, and the chiral spinor spaces $P_L(x),P_R(x)$ are the same as their duals under the twistor metric $P_L^*(x),P_R^*(x)$. In particular, the measure \eqref{eq:dual_measure} can be viewed as the boundary limit of the bulk spinor measures \eqref{eq:measure_chiral}.

At a bulk point $x$, an arbitrary twistor $U$ has a well-defined decomposition $U = u_L + u_R$. This is no longer true at a boundary point $\ell$: here, $U$ has a well-defined projection $u^*\in P^*(\ell)$, but its ``$P(\ell) $ component'' is ambiguous. However, one can \emph{span} the twistor space by first choosing $u^*$, and then spanning the equivalence class \eqref{eq:dual_space} by varying $u\in P(\ell)$. In this context, the two spinor measures \eqref{eq:measure},\eqref{eq:dual_measure} can be combined into the global twistor measure. From the identity $\epsilon_{abcd}\ell^{cd} = -2\ell_{ab}$, one can derive the explicit formula:
\begin{align}
 d^4U = -d^2u\,d^2u^* \ . \label{eq:measure_product_boundary}
\end{align}

\subsubsection{Boundary currents in spinor form} \label{sec:geometry:spinors_boundary:currents}

The spinor language is especially well-suited for describing boundary currents of arbitrary spin and their conservation laws. A spin-$s$ boundary object (i.e. a rank-$s$ totally traceless and symmetric tensor) can be described by a totally symmetric rank-$2s$ spinor $j_\ell$ with indices in $P(\ell)$:
\begin{align}
 \ell^b{}_{a_1} j_\ell^{a_1 a_2\dots a_{2s}} = 0 \ ,
\end{align}
or a totally symmetric rank-$2s$ spinor $j_*$ with indices in $P^*(\ell)$:
\begin{align}
 j_*^{a_1 a_2\dots a_{2s}} \cong j_*^{a_1 a_2\dots a_{2s}} + \ell^{(a_1}{}_c\,\lambda^{a_2\dots a_{2s})c} 
 \ \ \text{for any} \ \ \lambda^{a_2\dots a_{2s} c} \ . \label{eq:spinor_equivalence}
\end{align}
These two representations are related by the mapping \eqref{eq:u_u*}, i.e. by the spinor metric at $\ell$:
\begin{align}
 j_\ell^{a_1\dots a_{2s}} = \frac{1}{4^s}\,\ell^{a_1}{}_{b_1}\dots\ell^{a_{2s}}{}_{b_{2s}} j_*^{b_1\dots b_{2s}} \ . \label{eq:j_P_j*}
\end{align}
Note that neither $j_\ell$ nor $j_*$ is the direct translation into twistor indices of the boundary tensor $j^{\mu_1\dots\mu_s}$ from section \ref{sec:geometry:spacetime:currents}:
\begin{align}
 j^{a_1 b_1\dots a_s b_s} = \gamma_{\mu_1}^{a_1 b_1}\dots\gamma_{\mu_s}^{a_s b_s} j^{\mu_1\dots\mu_s} \ .
\end{align}
As opposed to $j_\ell^{a_1\dots a_{2s}}$ and $j_*^{a_1\dots a_{2s}}$, the twistor indices on $j^{a_1 b_1\dots a_s b_s}$ are \emph{not} totally symmetric. One can see from eqs. \eqref{eq:current_tensor_constraint}-\eqref{eq:current_tensor_equivalence} that $j$ is a sort of intermediate between $j_\ell$ and $j_*$, with one index in every $a_k b_k$ pair lying in $P(\ell)$, and the other in $P^*(\ell)$. The dictionary between $j_\ell$, $j$ and $j_*$ can be viewed as two successive applications of the mapping \eqref{eq:u_u*}:
\begin{align}
 j^{a_1 b_1\dots a_s b_s} &= \frac{1}{2^s}\delta^{[a_1}_{c_1} \ell^{b_1]}{}_{c_2}\dots\delta^{[a_s}_{c_{2s-1}} \ell^{b_s]}{}_{c_{2s}} j_*^{c_1 c_2\dots c_{2s-1}c_{2s}} \ ; \\ 
 j_\ell^{a_1 a_2\dots a_{2s-1} a_{2s}} &= \ell^{a_1}{}_{c_1}\dots\ell^{a_{2s-1}}{}_{c_s} j^{c_1 a_2\dots c_s a_{2s}} \ ,
\end{align}
or, restoring $j$ into tensor form:
\begin{align}
 j^{\mu_1\dots\mu_s} 
   &= \frac{1}{8^s}\ell_{\nu_1}\gamma^{\nu_1\mu_1}_{c_1 c_2}\dots\ell_{\nu_s}\gamma^{\nu_s\mu_s}_{c_{2s-1}c_{2s}} j_*^{c_1 c_2\dots c_{2s-1}c_{2s}} \ ; 
   \label{eq:j_j*} \\
 j_\ell^{a_1 a_2\dots a_{2s-1} a_{2s}}
   &= \ell^{\nu_1}\gamma_{\nu_1\mu_1}^{a_1 a_2}\dots\ell^{\nu_s}\gamma_{\nu_s\mu_s}^{a_{2s-1}a_{2s}} j^{\mu_1\dots\mu_s} \ .
\end{align}
One can also translate $j_\ell$ rather than $j$ into tensor indices. This yields the tensor:
\begin{align}
   j_\ell^{\mu_1\nu_1\dots\mu_s\nu_s} &= \frac{1}{4^s}\gamma^{\mu_1\nu_1}_{a_1 a_2}\dots\gamma^{\mu_s\nu_s}_{a_{2s-1} a_{2s}} \, j_\ell^{a_1 a_2\dots a_{2s-1} a_{2s}}
     = 2^s\ell^{[\mu_1}\delta^{\nu_1]}_{\rho_1}\dots\ell^{[\mu_s}\delta^{\nu_s]}_{\rho_s} j^{\rho_1\dots\rho_s} \ ,
\end{align}
which is invariant under \eqref{eq:current_tensor_equivalence}.

If $j$ has conformal weight $\Delta$, then $j_\ell$ and $j_*$ have weights $\Delta - s$ and $\Delta + s$, respectively. The conformally covariant divergence \eqref{eq:tensor_div}, which is well-defined for $\Delta = s+1$, is best expressed in spinor language in terms of $j_*$:
\begin{align}
 (\operatorname{div} j_*)^{a_1\dots a_{2s-2}} = \frac{1}{8}\gamma^{\mu\nu}_{bc}\ell_\mu\frac{\del}{\del\ell^\nu}\,j_*^{a_1 \dots a_{2s-2}bc} \ . \label{eq:spinor_div}
\end{align}
When $j_*$ has the correct conformal weight $\Delta + s = 2s+1$, one can show that this operation is consistent with the equivalence relation \eqref{eq:spinor_equivalence}.
With the particular numerical factor in \eqref{eq:spinor_div}, $\operatorname{div} j_*$ is related to the tensorial expression \eqref{eq:tensor_div} via the spin-$(s-1)$ version of the map \eqref{eq:j_j*}.

\subsubsection{More on the bulk-to-boundary limit} \label{sec:geometry:spinors_boundary:limit}

It is instructive to flesh out the limit \eqref{eq:limit_spinor_spaces} in some more detail. For this purpose, we will need to know the \emph{direction} from which the bulk point $x$ approaches the boundary point $\ell$. This direction can be encoded by a second boundary point $n$, where we normalize $\ell\cdot n = -1/2$ for convenience. We can then define the approach $x^\mu\rightarrow\ell^\mu/z$ as:
\begin{align}
x^\mu = \frac{1}{z}\ell^\mu + zn^\mu \ , \label{eq:approach}
\end{align}
such that $x\cdot x = -1$ is maintained throughout. The trajectory \eqref{eq:approach} is just the geodesic from the boundary point $n$ to the boundary point $\ell$, which approaches $\ell$ as $z\rightarrow 0$. The spacelike unit tangent to the trajectory \eqref{eq:approach} reads:
\begin{align}
 t^\mu = \frac{1}{z}\ell^\mu - zn^\mu \ . \label{eq:approach_t}
\end{align}

For simplicity, let us choose a frame such that:
\begin{align}
 \ell^\mu = \left(\frac{1}{2}, \mathbf{0}, \frac{1}{2} \right) \ ; \quad n^\mu = \left(\frac{1}{2}, \mathbf{0}, -\frac{1}{2} \right) \ . \label{eq:frame}
\end{align} 
Then the trajectory \eqref{eq:approach} is just the geodesic of changing $z$ at constant $\mathbf{r} = 0$ in the Poincare coordinates \eqref{eq:Poincare}. In the frame \eqref{eq:frame}, using the explicit gamma matrices \eqref{eq:gamma_null}, we can now observe the following. The spinor spaces $P(\ell)$ and $P_{L/R}(x)$ and are spanned by twistors of the form: 
\begin{align}
 P(\ell): \ U_\ell = \begin{pmatrix} u \\ 0 \end{pmatrix} \ ; \quad P_L(x): \ U_L = \begin{pmatrix} u \\ zu \end{pmatrix} \ ; \quad P_R(x): \ U_R = \begin{pmatrix} u \\ -zu \end{pmatrix} \ , \label{eq:P_bulk_boundary}
\end{align}
where $u$ is a 2-component spinor. This explicitly shows how $P_{L/R}(x)$ both converge towards $P(\ell)$. 

It will be useful to identify the three twistors \eqref{eq:P_bulk_boundary} for given $u$ as representing ``asymptotically the same'' boundary spinor. They can be mapped explicitly onto each other using the following operators:
\begin{align}
 U_R &= -tU_L = txU_L = (\ell n - n\ell)U_L \ ; \\
 U_L &= +tU_R = txU_R = (\ell n - n\ell)U_R \ ; \\
 U_\ell &= \frac{1}{2}(1 + tx)U_{L/R} = \ell nU_{L/R} \ .
\end{align}
Thus, the operator $tx = \ell n - n\ell$ asymptotically maps spinors in $P_L(x)$ to their ``asymptotically equal'' counterparts in $P_R(x)$ and vise versa, while the operator $(1+tx)/2 = \ell n$ maps them both to their ``asymptotically equal'' counterpart in $P(\ell)$. In other words, the projection $U\rightarrow \ell nU\in P(\ell)$ defines the ``boundary limit'' of a twistor $U$. In the language of section \ref{sec:spacetime_subgroup} below, the projector $\ell n$ can be interpreted as an infinite boost in the $\ell\wedge n$ plane. 

Finally, let us work out the action $\Gamma\rightarrow \ell n\Gamma n\ell$ of the projector $\ell n$ on a complete basis of twistor matrices $\Gamma$:
\begin{align}
 \begin{split}
   &1 \rightarrow 0 \ ; \quad \ell \rightarrow \ell \ ; \quad n \rightarrow 0 \ ; \quad \gamma_i \rightarrow 0 \ ; \\
   &\ell n - n\ell \rightarrow 0 \ ; \quad \ell\gamma_i \rightarrow \ell\gamma_i \ ; \quad n\gamma_i \rightarrow 0 \ ; \quad \gamma_{ij} \rightarrow 0 \ .
 \end{split} \label{eq:bilinear_limit_null}
\end{align}
Here, we defined $\gamma_i = e_i^\mu\gamma_\mu$, where the indices $(i,j,\dots)$ run over the values $1,2,3$, and the basis $e_i^\mu$ spans the 3d subspace orthogonal to both $\ell$ and $n$. Using a basis with $\{x,t\}$ in place of $\{\ell,n\}$, eqs. \eqref{eq:bilinear_limit_null} become:
\begin{align}
 \begin{split}
   &1 \rightarrow 0 \ ; \quad x \rightarrow \frac{1}{z}\ell \ ; \quad t \rightarrow \frac{1}{z}\ell \ ; \quad \gamma_i \rightarrow 0 \ ; \\
   &tx \rightarrow 0 \ ; \quad x\gamma_i \rightarrow \frac{1}{z}\ell\gamma_i \ ; \quad t\gamma_i \rightarrow \frac{1}{z}\ell\gamma_i \ ; \quad \gamma_{ij} \rightarrow 0 \ .
 \end{split} \label{eq:bilinear_limit}
\end{align}

\subsection{Bulk and boundary spinor spaces on an equal footing} \label{sec:geometry:spinors_equal}

For some purposes, in particular for the higher-spin two-point functions of section \ref{sec:algebra:two_point} below, one can avoid the distinction between bulk and boundary points. This feature is linked to covariance under the $O(1,5)$ group of bulk conformal transformations, though we will not pursue that angle explicitly.

Let us consider a 2-component spinor space, which may be either a boundary spinor space $P(\ell)$ or a bulk spinor space $P_L(x)$ or $P_R(x)$. This spinor space is spanned by a twistor matrix, which in index-free notation is again simply $P(\ell)$, $P_L(x)$ or $P_R(x)$. These can all be treated as special cases of:
\begin{align}
 P(\xi) = \frac{1}{2}\left(\sqrt{-\xi\cdot\xi} + \xi\right) \ , \label{eq:P_xi}
\end{align}
where the matrix $P^{ab}(\xi)$ is determined by a timelike or null vector $\xi^\mu\in\bbR^{1,4}$. The special cases of bulk and boundary spinor spaces correspond to:
\begin{align}
 \xi^\mu = \ell^\mu \ \Rightarrow \ P(\xi) = P(\ell) \ ; \quad \xi^\mu = \pm x^\mu \ \Rightarrow \ P(\xi) = P_{R/L}(x) \ . \label{eq:spinor_spaces}
\end{align}
Unifying eqs. \eqref{eq:measure_chiral_inverse} and \eqref{eq:measure}, we can define a metric and measure on the spinor space $P(\xi)$ via:
\begin{align}
 \frac{du^a du^b}{2\pi} \equiv P^{ab}(\xi)\, d^2u \ .
\end{align}
We note the identity:
\begin{align}
 \epsilon_{abcd}P^{cd}(\xi) = 2P_{ab}(-\xi) \ ,
\end{align}
which implies in particular that $P(-\xi)$ is the subspace orthogonal to $P(\xi)$. In other words, the space $P^*(\xi)$, i.e. the dual to $P(\xi)$ under the twistor metric, is just the quotient space of twistors modulo terms in $P(-\xi)$. For a boundary point, this reproduces the dual spinor space \eqref{eq:dual_space}, since the spaces $P(\ell)$ and $P(-\ell)$ coincide (with a factor of $-1$ between the corresponding matrices). For a bulk point, this means that the space $P^*_{L/R}(x)$ dual to $P_{L/R}(x)$ is the space of twistors modulo terms in $P_{R/L}(x)$, which can be identified with $P_{L/R}(x)$ itself.

Consider now a pair of spinor spaces $P(\xi)$ and $P(\xi')$, associated with a pair of bulk or boundary points. The relationship between these spaces is governed by two invariants:
\begin{align}
 P_{ab}(\xi)P^{ab}(\xi') &= \tr\left(P(\xi)P(\xi')\right) = \sqrt{(\xi\cdot\xi)(\xi'\cdot\xi')} - \xi\cdot\xi' \ ; \\
 \frac{1}{2}\epsilon_{abcd}P^{ab}(\xi)P^{cd}(\xi') &= \tr\left(P(\xi)P(-\xi')\right) = \sqrt{(\xi\cdot\xi)(\xi'\cdot\xi')} + \xi\cdot\xi' \ . 
\end{align}
An arbitrary twistor $U$ can be decomposed along $P(\xi)$ and $P(\xi')$ as follows:
\begin{align}
 U = u + u' \ ; \quad 
 u = \frac{2P(\xi)P(-\xi')U}{\tr\left(P(\xi)P(-\xi')\right)} \in P(\xi) \ ; \quad u' = \frac{2P(\xi')P(-\xi)U}{\tr\left(P(\xi')P(-\xi)\right)} \in P(\xi') \ ,
 \label{eq:U_u_u'}
\end{align}
where the scalar product of the two components $u,u'$ reads:
\begin{align}
 uu' = \frac{U\xi\xi'U}{2\left(\sqrt{(\xi\cdot\xi)(\xi'\cdot\xi')} + \xi\cdot\xi'\right)} \ . \label{eq:uu'}
\end{align}
The twistor measure decomposes under \eqref{eq:U_u_u'} as:
\begin{align}
 d^4U = \frac{1}{4}\epsilon_{abcd}P^{ab}(\xi)P^{cd}(\xi')\,d^2u\,d^2u' = \frac{1}{2}\left(\sqrt{(\xi\cdot\xi)(\xi'\cdot\xi')} + \xi\cdot\xi'\right) d^2u\,d^2u' \ . \label{eq:U_u_u'_measure}
\end{align}
The chiral decomposition $U = u_L + u_R$ at a single bulk point $x$ can be viewed as a special case of \eqref{eq:U_u_u'}, with eqs. \eqref{eq:uu'}-\eqref{eq:U_u_u'_measure} reproducing the identities $u_L u_R = 0$ and $d^4U = d^2u_L d^2u_R$.

\subsubsection{Integrals over spinor spaces} \label{sec:geometry:spinors_equal:integrals}

In calculations below, we will need the 2-component spinor versions of the 4-component twistor integrals \eqref{eq:delta_integral_raw}-\eqref{eq:Gaussian}. Consider a general (bulk or boundary) spinor space $P(\xi)$ as above. A Gaussian integral over $P(\xi)$ can be calculated as:
\begin{align}
 \int_{P(\xi)} d^2u\, e^{uAu/2} = \frac{\pm 1}{\sqrt{\det_{P(\xi)}(A)}} \ ; \quad \det\nolimits_{P(\xi)}(A) = -\frac{1}{2}\tr\left(P(\xi)A\right)^2 \ .
 \label{eq:Gaussian_spinor}
\end{align}
Here, $A$ is a symmetric twistor matrix, while $\det\nolimits_{P(\xi)}(A)$ is the determinant of $A$, viewed as a $2\times 2$ quadratic form over the spinor space $P(\xi)$.

The generic analog of the delta-function-type integral \eqref{eq:delta_integral_raw} involves a pair of spinor spaces $P(\xi),P(\xi')$. The integral reads:
\begin{align}
 \int_{P(\xi)} d^2u \int_{P(\xi')} d^2u' f(u)\, e^{iuu'} = \frac{2}{P_{ab}(\xi)P^{ab}(\xi')}\,f(0) 
   = \frac{2}{\sqrt{(\xi\cdot\xi)(\xi'\cdot\xi')} - \xi\cdot\xi'}\,f(0) \ . \label{eq:delta_integral_spinor} 
\end{align}
In addition, at a single boundary point $\ell$, one can write the following delta-function-type integrals over the spinor space $P(\ell)$ and its dual space $P^*(\ell)$:
\begin{align}
 \int_{P(\ell)} d^2u \int_{P^*(\ell)} d^2u^* f(u)\, e^{iuu^*} = f(0) \ ; \quad \int_{P(\ell)} d^2u \int_{P^*(\ell)} d^2u^* f(u^*)\, e^{iuu^*} = f(0) \ .
 \label{eq:delta_integral_boundary}
\end{align}

The integral \eqref{eq:delta_integral_spinor} can be written explicitly in terms of a delta function as follows:
\begin{align}
 \int_{P(\xi)} d^2u\,\delta_{\xi'}(u) f(u) = \frac{2}{\sqrt{(\xi\cdot\xi)(\xi'\cdot\xi')} - \xi\cdot\xi'}\,f(0) \ ,
\end{align}
where the spinor delta function is defined as:
\begin{align}
 \delta_\xi(U) = \int_{P(\xi)} d^2v\,e^{ivU} \ . \label{eq:delta_xi}
\end{align}
In the particular cases \eqref{eq:spinor_spaces} of bulk and boundary spinor spaces, we denote these delta functions as:
\begin{align}
 \xi^\mu = \ell^\mu \ \Rightarrow \ \delta_\xi(U) = \delta_\ell(U) \ ; \quad \xi^\mu = \pm x^\mu \ \Rightarrow \ \delta_\xi(U) \equiv \delta_x^{R/L}(U) \ . \label{eq:spinor_deltas}
\end{align}
The notation is meant to signify that $\delta_\xi(U)$ is a delta function with respect to $U$, with support on a 2d spinor space determined by $\xi$. Specifically, it has support on the subspace $P(-\xi)$ which is orthogonal to $P(\xi)$. For a boundary point, this means that $\delta_\ell(U)$ has support on $P(\ell)$, forcing the $P^*(\ell)$ component of $U$ to vanish. For a bulk point, it means that $\delta_x^{R/L}(U)$ has support on $P_{L/R}(x)$, forcing the $P_{R/L}(x)$ component to vanish. The boundary delta function $\delta_\ell(U)$ has conformal weight $\Delta = 1$, and can be used to rewrite the second integral in \eqref{eq:delta_integral_boundary} as:
\begin{align}
 \int_{P^*(\ell)} d^2u^*\,\delta_\ell(u^*)\, f(u^*) = f(0) \ .
\end{align}

The comments from section \ref{sec:geometry:twistors:integrals} regarding sign ambiguities in twistor integrals apply equally well to the spinor case. Gaussians are well-defined functions, but their integrals have a sign ambiguity that cannot be globally fixed. Conversely, delta functions have well-defined integrals, but they themselves are defined as limits of ordinary functions only up to sign. An additional subtlety arises when adding or comparing integrals over different spinor spaces, associated with different spacetime points. In that case, one must make a separate contour choice for every integral, and this choice may fail to be consistent across a large enough spacetime region.

\section{Higher-spin algebra} \label{sec:algebra}

\subsection{Spacetime-independent structure}

In higher-spin theory, one introduces (spacetime-independent) twistor coordinates $Y^a$, which are acted on by the non-commutative star product:
\begin{align}
 Y^a\star Y^b = Y^a Y^b + iI^{ab} \ . \label{eq:star_basic}
\end{align}
By associativity, this extends into a product on polynomials in $Y$:
\begin{align}
 f(Y)\star g(Y) = f \exp\left(iI^{ab}\overleftarrow{\frac{\del}{\del Y^a}}\overrightarrow{\frac{\del}{\del Y^b}}\right) g \ . \label{eq:star_diff}
\end{align}
In practical calculations, it is convenient to use the index-free notation of section \ref{sec:geometry:twistors:index_free}, where some twistors are implicitly lower-index and some are upper-index. One can then use the formulas:
\begin{align}
 I^{ab}\overleftarrow{\frac{\del}{\del Y^a}}\overrightarrow{\frac{\del}{\del Y^b}} = \overleftarrow{\frac{\del}{\del Y^a}}\overrightarrow{\frac{\del}{\del Y_a}}
  = - \overleftarrow{\frac{\del}{\del Y_a}}\overrightarrow{\frac{\del}{\del Y^a}} \ , \label{eq:ddY_ddY}
\end{align}
where it is important that $\del/\del Y_a$ is \emph{minus} the raised-index version of $\del/\del Y^a$. Together with rearrangements of the form \eqref{eq:reverse}, one can reduce calculations to convenient index-free expressions such as:
\begin{align}
 \begin{split}
   U\Gamma_1\dots\Gamma_mY \left(I^{ab}\overleftarrow{\frac{\del}{\del Y^a}}\overrightarrow{\frac{\del}{\del Y^b}}\right) Y\Gamma_{m+1}\dots\Gamma_nV
     &= U\Gamma_1\dots\Gamma_n V \ ; \\
   Y\Gamma_1\dots\Gamma_m \left(I^{ab}\overleftarrow{\frac{\del}{\del Y^a}}\overrightarrow{\frac{\del}{\del Y^b}}\right) \Gamma_{m+1}\dots\Gamma_nY
     &= -\tr(\Gamma_1\dots\Gamma_n) \ .
 \end{split}
\end{align}
The star product also extends to non-polynomial functions, where one must resort to an integral formula:
\begin{align}
 f(Y)\star g(Y) = f \exp\left(iI^{ab}\overleftarrow{\frac{\del}{\del Y^a}}\overrightarrow{\frac{\del}{\del Y^b}}\right) g
   = \int d^4U d^4V f(Y+U)\, g(Y+V)\, e^{-iUV} \ . \label{eq:star_int}
\end{align}

The higher-spin symmetry algebra is the infinite-dimensional Lie algebra of even (i.e. integer-spin) functions $f(Y)$ with the associative product \eqref{eq:star_int}. It contains as a subalgebra the generators of the $EAdS_4$ isometry group $O(1,4)$:
\begin{align}
 M_{\mu\nu} = -\frac{i}{8} Y\gamma_{\mu\nu}Y \quad ; \quad [M^{\mu\nu}, M_{\rho\sigma}]_\star = 4\delta^{[\mu}_{[\rho}\, M^{\nu]}{}_{\sigma]} \ . \label{eq:generators}
\end{align}
The product \eqref{eq:star_int} respects a trace operation, defined simply by evaluating $f(Y)$ at $Y=0$:
\begin{align}
 \tr_\star f(Y) = f(0) \ ; \quad \tr_\star(f\star g) = \tr_\star(g\star f) =  \int d^4U d^4V f(U)\,g(V)\,e^{-iUV} \ . \label{eq:str}
\end{align}
Here, the equality $\tr_\star(f\star g) = \tr_\star(g\star f)$ relies on $f(Y),g(Y)$ being even functions. The $\tr_\star$ operation is usually denoted in the literature by ``$\str$'', since in certain generalizations of the algebra \eqref{eq:star_int}, the trace \eqref{eq:str} becomes a supertrace. 

Another important object is the delta function \eqref{eq:delta_U}:
\begin{align}
 \delta(Y) = \int d^4U\,e^{iUY} \ . \label{eq:delta}
\end{align}
A star product with $\delta(Y)$ implements the Fourier transform:
\begin{align}
 f(Y)\star\delta(Y) = \int d^4U f(U)\,e^{iUY} \ ; \quad \delta(Y)\star f(Y) = \int d^4U f(U)\,e^{-iUY} \ . \label{eq:delta_Fourier}
\end{align}
The following properties establish $\delta(Y)$ as a Klein operator of the algebra \eqref{eq:star_int}:
\begin{align}
 \delta(Y)\star\delta(Y) = 1 \ ; \quad \delta(Y)\star f(Y)\star\delta(Y) = f(-Y) \ , \label{eq:delta_Klein}
\end{align}
i.e. $\delta(Y)$ (anti)commutes with even (odd) functions $f(Y)$. This implies that $\delta(Y)$ is invariant in the adjoint representation of the higher-spin symmetry group (recall that the symmetry includes only integer spins, i.e. only generators even in $Y$). 

The star product $f\star g$, the trace $\tr_\star f$ and the invariant Klein operator $\delta(Y)$ are the only allowed ingredients in an expression that preserves (undeformed) higher-spin symmetry. The role of $\delta(Y)$ in this list is somewhat subtle. The issue is the contour ambiguity of the integral formula \eqref{eq:star_int}, which arises whenever we do higher-spin algebra with non-polynomial functions. As discussed in section \ref{sec:geometry:twistors:integrals}, even the simplest cases - delta functions and Gaussians - are associated with a sign ambiguity. In particular, one should be careful with assigning meaning to the sign of $\delta(Y)$ and its star products. While this sign ambiguity may not look like much, there is a sense in which it is the \emph{only} information carried in star products with $\delta(Y)$. Indeed, since $\delta(Y)$ squares to unity, one may think of decomposing the space of functions $f(Y)$ into eigenspaces with eigenvalues $\pm 1$ under star-multiplication by $\delta(Y)$. Formally, this decomposition is accomplished by the pair of projectors:
\begin{align}
 \calP_\pm(Y) = \frac{1 \pm \delta(Y)}{2} \  . \label{eq:projectors_delta}
\end{align}
Conceptually, these projectors play an important role in the theory: as we will see, they are related to bulk antipodal symmetry, as well as to the two types of asymptotic boundary data (Neumann vs. Dirichlet or magnetic vs. electric). However, since the sign of $\delta(Y)$ is a priori ambiguous, one shouldn't take the projectors \eqref{eq:projectors_delta} too seriously. In particular, in section \ref{sec:linear_HS:antipodal}, we will see in detail how they fail to be well-defined linear operators. In the present section, we will continue to list useful formal identities involving $\delta(Y)$. Later in the paper, we will make use of $\delta(Y)$ and the projectors $\calP_\pm(Y)$, but with a dose of care and self-consciousness.

\subsection{Structure at a bulk point}

Choosing a bulk point $x\in EAdS_4$ picks out a preferred rotation group $SO(4) = SO(3)_L\times SO(3)_R$ out of the isometry group $SO(1,4)$. In the star-product language, the two chiral $SO(3)$'s are generated by bilinears $y^a_L y^b_L$ and $y^a_R y^b_R$, where we used the chiral decomposition $Y = y_L+y_R$ of the twistor $Y$ into Weyl spinors at $x$. Each of the chiral $SO(3)$'s can be extended into its own higher-spin subalgebra, given respectively by chiral functions $f(y_L)$ and $f(y_R)$. Since left-handed and right-handed spinors are orthogonal under the twistor metric, the two subalgebras commute. Explicitly, the chiral decomposition of the star product \eqref{eq:star_basic} reads:
\begin{align}
 y_L^a\star y_L^b = y_L^a y_L^b + iP_L^{ab} \ ; \quad y_R^a\star y_R^b = y_R^a y_R^b + iP_R^{ab} \ ; \quad y_L^a\star y_R^b = y_R^b\star y_L^a = y_L^a y_R^b \ ,
\end{align}
where we must keep in mind that the projectors $P_{L/R}$ and the Weyl spinors $y_{L/R}$ depend on the bulk point $x$.

Analogously to the role of $\delta(Y)$, delta functions with respect to $y_L$ and $y_R$ play the role of Klein operators for the left-handed and right-handed higher-spin subalgebras. We've already encountered these spinor delta functions in eqs. \eqref{eq:delta_xi}-\eqref{eq:spinor_deltas}:
\begin{align}
 \delta^L_x(Y) = \int_{P_L(x)} d^2u_L \,e^{iu_L Y} \ ; \quad \delta^R_x(Y) = \int_{P_R(x)} d^2u_R \,e^{iu_R Y} \ . \label{eq:delta_chiral}
\end{align}
The delta function $\delta^{L/R}_x(Y)$ depends on the twistor $Y$ only through the spinor component $y_{L/R}$. These delta functions have star-product properties \cite{Didenko:2009td} analogous to eqs. \eqref{eq:delta_Fourier}-\eqref{eq:delta_Klein}:
\begin{align}
 \begin{split}
   f(y_L+y_R)\star\delta^L_x(y_L) &= \int d^2u_L\,f(u_L+y_R)\,e^{iu_L y_L} \ ; \\
   \delta^L_x(y_L)\star f(y_L+y_R) &= \int d^2u_L\,f(u_L+y_R)\,e^{-iu_L y_L} \ ; \\
   \delta^L_x(y_L)\star f(y_L+y_R)\star\delta^L_x(y_L) &= f(-y_L+y_R) \ ,
 \end{split} \label{eq:delta_Fourier_bulk_raw}
\end{align}
and similarly for $\delta^R_x(y_R)$. These can be written more covariantly as:
\begin{align}
 &f(Y)\star\delta^{L/R}_x(Y) = \int d^2u_{L/R}\,f(Y + u_{L/R})\,e^{iu_{L/R} Y} \ ; \label{eq:delta_Fourier_bulk_right} \\
 &\delta^{L/R}_x(Y)\star f(Y) = \int d^2u_{L/R}\,f(Y + u_{L/R})\,e^{-iu_{L/R} Y} \ ; \label{eq:delta_Fourier_bulk_left} \\
 &\delta^L_x(Y)\star f(Y)\star\delta^L_x(Y) = f(xY) \ ; \quad \delta^R_x(Y)\star f(Y)\star\delta^R_x(Y) = f(-xY) \label{eq:delta_Klein_bulk} \ .
\end{align}
As a special case, we have:
\begin{align}
 \delta^{L/R}_x(Y)\star\delta(Y) = \delta(Y)\star\delta^{L/R}_x(Y) = \delta^{R/L}_x(Y) \ . \label{eq:bulk_delta_odd}
\end{align}
The products of the chiral delta functions are $x$-independent:
\begin{align}
 \begin{split}
   \delta^L_x(Y)\star\delta^L_x(Y) &= \delta^R_x(Y)\star\delta^R_x(Y) = 1 \ ; \\
   \delta^L_x(Y)\star\delta^R_x(Y) &= \delta^R_x(Y)\star\delta^L_x(Y) = \delta^L_x(Y)\delta^R_x(Y) = \delta(Y) \ .
 \end{split} \label{eq:bulk_delta_products}
\end{align}
Finally, it will be helpful to explicitly express the $x$-dependence of $\delta^L_x(Y)$ and $\delta^R_x(Y)$. Taking $x$ derivatives of integrals such as \eqref{eq:delta_chiral} is subtle, since the subspace over which we are integrating is itself a function of $x$. A useful workaround (e.g. for the $d^2u_L$ integral) is to perform a change of variables $u_L = P_L(x)u_L'$, where $u_L'$ can now be integrated over the left-handed spinor space $P_L(x')$ at an arbitrary \emph{fixed} point $x'$. After taking the desired $x$ derivatives, we can replace $x'\rightarrow x$. By this method, we find:
\begin{align}
 \begin{split}
   \nabla_\mu\delta^L_x &= -\frac{1}{2}(\gamma_\mu)^a{}_b\,y_R^b\,\frac{\del\delta^L_x}{\del y_L^a} = -\frac{i}{2}(y_L\gamma_\mu y_R)\star\delta^L_x 
      = \frac{i}{2}\,\delta^L_x\star(y_L\gamma_\mu y_R) \ ; \\
   \nabla_\mu\delta^R_x &= \frac{1}{2}(\gamma_\mu)^a{}_b\,y_L^b\,\frac{\del\delta^R_x}{\del y_R^a} = -\frac{i}{2}(y_L\gamma_\mu y_R)\star\delta^R_x
      = \frac{i}{2}\,\delta^R_x\star(y_L\gamma_\mu y_R) \ ,
 \end{split}
\end{align}
or, in more covariant notation:
\begin{align}
 \begin{split}
   \nabla_\mu\delta^L_x &= -\frac{i}{4}(Y\gamma_\mu xY)\star\delta^L_x = \frac{i}{4}\,\delta^L_x\star(Y\gamma_\mu xY) \ ; \\
   \nabla_\mu\delta^R_x &= -\frac{i}{4}(Y\gamma_\mu xY)\star\delta^R_x = \frac{i}{4}\,\delta^R_x\star(Y\gamma_\mu xY) \ . \label{eq:nabla_delta}
 \end{split} 
\end{align}

\subsection{Structure at a boundary point} \label{sec:algebra:boundary}

Now, instead of a bulk point $x$, let us fix a boundary point $\ell$. The isometry group $SO(1,4)$, now viewed as the boundary conformal group, acquires three preferred subgroups, nested within each other:
\begin{enumerate}
 \item Special conformal transformations around $\ell$ (or, equivalently, translations in a frame where $\ell$ is the point at infinity). These are generated by bilinears $YAY$, where the symmetric twistor matrix $A$ satisfies $\ell A = 0$. 
 \item Special conformal transformations and rotations around $\ell$. These are generated by bilinears $YAY$ where $\ell A - A\ell = 0$.
 \item Special conformal transformations, rotations and dilatations around $\ell$. These are generated by bilinears $YAY$ where $\ell A - A\ell = \lambda\ell$ for some scalar $\lambda$.
\end{enumerate}
Neither of these subgroups extends into an interesting higher-spin subalgebra. The only subgroup that extends at all is the first one. The corresponding higher-spin subalgebra $\calA_0(\ell)$ consists of functions $f(Y)$ that satisfy $f(Y+u) = f(Y)$ for any $u\in P(\ell)$, i.e. of functions $f(y^*)$ over the boundary spinor space $P^*(\ell)$:
\begin{align}
 f(Y)\in\calA_0(\ell) \quad \Longleftrightarrow \quad f(Y+u) = f(Y) \ \ \text{for} \ \ u\in P(\ell) \quad \Longleftrightarrow \quad f(Y) = f(y^*) \ .
\end{align}
The special conformal transformations around $\ell$ are generated by the quadratic piece of this subalgebra. The entire subalgebra is commuting, and the star product is simply:
\begin{align}
 f,g\in\calA_0(\ell) \quad \Longrightarrow \quad f(Y)\star g(Y) = f(Y)g(Y) \ .
\end{align}

A special element of the subalgebra $\calA_0(\ell)$ is the delta function with respect to $y^*$, which we've encountered in eqs. \eqref{eq:delta_xi}-\eqref{eq:spinor_deltas}:
\begin{align}
 \delta_\ell(Y) = \int_{P(\ell)} d^2u\, e^{iuY} \ . \label{eq:delta_ell}
\end{align}
In the bulk-to-boundary limiting procedure \eqref{eq:limit}, the delta function \eqref{eq:delta_ell} can be expressed as a rescaled limit of the bulk delta functions \eqref{eq:delta_chiral}:
\begin{align}
 \delta_\ell(Y) = \lim_{x\rightarrow\ell/z} \frac{1}{z}\,\delta^R_x(Y) = -\lim_{x\rightarrow\ell/z} \frac{1}{z}\,\delta^L_x(Y) \ . \label{eq:delta_limit}
\end{align}
However, unlike its bulk counterparts, $\delta_\ell(Y)$ is \emph{not} a Klein operator. In particular, the star product $\delta_\ell(Y)\star\delta_\ell(Y)$ is divergent. The star product of $\delta_\ell(Y)$ with the global delta function $\delta(Y)$ reads:
\begin{align}
 \delta_\ell(Y)\star\delta(Y) = \delta(Y)\star\delta_\ell(Y) = -\delta_\ell(Y) \ . \label{eq:boundary_delta_odd}
\end{align}
From the point of view of the bulk-to-boundary limit \eqref{eq:delta_limit}, these identities can be viewed as a limiting case of \eqref{eq:bulk_delta_odd}.

The delta function $\delta_\ell(Y)$ is a member not only of the subalgebra $\calA_0(\ell)$, but of two additional (also degenerate) higher-spin subalgebras. These subalgebras, which we denote as $\calA_\pm(\ell)$, consist of functions with the property:
\begin{align}
 f(Y)\in\calA_\pm(\ell) \quad \Longleftrightarrow \quad f(Y+u) = e^{\pm iuY}f(Y) \ \ \text{for} \ \ u\in P(\ell) \ , \label{eq:A+-}
\end{align}
Functions of the form \eqref{eq:A+-} can be thought of as ``twisted'' functions on $P^*(\ell)$; like the true functions on $P^*(\ell)$ that make up the subalgebra $\calA_0(\ell)$, they depend freely only on a single two-component spinor. The star product in the subalgebras $\calA_\pm(\ell)$ reads:
\begin{align}
 \begin{split}
   f,g\in\calA_-(\ell) \quad \Longrightarrow \quad f(Y)\star g(Y) = g(0)f(Y) = (\tr_\star g)f(Y) \ ; \\
   f,g\in\calA_+(\ell) \quad \Longrightarrow \quad f(Y)\star g(Y) = f(0)g(Y) = (\tr_\star f)g(Y) \ .
 \end{split}
\end{align}

The definition \eqref{eq:A+-} of the subalgebras $\calA_\pm(\ell)$ can be expressed concisely in star-product form:
\begin{align}
 \begin{split}
   f(Y)\in\calA_-(\ell) \quad &\Longleftrightarrow \quad f(Y)\star \ell\,Y = 0 \ ; \\   
   f(Y)\in\calA_+(\ell) \quad &\Longleftrightarrow \quad \ell\,Y\star f(Y) = 0 \ .
 \end{split} \label{eq:A+-_star}
\end{align}
From here, it follows that multiplication on the right (left) by a function in $\calA_-(\ell)$ ($\calA_+(\ell)$) projects \emph{any} function into the corresponding subalgebra:
\begin{align}
 f\in\calA_-(\ell) \ \Rightarrow \ g\star f \in \calA_-(\ell) \ ; \quad f\in\calA_+(\ell) \ \Rightarrow \ f\star g \in \calA_+(\ell) \ .
\end{align}
In particular, since $\delta_\ell(Y)$ is an element of both $\calA_-(\ell)$ and $\calA_+(\ell)$, we have, for any $f(Y)$:
\begin{align}
 \begin{split}
   f(Y)\star\delta_\ell(Y) &= \int_{P(\ell)} d^2u\,f(Y + u)\,e^{iuY} \in \calA_-(\ell) \ ; \\
   \delta_\ell(Y)\star f(Y) &= \int_{P(\ell)} d^2u\,f(Y + u)\,e^{-iuY} \in \calA_+(\ell) \ ,
 \end{split} \label{eq:delta_Fourier_boundary}
\end{align}
while the product $\delta_\ell(Y)\star f(Y)\star\delta_\ell(Y)$ is divergent. The formulas \eqref{eq:delta_Fourier_boundary} can be recognized as boundary limits of \eqref{eq:delta_Fourier_bulk_right}-\eqref{eq:delta_Fourier_bulk_left}. There is no boundary analog of the Fourier-transform formulas \eqref{eq:delta_Fourier_bulk_raw}, because at a boundary point $\ell$, twistor space does not decompose into an orthogonal pair of spinor spaces.

\subsection{Structure at two and more points} \label{sec:algebra:two_point}

A key object in our analysis will be the star product of two spinor delta functions $\delta^L_x(Y)$, $\delta^R_x(Y)$ or $\delta_\ell(Y)$ at a pair of bulk or boundary points. In this section, we will compute these products and discuss their properties. These two-point products are closely related to various propagators in the HS literature, such as the $\calD$-functions of \cite{Gelfond:2008ur,Gelfond:2008td} and the boundary-to-bulk propagators of \cite{Didenko:2012tv}, and are quite similar in spirit to propagators in ordinary field theory. However, one should keep in mind an important detail: while the products $\delta\star\delta$ depend on two spacetime points, they depend on only \emph{one} twistor variable $Y$, which is not associated with either point in particular. In section \ref{sec:spacetime_subgroup:null}, we will discuss these two-point products from a different viewpoint, as group elements of the spacetime symmetry $SO(1,4)$.

\subsubsection{The general two-point product}

The different kinds of two-point products can all be computed together, using the machinery of section \ref{sec:geometry:spinors_equal}. Recall from \eqref{eq:spinor_deltas} that the delta functions $\delta^L_x(Y),\delta^R_x(Y),\delta_\ell(Y)$ are all special cases of the general spinor delta function \eqref{eq:delta_xi}:
\begin{align}
 \delta_\xi(Y) = \int_{P(\xi)} d^2u\,e^{iuY} \ . \label{eq:delta_xi_again}
\end{align}
The star-product formulas \eqref{eq:delta_Fourier_bulk_right}-\eqref{eq:delta_Fourier_bulk_left},\eqref{eq:bulk_delta_odd},\eqref{eq:boundary_delta_odd},\eqref{eq:delta_Fourier_boundary} are all particular cases of the identities:
\begin{align}
 &f(Y)\star\delta_\xi(Y) = \int_{P(\xi)} d^2u\,f(Y+u)\,e^{iuY} \ ; \label{eq:delta_Fourier_xi_first} \\ 
 &\delta_\xi(Y)\star f(Y) = \int_{P(\xi)} d^2u\,f(Y+u)\,e^{-iuY} \ ; \\
 &\delta_\xi(Y)\star\delta(Y) = \delta(Y)\star\delta_\xi(Y) = \delta_{-\xi}(Y) \ .
\end{align}
We can now compute the star product of two delta functions of the general type \eqref{eq:delta_xi_again}. First, using eq. \eqref{eq:delta_Fourier_xi_first}, we get the integral expression: 
\begin{align}
 \delta_\xi(Y)\star\delta_{\xi'}(Y) = \int_{P(\xi)} d^2u \int_{P(\xi')} d^2u'\,e^{i(uY + u'Y + uu')} \ .
\end{align}
With some work, this integral can be brought to the form \eqref{eq:delta_integral_spinor}. To do this, we decompose the twistor $Y$ into a pair of spinors as in eq. \eqref{eq:U_u_u'}:
\begin{align}
 Y = y + \bar y' \ ; \quad y \in P(\xi) \ ; \quad \bar y'\in P(-\xi') \ .
\end{align}
The $\bar y'$ piece is identically orthogonal to $u'\in P(\xi')$, while the $y$ piece can be used to shift the integration variable $u\in P(\xi)$. This brings the integral into the form:
\begin{align}
 \begin{split}
  \delta_\xi(Y)\star\delta_{\xi'}(Y) &= \int_{P(\xi)} d^2u \int_{P(\xi')} d^2u'\,e^{i(uY + y\bar y')} e^{iuu'} 
    = \frac{2e^{iy\bar y'}}{\sqrt{(\xi\cdot\xi)(\xi'\cdot\xi')} - \xi\cdot\xi'} \\
    &= \frac{2}{\sqrt{(\xi\cdot\xi)(\xi'\cdot\xi')} - \xi\cdot\xi'}\,\exp\left(\frac{-iY\xi\xi'Y/2}{\sqrt{(\xi\cdot\xi)(\xi'\cdot\xi')} - \xi\cdot\xi'}\right) \ .
 \end{split} \label{eq:general_two_point}
\end{align}
where we used eq. \eqref{eq:delta_integral_spinor} in the first line and eq. \eqref{eq:uu'} in the second line. For particular cases of bulk/boundary points, the result \eqref{eq:general_two_point} reads:
\begin{align}
 \delta_\ell(Y)\star\delta_{\ell'}(Y) &= -\frac{2}{\ell\cdot\ell'}\exp\frac{iY\ell\ell' Y}{2\ell\cdot\ell'} \ ; \label{eq:delta_ell_ell} \\
 \delta_\ell(Y)\star\delta^R_x(Y) &= -\frac{2}{\ell\cdot x}\exp\frac{iY\ell x Y}{2\ell\cdot x} \ ; \quad 
 \delta^R_x(Y)\star\delta_\ell(Y) = -\frac{2}{\ell\cdot x}\exp\frac{iYx\ell Y}{2\ell\cdot x} \ ; \label{eq:delta_ell_x} \\
 \delta^R_x(Y)\star\delta^R_{x'}(Y) &= \frac{2}{1 - x\cdot x'}\exp\frac{iYxx'Y}{2(x\cdot x' - 1)} \ , \label{eq:delta_x_x}
\end{align}
where one can substitute $\delta^R_x(Y)\rightarrow\delta^L_x(Y)$ via the antipodal map $x\rightarrow -x$, and likewise for $x'$.

\subsubsection{Properties of the boundary-boundary product}

We now focus on the boundary-boundary two-point product \eqref{eq:delta_ell_ell}, which possesses some remarkable properties. First, if we multiply \eqref{eq:delta_ell_ell} by $\sqrt{-\ell\cdot\ell'}$, the result has the conformal weight $\Delta = 1/2$ of a 3d free massless scalar with respect to both boundary points. We can then evaluate the 3d conformal Laplacian \eqref{eq:Laplacian}, only to find that the massless wave equation is satisfied:
\begin{align}
 \sqrt{-\ell\cdot\ell'}\,\delta_\ell(Y)\star\delta_{\ell'}(Y) &= \frac{2}{\sqrt{-\ell\cdot\ell'}}\exp\frac{iY\ell\ell' Y}{2\ell\cdot\ell'} \ ; \label{eq:K_raw} \\
 \Box_\ell\left(\sqrt{-\ell\cdot\ell'}\,\delta_\ell(Y)\star\delta_{\ell'}(Y)\right) 
   &= \Box_{\ell'}\left(\sqrt{-\ell\cdot\ell'}\,\delta_\ell(Y)\star\delta_{\ell'}(Y)\right) = 0 \quad \forall \ell\neq\ell' \ . \label{eq:ell_ell_wave}
\end{align}
At $\ell=\ell'$, the two-point product has a singularity, and the wave equation \eqref{eq:ell_ell_wave} picks up a source term. Moreover, since this is an \emph{essential} singularity, the source term will contain not just a delta distribution, but also an infinite tower of its derivatives. Thus, upon integration over $\ell$ or $\ell'$, the source will \emph{not} appear localized at $\ell=\ell'$, as we will see explicitly in section \ref{sec:CFT:twistor:currents}.

Upon taking the higher-spin trace \eqref{eq:str}, the essential singularity in \eqref{eq:K_raw} becomes a simple pole. In fact, up to a numerical factor, this trace is just the $\Delta = 1/2$ boundary-to-boundary propagator, which satisfies the wave equation with a point source:
\begin{align}
 \tr_\star\left(\sqrt{-\ell\cdot\ell'}\,\delta_\ell(Y)\star\delta_{\ell'}(Y)\right) &= \frac{2}{\sqrt{-\ell\cdot\ell'}} \ ; \label{eq:str_ell_ell} \\
 \Box_\ell\tr_\star\left(\sqrt{-\ell\cdot\ell'}\,\delta_\ell(Y)\star\delta_{\ell'}(Y)\right) &= -8\pi\sqrt{2}\,\delta^{5/2,1/2}(\ell,\ell') \ . \label{eq:str_ell_ell_wave}
\end{align}
Here, the superscripts on the boundary delta function $\delta(\ell,\ell')$ denote its conformal weight with respect to each argument. To derive the wave equation \eqref{eq:str_ell_ell_wave}, we recall that in the flat frame \eqref{eq:flat}, $\sqrt{-\ell\cdot\ell'}$ is just the 3d Euclidean distance $|\mathbf{r} - \mathbf{r'}|/\sqrt{2}$. The full Gaussian \eqref{eq:K_raw} can now be understood as a Taylor series of derivatives of the propagator \eqref{eq:str_ell_ell}. This can be seen explicitly by converting the coefficients of different powers of $Y$ into boundary tensors, or more abstractly from eqs. \eqref{eq:K_raw2},\eqref{eq:K_sqrt_infinitesimal} below.

Another feature of the boundary-boundary product \eqref{eq:delta_ell_ell} is that it belongs simultaneously to the higher-spin subalgebras $\calA_+(\ell)$ and $\calA_-(\ell')$. As we can see from \eqref{eq:delta_Fourier_boundary}, the same is true for the more general product $\delta_\ell(Y)\star f(Y)\star\delta_{\ell'}(Y)$, where $f(Y)$ is an arbitrary function. On the other hand, it's clear from the definition \eqref{eq:A+-} or \eqref{eq:A+-_star} that the intersection $\calA_+(\ell)\cap\calA_-(\ell')$ is one-dimensional. Therefore, for any function $f(Y)$, we must have:
\begin{align}
 \delta_\ell(Y)\star f(Y)\star\delta_{\ell'}(Y) = \lambda\,\delta_\ell(Y)\star\delta_{\ell'}(Y) \ , \label{eq:forgetful}
\end{align}
for some $Y$-independent coefficient $\lambda$. Taking the higher-spin trace of both sides, we can express this coefficient as:
\begin{align}
 \lambda = -\frac{\ell\cdot\ell'}{2}\tr_\star\left(\delta_\ell(Y)\star f(Y)\star\delta_{\ell'}(Y)\right) \ . \label{eq:forgetful_coeff}
\end{align}
Eq. \eqref{eq:forgetful} is the underlying root of the ``forgetful property'' \cite{Didenko:2012tv} of higher-spin propagators. 

As a special case of \eqref{eq:forgetful}, we evaluate the three-point product:
\begin{align}
 \delta_\ell(Y)\star\delta_{\ell'}(Y)\star\delta_{\ell''}(Y) = \pm i\sqrt{-\frac{\ell\cdot\ell''}{2(\ell\cdot\ell')(\ell'\cdot\ell'')}}\,
   \delta_\ell(Y)\star\delta_{\ell''}(Y) \ . \label{eq:delta_ell_ell_ell}
\end{align}
The sign ambiguity is due to a Gaussian integration of the form \eqref{eq:Gaussian_spinor}. An efficient way to derive eq. \eqref{eq:delta_ell_ell_ell} is to use the result \eqref{eq:delta_ell_ell} for the product of two delta functions, and then factor in the third delta function via \eqref{eq:delta_Fourier_boundary}; thanks to eq. \eqref{eq:forgetful}, it suffices to evaluate the result at $Y=0$.

\section{Linearized higher-spin gravity} \label{sec:linear_HS}

In this section, we formulate linearized higher-spin gravity on $EAdS_4$, along with its solution via the Penrose transform. The formulas that appear here will receive a more geometric interpretation in section \ref{sec:spacetime_subgroup}.  In section \ref{sec:linear_HS:fields}, we describe free massless fields of arbitrary integer spin. In section \ref{sec:linear_HS:Penrose}, we review the Penrose transform in (A)dS\textsubscript{4}. In section \ref{sec:linear_HS:unfolded}, we introduce the unfolded formulation, which recasts both the field equations and the Penrose transform into HS-covariant star-product expressions. Finally, in section \ref{sec:linear_HS:antipodal}, we discuss antipodal symmetry $x^\mu \leftrightarrow -x^\mu$ and its analogue in the twistor language.

\subsection{Free massless fields in $EAdS_4$} \label{sec:linear_HS:fields}

Our starting point is a set of free massless fields, one for each integer spin. A field with spin $s>0$ is described by the self-dual and anti-self-dual parts of its field strength (i.e. the higher-spin generalization of the Maxwell tensor and the linearized Weyl tensor). These are encoded by purely left-handed and purely right-handed totally symmetric spinors with $2s$ indices. The field content is thus:
\begin{align}
\text{Spin 0:} \quad C^{(0,0)} \ , \quad \text{Spin 1:} \quad C^{(2,0)}_{\alpha\beta}, C^{(0,2)}_{\dot\alpha\dot\beta} \ , \quad 
\text{Spin 2:} \quad C^{(4,0)}_{\alpha\beta\gamma\delta}, C^{(0,4)}_{\dot\alpha\dot\beta\dot\gamma\dot\delta} \ , \quad \text{etc.} \ , \label{eq:HS_fields}
\end{align}
where the numbers in parentheses signify the number of left-handed and right-handed spinor indices. We are temporarily introducing designated indices $(\alpha,\beta,\dots)$ and $(\dot\alpha,\dot\beta,\dots)$ respectively for left-handed and right-handed Weyl spinors at a bulk point $x$. These are the same as twistor indices $(a,b,\dots)$, but with $P_{L/R}(x)$ chiral projections implied. The spinor fields \eqref{eq:HS_fields} can also be expressed in tensor form, using the convention \eqref{eq:conversion_bivectors} for converting a symmetric pair of twistor indices into an antisymmetric pair of tensor indices. The left-handed and right-handed parts of the spin-$s$ field strength combine into a single tensor, via:
\begin{align}
\begin{split}
C_{\mu_1\nu_1\dots\mu_s\nu_s} &= C^L_{\mu_1\nu_1\dots\mu_s\nu_s} + C^R_{\mu_1\nu_1\dots\mu_s\nu_s} \ ; \\
C^L_{\mu_1\nu_1\dots\mu_s\nu_s} &= \frac{1}{4^s}\gamma_{\mu_1\nu_1}^{\alpha_1\beta_1}\dots \gamma_{\mu_s\nu_s}^{\alpha_s\beta_s} C^{(2s,0)}_{\alpha_1\beta_1\dots\alpha_s\beta_s} \ ; \\
C^R_{\mu_1\nu_1\dots\mu_s\nu_s} &= \frac{1}{4^s}\gamma_{\mu_1\nu_1}^{\dot\alpha_1\dot\beta_1}\dots \gamma_{\mu_s\nu_s}^{\dot\alpha_s\dot\beta_s} C^{(0,2s)}_{\dot\alpha_1\dot\beta_1\dots\dot\alpha_s\dot\beta_s} \ .
\end{split} \label{eq:C_tensor}
\end{align}
The tensor field $C_{\mu_1\nu_1\dots\mu_s\nu_s}$ has the symmetries of a generalized Weyl tensor: it is totally traceless, antisymmetric within each $\mu_k\nu_k$ index pair, symmetric under the exchange of any two such pairs, and vanishes when antisymmetrized over any three indices. The right-handed and left-handed parts of $C_{\mu_1\nu_1\dots\mu_s\nu_s}$ are distinguished by their eigenvalues $\pm 1$ under a Hodge dualization of any $\mu_k\nu_k$ index pair:
\begin{align}
-\frac{1}{2}\epsilon_{\mu_1\nu_1}{}^{\lambda\rho\sigma} x_\lambda C^{R/L}_{\rho\sigma\mu_2\nu_2\dots\mu_s\nu_s} = \pm C^{R/L}_{\rho\sigma\mu_2\nu_2\dots\mu_s\nu_s} \ , \label{eq:self_dual}
\end{align}
where the minus sign on the LHS arises from the fact that the time component of $x_\lambda$ is negative.

Let us now write the field equations satisfied by the field strengths \eqref{eq:HS_fields}. The scalar field $C^{(0,0)}$ satisfies the wave equation for a conformally coupled massless scalar:
\begin{align}
\nabla_\mu\nabla^\mu C^{(0,0)} = -2C^{(0,0)} \ , \label{eq:scalar_equation}
\end{align}
while the fields with spin $s>0$ satisfy the free massless equations:
\begin{align}
\nabla^{\alpha_1}{}_{\dot\beta}\,C^{(2s,0)}_{\alpha_1\alpha_2\dots\alpha_{2s}} = 0 \ ; \quad 
\nabla_\beta{}^{\dot\alpha_1}\,C^{(0,2s)}_{\dot\alpha_1\dot\alpha_2\dots\dot\alpha_{2s}} = 0 \ . \label{eq:spinning_equation}
\end{align}

\subsection{The Penrose transform} \label{sec:linear_HS:Penrose}

The Penrose transform \cite{Penrose:1986ca,Ward:1990vs} is a closed-form general solution to the field equations \eqref{eq:scalar_equation}-\eqref{eq:spinning_equation} in terms of an even (but otherwise unconstrained) twistor function $F(Y)$. It is important that $F(Y)$ is a holomorphic function, i.e. without additional dependence on the complex-conjugate variable $\overline Y$; throughout this paper, we are taking this property of twistor functions for granted.

More specifically, each of the individual fields \eqref{eq:HS_fields}, i.e. each separate helicity, is captured by a twistor function $F(Y)$ of a particular \emph{degree of homogeneity} $-2\pm 2s$. A general even function can be decomposed into eigenfunctions of the homogeneity operator $Y^a(\del/\del Y^a)$, with even integer eigenvalues. In this way, a general even function $F(Y)$ contains a single free massless field of each helicity, i.e. precisely the higher-spin multiplet \eqref{eq:HS_fields}.

In the notations of this paper, the Penrose transform for each of the fields \eqref{eq:HS_fields} reads:
\begin{align}
 C^{(2s,0)}_{\alpha_1\dots\alpha_{2s}}(x) &= i\int_{P_R(x)} d^2u_R \left.\frac{\del^s F_R(u_L+u_R)}{\del u_L^{\alpha_1}\dots\del u_L^{\alpha_{2s}}}\right|_{u_L = 0} \ ; \label{eq:Penrose_R_to_L} \\ 
 C_{(0,2s)}^{\dot\alpha_1\dots\dot\alpha_{2s}}(x) &= i(-1)^s\int_{P_R(x)} d^2u_R\, u_R^{\dot\alpha_1}\dots u_R^{\dot\alpha_{2s}} F_R(u_R) \ , \label{eq:Penrose_R_to_R}
\end{align}
where $F_R(Y)$ is an arbitrary even twistor function, and the factors of $i$ and $(-1)^s$ are for later convenience. The spin-0 field $C^{(0,0)}$ is contained in \eqref{eq:Penrose_R_to_L}-\eqref{eq:Penrose_R_to_R} a shared special case:
\begin{align}
 C^{(0,0)}(x) &= i\int_{P_R(x)} d^2u_R\, F_R(u_R) \ . \label{eq:Penrose_R_to_scalar}
\end{align}
The $R$ subscript in $F_R(Y)$ is to indicate that the integrals in \eqref{eq:Penrose_R_to_L}-\eqref{eq:Penrose_R_to_scalar} are over the 2d spinor subspace $P_R(x)$. An alternative transform, using $P_L(x)$ instead, reads:
\begin{align}
 C_{(2s,0)}^{\alpha_1\dots\alpha_{2s}}(x) &= -i(-1)^s\int_{P_L(x)} d^2u_L\, u_L^{\alpha_1}\dots u_L^{\alpha_{2s}} F_L(u_L) \ ; \label{eq:Penrose_L_to_L} \\ 
 C^{(0,2s)}_{\dot\alpha_1\dots\dot\alpha_{2s}}(x) &= -i\int_{P_L(x)} d^2u_L \left.\frac{\del^s F_L(u_L+u_R)}{\del u_R^{\dot\alpha_1}\dots\del u_R^{\dot\alpha_{2s}}}\right|_{u_R = 0} \ , \label{eq:Penrose_L_to_R}
\end{align}
where $F_L(Y)$ is again an arbitrary even twistor function, and we introduced an extra sign factor for later convenience. The transforms \eqref{eq:Penrose_R_to_L}-\eqref{eq:Penrose_L_to_R} can also be written in Dirac-spinor (i.e. twistor) indices, as:
\begin{align}
 \begin{split}
   C^{(2s,0)}_{a_1\dots a_{2s}}(x) = i\int_{P_R(x)} d^2u \left.\frac{\del^s F_R(U)}{\del U^{a_1}\dots\del U^{a_{2s}}}\right|_{U = u} = -i(-1)^s\int_{P_L(x)} d^2u\, u_{a_1}\dots u_{a_{2s}} F_L(u) \ ; \\
   C^{(0,2s)}_{a_1\dots a_{2s}}(x) = i(-1)^s\int_{P_R(x)} d^2u\, u_{a_1}\dots u_{a_{2s}} F_R(u) = -i\int_{P_L(x)} d^2u \left.\frac{\del^s F_L(U)}{\del U^{a_1}\dots\del U^{a_{2s}}}\right|_{U = u} \ .
 \end{split}
\end{align}
Here, the integrals automatically project the Dirac indices into the correct Weyl subspace in each case.

Proving that the fields \eqref{eq:Penrose_R_to_L}-\eqref{eq:Penrose_L_to_R} indeed satisfy the field equations \eqref{eq:scalar_equation}-\eqref{eq:spinning_equation} is rather straightforward. The main subtlety is the $x$-dependence of the spinor integration range, which must be taken into account when taking spacetime derivatives. This can be dealt with by same method as when deriving eq. \eqref{eq:nabla_delta}, i.e. by performing a change of variables that shifts the $x$-dependence into the integrand. The details, in a slightly different language, can be found e.g. in \cite{Neiman:2013hca}. 

The above presentation of the Penrose transform differs somewhat from the one normally given in a twistor-theory textbook. The first difference is that that we're starting in (A)dS spacetime, and treating twistors as the spinors of the isometry group $SO(1,4)$. Normally, one starts instead in flat spacetime, and treats twistors as the spinors of the conformal group $SO(2,4)$ (which, with our $EAdS_4$ signature, would actually be $SO(1,5)$). As far as the Penrose transform is concerned, this difference is merely superficial: both the transform and the free massless field equations are conformally covariant, so that Minkowski and (A)dS are equally good starting points. That being said, the unfolded, star-product-based formalism of the next subsection is \emph{not} covariant under the 4d conformal group; there, the non-vanishing cosmological constant will be crucial.

Another difference between our presentation and the standard one is that the integrals in \eqref{eq:Penrose_R_to_L}-\eqref{eq:Penrose_L_to_R} are over $\bbC^2$ spinor subspaces (with measure $d^2u$), as opposed to their projective $\bbC\bbP^1$ versions (with measure $udu$). Thus, we are using the (well-known, but not as common) ``non-projective'' version of the transform. The projective vs. non-projective integrals are very closely related. In particular, the non-projective integrals \eqref{eq:Penrose_R_to_L}-\eqref{eq:Penrose_R_to_R} pick out the component of the twistor function $F_R(Y)$ with homogeneity $-2\pm 2s$ respectively, as one can show by rescaling the integration variable. This is the already-mentioned relation between helicity and the homogeneity of the twistor function. For a function $F_R(Y)$ of the ``correct'' homogeneity, the projective integral $u_R du_R$ will agree with the non-projective one up to numerical factors; essentially, the extra 1d integral in the non-projective case can be treated as $\int d\alpha/\alpha = \pm 2\pi i$. For a function $F_R(Y)$ of the ``wrong'' homogeneity, the projective integral is ill-defined, while the non-projective one evaluates to zero. Thus, the non-projective Penrose transform \eqref{eq:Penrose_R_to_L}-\eqref{eq:Penrose_L_to_R} is the same as the projective one, except that it allows us not to worry about mixing different spins/homogeneities in the integrand.

Finally, as we repeatedly discuss in this paper, integrals of the form \eqref{eq:Penrose_R_to_L}-\eqref{eq:Penrose_L_to_R} suffer from contour ambiguities. From the HS point of view, these are directly related to the analogous ambiguity in the integral definition \eqref{eq:star_int} of the star product. Due to these contour ambiguities, the Penrose transform is more properly defined in terms of sheaf cohomology. However, in keeping with the HS literature, we do not follow that more rigorous path, and instead continue working with ordinary functions, while keeping the ambiguity in mind.  The advantage of this ``naive'' approach is that it allows us to treat $F_{L/R}(Y)$ and $C(x;Y)$ on an equal footing, as ordinary functions of the $Y$ variable.

\subsection{Unfolded formulation and the higher-spin-covariant perspective} \label{sec:linear_HS:unfolded}

The next step is to rephrase the dynamics of our free massless fields in unfolded form. Let us introduce the full set of on-shell-inequivalent derivatives of the fields $C^{(2s,0)},C^{(0,2s)}$ for $s\geq 0$:
\begin{align}
 \begin{split}
   \big(C^{(2s+k,k)}\big)_{\alpha_1\dots\alpha_{2s}\beta_1\dots\beta_k}{}^{\dot\beta_1\dots\dot\beta_k} 
    &= i^k\nabla_{(\beta_1}{}^{(\dot\beta_1}\dots\nabla_{\beta_k}{}^{\dot\beta_k)}\,C^{(2s,0)}_{\alpha_1\dots\alpha_{2s})} \ ; \\
   \big(C^{(k,2s+k)}\big)^{\beta_1\dots\beta_k}{}_{\dot\beta_1\dots\dot\beta_k\dot\alpha_1\dots\dot\alpha_{2s}} 
    &= i^k\nabla^{(\beta_1}{}_{(\dot\beta_1}\dots\nabla^{\beta_k)}{}_{\dot\beta_k}\,C^{(0,2s)}_{\dot\alpha_1\dots\dot\alpha_{2s})} \ ,
 \end{split} \label{eq:unfolding}
\end{align}
where the factors of $i$ are for later convenience. We now have a field $C^{(m,n)}$ for every pair of integers $m,n$ such that $m+n$ is even, i.e. one field for every integer-spin representation of the bulk rotation group $SO(4)$. We can neatly package these into a single scalar master field $C(x;Y)$, which is an even function of the twistor coordinate $Y$: 
\begin{align}
 \begin{split}
   C(x;Y) &= \sum_{m,n} \frac{1}{m!n!}\, C^{(m,n)}_{\alpha_1\dots\alpha_m\dot\alpha_1\dots\dot\alpha_n}\,
     y_L^{\alpha_1}\dots y_L^{\alpha_m}y_R^{\dot\alpha_1}\dots y_R^{\dot\alpha_n} \ ; \\
   C^{(m,n)}_{\alpha_1\dots\alpha_m\dot\alpha_1\dots\dot\alpha_n} 
     &= (P_L)^{a_1}{}_{\alpha_1}\dots (P_L)^{a_m}{}_{\alpha_m}(P_R)^{a_{m+1}}{}_{\dot\alpha_1}\dots (P_R)^{a_{m+n}}{}_{\dot\alpha_n}\,
       \left.\frac{\del^{m+n} C}{\del Y^{a_1}\dots\del Y^{a_{m+n}}}\right|_{Y=0} \ ,
 \end{split} \label{eq:master_field}
\end{align}
where $y_{L/R} = P_{L/R}(x)Y$ are the chiral components of $Y$ at the point $x$.  The field equations \eqref{eq:scalar_equation}-\eqref{eq:spinning_equation} and the definitions \eqref{eq:unfolding} are all encapsulated in the following unfolded equation:
\begin{align}
 \nabla_\mu C = \frac{i}{2}\,C\star (y_L\gamma_\mu y_R) \ ,
\end{align}
or, expressing $y_L$ and $y_R$ explicitly in terms of $Y$ and $x$:
\begin{align}
 \nabla_\mu C = \frac{i}{4}\,C\star (Y\gamma_\mu xY) \ . \label{eq:nabla_C}
\end{align}
The star product in \eqref{eq:nabla_C} breaks down into three terms of the form $CYY$, $(\del C/\del Y)Y$ and $\del^2 C/\del Y^2$. Among these, the $(\del C/\del Y)Y$ piece accounts for the $x$ dependence of the chiral decomposition $Y = y_L+y_R$ in \eqref{eq:master_field}, while the other two account for the $x$ dependence of the component fields $C^{(m,n)}$ themselves. Specifically, the $\del^2 C/\del Y^2$ term encodes the flat-spacetime version of eqs. \eqref{eq:scalar_equation}-\eqref{eq:unfolding}, while the $CYY$ term corrects the second derivatives to account for the curvature of $EAdS_4$.

Having written the unfolded equation in the form \eqref{eq:nabla_C}, we recognize from \eqref{eq:nabla_delta} that it is solved by the chiral delta functions $\delta^{L/R}_x(Y)$. Moreover, we see that the general solution can be expressed as:
\begin{align}
 C(x;Y) = F_R(Y)\star i\delta^R_x(Y) = i\int_{P_R(x)} d^2u_R\, F_R(Y + u_R)\,e^{iu_R Y} \ , \label{eq:C_solution_R}
\end{align}
or, equivalently:
\begin{align}
C(x;Y) = -F_L(Y)\star i\delta^L_x(Y) = -i\int_{P_L(x)} d^2u_L\, F_R(Y + u_L)\,e^{iu_L Y} \ , \label{eq:C_solution_L}
\end{align}
where the $\pm i$ factors are chosen for later convenience, and we used eqs. \eqref{eq:delta_Fourier_bulk_right}-\eqref{eq:delta_Fourier_bulk_left} to obtain the explicit integral expressions. The spacetime-independent functions $F_{L/R}(Y)$ are Fourier transforms of each other:
\begin{align}
F_R(Y) = -F_L(Y)\star\delta(Y) \ . \label{eq:F_LR}
\end{align}
Moreover, using the decomposition \eqref{eq:master_field}, one can see that these functions are the same as the $F_{L/R}(Y)$ from section \ref{sec:linear_HS:Penrose}. Thus, we recognize eqs. \eqref{eq:C_solution_R}-\eqref{eq:C_solution_L} as the unfolded, HS-covariant formulation of the Penrose transform!

It may seem strange that the unfolded equation \eqref{eq:nabla_C} prefers $C\star(Y\gamma_\mu xY)$ over $(Y\gamma_\mu xY)\star C$. It turns out that the second possibility is in fact realized, if we replace $i\rightarrow -i$ in the definition \eqref{eq:unfolding} of the unfolded fields. Equivalently, we can define an alternative master field $\tilde C$ as:
\begin{align}
 \tilde C(x;Y) &= \sum_{m,n} \frac{(-1)^m}{m!n!}\, C^{(m,n)}_{\alpha_1\dots\alpha_m\dot\alpha_1\dots\dot\alpha_n}\,
     y_L^{\alpha_1}\dots y_L^{\alpha_m}y_R^{\dot\alpha_1}\dots y_R^{\dot\alpha_n} = C(x;y_R - y_L) = C(x;xY) \ ,
\end{align}
for which the field equation and its solution read:
\begin{align}
 \nabla_\mu\tilde C &= -\frac{i}{4}(Y\gamma_\mu xY)\star\tilde C \ ; \label{eq:nabla_C_tilde} \\
 \tilde C(x;Y) &= -i\delta^R_x(Y)\star F_R(Y) = i\delta^L_x(Y)\star F_L(Y) \ .
\end{align}
In fact, it follows from \eqref{eq:delta_Klein_bulk} that the Penrose transform $F_{L/R}(Y)$ is the same as in \eqref{eq:C_solution_R}-\eqref{eq:C_solution_L}.

Since $\delta^R_x(Y)$ and $\delta^L_x(Y)$ square to unity, we can explicitly solve for the master field at a point $x'$ in terms of the master field at a point $x$ via:
\begin{align}
 C(x';Y) = C(x;Y)\star\delta^R_x(Y)\star\delta^R_{x'}(Y) \ ,
\end{align}
where the two-point product $\delta^R_x\star\delta^R_{x'} = \delta^L_x\star\delta^L_{x'}$ is given by the Gaussian \eqref{eq:delta_x_x}:
\begin{align}
 \delta^R_x(Y)\star\delta^R_{x'}(Y) = \delta^L_x(Y)\star\delta^L_{x'}(Y) = \frac{2}{1 - x\cdot x'}\exp\frac{iYxx'Y}{2(x\cdot x' - 1)} \ . \label{eq:bulk_bulk_master}
\end{align} 
It is simultaneously a solution to the unfolded equation \eqref{eq:nabla_C} in $x'$ and to the ``flipped'' equation \eqref{eq:nabla_C_tilde} in $x$. The fact that the master field at $x'$ can be deduced from its value at a \emph{single} point $x$ is a feature of the unfolded formalism.

Note that in all of the above, we did not require a higher-spin gauge connection. Instead, we directly wrote the linear field equations and their solutions in terms of gauge-invariant field strengths on the background $EAdS_4$ geometry. HS symmetry appears only as a \emph{global} symmetry of the equations, parameterized by a spacetime-independent even function $\varepsilon(Y)$. The Penrose transform $F_R(Y)$ transforms under this symmetry in the adjoint:
\begin{align}
 \delta F_R = \varepsilon\star F_R - F_R\star\varepsilon \ ,
\end{align}
and likewise for $F_L(Y)$. The master field $C(x;Y)$ transforms in the ``twisted adjoint'':
\begin{align}
 \delta C = \varepsilon\star C - C\star\delta^R_x\star\varepsilon\star\delta^R_x \ ,
\end{align}
where the product $\delta^R_x\star\varepsilon\star\delta^R_x = \delta^L_x\star\varepsilon\star\delta^L_x$ can be evaluated as in \eqref{eq:delta_Klein_bulk}.

In section \ref{sec:CFT}, we will similarly describe the free $U(N)$ vector model (with external sources) in a language that renders global higher-spin symmetry manifest, while avoiding any gauge redundancy.

\subsection{Antipodal symmetry} \label{sec:linear_HS:antipodal}

A special role is played by solutions with the antipodal symmetry:
\begin{align}
 C(-x;Y) = \pm C(x;Y) \ , \label{eq:C_antipodal}
\end{align}
The antipodal map $x^\mu\rightarrow -x^\mu$ is the central element of the spacetime symmetry group $O(1,4)$. Under this map, the $EAdS_4$ hyperboloid \eqref{eq:EAdS} is sent into its $x^0<0$ counterpart. Thus, strictly speaking, $C(-x;Y)$ is an analytic continuation of the solution $C(x;Y)$ into the antipodal $EAdS_4$. In the Poincare coordinates \eqref{eq:Poincare}, the antipodal map corresponds to the operation $z\rightarrow -z$, which was invoked in the discussion \cite{Vasiliev:2012vf} of higher-spin holography. Indeed, as we will see in section \ref{sec:holography:asymptotics}, the two antipodal parities in \eqref{eq:C_antipodal} directly correspond to the two types of asymptotic boundary data for each of the component fields in $C(x;Y)$ \cite{Vasiliev:2012vf}. A detailed analysis of this relation in the language of individual fields was carried out in \cite{Neiman:2014npa,Halpern:2015zia} (see also \cite{Ng:2012xp}). The antipodal symmetry is also of significance in the de Sitter context \cite{Halpern:2015zia}, as we will review in section \ref{sec:discuss}. 

Let us now see how the symmetry \eqref{eq:C_antipodal} is expressed at the level of spacetime-independent twistor functions. Plugging the identity $\delta^L_x(Y) = \delta^R_{-x}(Y)$ into the Penrose transform \eqref{eq:C_solution_R}-\eqref{eq:F_LR}, we obtain that \eqref{eq:C_antipodal} is equivalent to any of the following:
\begin{align}
 F_L(Y) = \mp F_R(Y) \  \Longleftrightarrow \ F_R(Y)\star\delta(Y) = \pm F_R(Y) \ \Longleftrightarrow \ F_L(Y)\star\delta(Y) = \pm F_L(Y) \ . \label{eq:F_antipodal}
\end{align}
Taking the Penrose transform of \eqref{eq:F_antipodal}, we can also express \eqref{eq:C_antipodal} as a star-product symmetry of $C(x;Y)$ at a single point $x$:
\begin{align}
 C(x;Y)\star\delta(Y) = \pm C(x;Y) \ . \label{eq:C_antipodal_delta}
\end{align}
An arbitrary bulk solution $C(x;Y)$ can be decomposed into antipodally even and odd pieces in the sense of \eqref{eq:C_antipodal}-\eqref{eq:C_antipodal_delta}. For the twistor functions $F_{L/R}(Y)$, as well as for $C(x;Y)$ viewed as a function of $Y$ at fixed $x$, the corresponding decomposition is accomplished by the projectors $\calP_\pm(Y)$ from \eqref{eq:projectors_delta}. 

That being said, we must emphasize that conditions such as \eqref{eq:F_antipodal}-\eqref{eq:C_antipodal_delta}, as well as the projectors $\calP_\pm(Y)$, should be handled with caution, due to contour ambiguities in the star product, as well as in the delta functions $\delta(Y),\delta^{L/R}_x(Y)$ themselves. When in doubt, it is helpful to look back to the original condition \eqref{eq:C_antipodal} in spacetime. We will now present a simple example that shows how \eqref{eq:F_antipodal}-\eqref{eq:C_antipodal_delta} can fail to be well-defined linear properties, or, equivalently, how $\calP_\pm(Y)$ can fail to be well-defined projectors.

 Consider a conformally-coupled massless scalar field, with field equation \eqref{eq:scalar_equation}. An important solution to this field equation is the boundary-to-bulk propagator $1/(\ell\cdot x)$. For $x^\mu$ timelike, i.e. on $EAdS_4$ and its antipodal image, this propagator is non-singular, and is odd under the antipodal map $x\rightarrow -x$. Now, consider a superposition of such propagators, obtained by integrating $\ell^\mu$ over an $S_3$ section of the $\bbR^{1,4}$ lightcone, i.e. over the 3-sphere $(\ell\cdot\ell = 0, \ \ell\cdot x_0 = -1)$, where $x_0$ is some future-pointing unit vector. This can be expressed as a conformally covariant $d^3\ell$ integral by inserting $1/(\ell\cdot x_0)^2$ into the integrand. The result reads:
\begin{align}
 \int \frac{d^3\ell}{(\ell\cdot x_0)^2}\,\frac{1}{(\ell\cdot x)} = 4\pi^2 \times \left\{
    \begin{array}{ll}
       \displaystyle 1/(x\cdot x_0 - 1) & \quad x\text{ future-pointing} \\
       \displaystyle 1/(x\cdot x_0 + 1) & \quad x\text{ past-pointing}
    \end{array} \right. \ , \label{eq:nonanalytic}
\end{align}
where we recall that future-pointing vs. past-pointing $x^\mu$ correspond to points on the original $EAdS_4$ vs. the antipodal one. The key property of the bulk solution \eqref{eq:nonanalytic} is that it's still odd under $x\rightarrow -x$, but this is accomplished non-analytically: if we were to analytically continue the solution from future-pointing $x$ to past-pointing $x$, the result wouldn't have a definite antipodal parity. As an aside, note that the RHS of \eqref{eq:nonanalytic} is just a bulk-to-bulk propagator between $x_0$ and $x$. Therefore, \eqref{eq:nonanalytic} is a simple example of the split representation \cite{Costa:2014kfa} of bulk-to-bulk propagators as boundary integrals. 

Now, let us upgrade the statement \eqref{eq:nonanalytic} to the master-field level. The master field for the boundary-to-bulk propagator $1/(\ell\cdot x)$ reads:
\begin{align}
 C_\ell(x;Y) = \frac{1}{2}\,\delta_\ell(Y)\star\delta_x^L(Y) = -\frac{1}{2}\,\delta_\ell(Y)\star\delta_x^R(Y) = \frac{1}{\ell\cdot x}\exp\frac{iY\ell xY}{2\ell\cdot x} \ .
\end{align}
This is clearly antipodally odd, both in the spacetime sense of \eqref{eq:C_antipodal} and in the star-product sense of \eqref{eq:C_antipodal_delta}:
\begin{align}
 C_\ell(x;Y)\star\delta(Y) = C_\ell(-x;Y) = -C_\ell(x;Y) \ . 
\end{align}
However, taking the linear superposition \eqref{eq:nonanalytic}, we find:
\begin{align}
\int \frac{d^3\ell}{(\ell\cdot x_0)^2}\,C_\ell(x;Y) = 4\pi^2 \times \left\{
  \begin{array}{ll}
     \displaystyle C^{(-)}_{x_0}(x;Y) & \quad x\text{ future-pointing} \\
     \displaystyle C^{(+)}_{x_0}(x;Y) & \quad x\text{ past-pointing}
  \end{array} \right. \ , \label{eq:nonanalytic_C}
\end{align}
where $C^{(\pm)}_{x_0}(x;Y)$ is the master field corresponding to the bulk-to-bulk propagator $1/(x\cdot x_0 \pm 1)$, which we encountered in \eqref{eq:bulk_bulk_master}:
\begin{align}
 \begin{split}
   C^{(-)}_{x_0}(x;Y) &= -\frac{1}{2}\,\delta^L_{x_0}(Y)\star\delta^L_x(Y) = -\frac{1}{2}\,\delta^R_{x_0}(Y)\star\delta^R_x(Y) = \frac{1}{x\cdot x_0 - 1}\exp\frac{iYx_0 xY}{2(x\cdot x_0 - 1)} \ ; \\
   C^{(+)}_{x_0}(x;Y) &= +\frac{1}{2}\,\delta^R_{x_0}(Y)\star\delta^L_x(Y) = +\frac{1}{2}\,\delta^L_{x_0}(Y)\star\delta^R_x(Y) = \frac{1}{x\cdot x_0 + 1}\exp\frac{iYx_0 xY}{2(x\cdot x_0 + 1)} \ .
 \end{split} \label{eq:nonanalytic_C_details}
\end{align}
We now see that while the spacetime antipodal symmetry $C(-x;Y) = -C(x;Y)$ is preserved by the superposition \eqref{eq:nonanalytic_C}, its star-product analogue $C(x;Y)\star\delta(Y) = -C(x;Y)$ is not. We also see exactly why this happens: the master field $C^{(\pm)}_{x_0}(x;Y)$ on each branch contains only the local Taylor series of the bulk solution. Therefore, the master field at each $x$ ``sees'' the analytic continuation of the bulk solution from the neighborhood of $x$, which does \emph{not} have definite antipodal parity, instead of seeing the antipodally odd, but nonanalytic, global superposition \eqref{eq:nonanalytic_C}-\eqref{eq:nonanalytic_C_details}.

\section{Higher-spin representation of spacetime symmetries and the Penrose transform} \label{sec:spacetime_subgroup}

As we've seen in eq. \eqref{eq:generators}, the quadratic elements $Y_a Y_b$ of the higher-spin algebra generate the spacetime symmetry group $SO(1,4)$. In this section, we consider the finite group elements that arise by exponentiating these generators (the completion of $SO(1,4)$ into $O(1,4)$ will be addressed in section \ref{sec:spacetime_subgroup:HS:antipodal}). In the process, we will clarify the role of the delta functions $\delta(Y),\delta^{L/R}_x(Y),\delta_\ell(Y)$ with respect to spacetime symmetries. This in turn will lead us to the geometric interpretation \eqref{eq:CPT}-\eqref{eq:sqrt_CPT} of the Penrose transform as a square root of CPT.

\subsection{Clifford algebra} \label{sec:spacetime_subgroup:clifford}

As mentioned in eq. \eqref{eq:Clifford_HS}, HS algebra is just a simple variation on Clifford algebra, where the vector $\gamma_\mu$ subject to anticommutation relations is replaced with a twistor $Y_a$ subject to commutation relations. Correspondingly, our analysis below will closely mirror the well-known geometric properties of Clifford algebra (for a particularly spirited review of these, see \cite{GeometricAlgebra}). In Clifford algebra, commutation with the infinitesimal generators $\gamma_{[\mu}\gamma_{\nu]}/2$ realizes the standard action of the orthogonal group -- in our case, $SO(1,4)$:
\begin{align}
 \left[\frac{1}{2}\gamma_{[\nu}\gamma_{\rho]}, \gamma^{\mu_1}\dots\gamma^{\mu_n} \right] 
   = 2\left(\delta^{\mu_1}_{[\nu}\gamma_{\rho]}\gamma^{\mu_2}\!\dots\gamma^{\mu_n} + \ldots + \gamma^{\mu_1}\!\dots\gamma^{\mu_{n-1}}\delta^{\mu_n}_{[\nu}\gamma_{\rho]}\right) \ . \label{eq:infinitesimal_Clifford}
\end{align}
Alternatively, instead of starting with infinitesimal generators, one can construct $SO(1,4)$ out of some fundamental \emph{finite} transformations. In particular, the adjoint action of $x = x^\mu\gamma_\mu$ is a reversal of the subspace orthogonal to a unit vector $x^\mu$. In our conventions, with $x^\mu$ timelike, this reads explicitly as:
\begin{align}
 x\,\gamma^{\mu_1}\!\dots\gamma^{\mu_n} x = \tilde\gamma^{\mu_1}\!\dots\tilde\gamma^{\mu_n} \ , \quad \text{where} \quad \tilde\gamma^\mu \equiv -(\delta^\mu_\nu + 2x^\mu x_\nu)\gamma^\nu \ . \label{eq:reflection_Clifford}
\end{align}
By combining two such reversals with respect to a pair of axes $x,x'$, one obtains a finite rotation (or boost) by \emph{twice the angle} between $x$ and $x'$:
\begin{align}
 xx'\gamma^{\mu_1}\!\dots\gamma^{\mu_n} x'x = \tilde\gamma^{\mu_1}\!\dots\tilde\gamma^{\mu_n} \ , \quad \text{where} \quad 
 \tilde\gamma^\mu \equiv \left(\delta^\mu_\nu + 2x^\mu x_\nu + 2x'^\mu x'_\nu + 4(x\cdot x')x^\mu x'_\nu \right)\gamma^\nu \ . \label{eq:finite_rotation_Clifford}
\end{align}
In particular, a rotation by $\pi$ (in a spacelike plane) can be represented by $xx'$ with $x'$ perpendicular to $x$. A rotation by $2\pi$, represented by the algebra element $xx' = -1$, is obtained via $x' = -x$. The infinitesimal generators $\gamma_{[\mu}\gamma_{\nu]}/2$ can be obtained by expanding \eqref{eq:finite_rotation_Clifford} around $x=x'$. 

In odd dimensions, such as our case with the embedding space $\bbR^{1,4}$, one can also go in the opposite direction, and derive the reflection \eqref{eq:reflection_Clifford} by exponentiating the infinitesimal generators. The way to do this in $\bbR^{1,4}$ is to rotate by $\pi$ in a pair of planes orthogonal both to $x^\mu$ and to each other. If the rotation is performed in both planes at once, then, depending on the planes' orientation, it will belong to either the left-handed or the right-handed subgroup of the 4d rotations $SO(4) = SO(3)_L\times SO(3)_R$ around $x^\mu$. We then obtain either $x$ or $-x$ as the reflection operator. When used in the adjoint, both $x$ and $-x$ produce the same reflection \eqref{eq:reflection_Clifford}. 

For comparison with the higher-spin case below, let us perform this calculation explicitly. We choose a frame such that $x^\mu = e_0^\mu$, and use the representation \eqref{eq:gamma_timelike} for the gamma matrices. Now, consider e.g. a right-handed rotation along the bivector $e_1\wedge e_2 + e_3\wedge e_4$. A rotation by an infinitesimal angle $\varepsilon$ in each of the two planes $e_1\wedge e_2$ and $e_3\wedge e_4$ is represented in Clifford algebra by:
\begin{align}
  1 + \frac{\varepsilon}{2}(\gamma_1\gamma_2 + \gamma_3\gamma_4) = 1 - i\varepsilon\begin{pmatrix} 0 & 0 \\ 0 & \sigma_3 \end{pmatrix} \ .
\end{align}
Exponentiating, we obtain the operator for rotation by a finite angle $\theta$:
\begin{align}
g_\theta = \exp\left(-i\theta\begin{pmatrix} 0 & 0 \\ 0 & \sigma_3 \end{pmatrix}\right) = \begin{pmatrix} 1 & 0 \\ 0 & \cos\theta - i\sin\theta\,\sigma_3 \end{pmatrix} \ . \label{eq:RH_Clifford}
\end{align}
In particular, for $\theta = \pi$, we get the operator:
\begin{align}
g_\pi = \begin{pmatrix} 1 & 0 \\ 0 & -1 \end{pmatrix} = -\gamma_0 = -x \ .
\end{align}
Similarly, a left-handed rotation along the bivector $e_1\wedge e_2 - e_3\wedge e_4$ will produce $+x$ as the reflection operator. Note that $g_{2\pi} = 1$, as it should be: while a single $2\pi$ rotation need not take us back to the identity, a combination of $2\pi$ rotations in a \emph{pair} of planes must always do so.

So far, we considered the \emph{adjoint} action $g\Gamma g^{-1}$ in Clifford algebra, where $g$ represents an $SO(1,4)$ group element. As we saw in \eqref{eq:infinitesimal_Clifford} or \eqref{eq:reflection_Clifford}, this realizes the standard action of $SO(1,4)$ on the algebra element $\Gamma$, which consists of spin-0 and spin-1 pieces. The next natural question is what happens if one acts instead in the fundamental, i.e. simply via $g\Gamma$. The answer, of course, is that this transformation law describes spinors (or, in our case, twistors). In particular, the reflection \eqref{eq:reflection_Clifford} is realized on twistors as $U\rightarrow \pm xU$. 

When describing spinors from within the Clifford algebra itself (as Cartan had done originally), the geometric structure of $SO(1,4)$ becomes obscured. That is why it's better to introduce separate indices for spinors, and to develop geometric intuition about spinor space in its own right. As we will see below, the situation in higher-spin algebra is different: there, the $SO(1,4)$ can be made manifest not only in the algebra's adjoint representation, but also in the fundamental. In both the Clifford and HS cases, while the adjoint action of $SO(1,4)$ can be formulated on the individual vector $\gamma_\mu$ or twistor $Y_a$, the fundamental action mixes different powers of these objects.

\subsection{Higher-spin algebra} \label{sec:spacetime_subgroup:HS}

Let us now perform the analogous analysis for HS algebra in place of Clifford algebra. Since infinitesimal $SO(1,4)$ rotations are generated by $Y_a Y_b$, finite rotations will be generated, via exponentiation, by Gaussian functions. In addition, we will find that various reflections are represented by delta functions. 

As a first step, let us identify the higher-spin analog of $x^\mu\gamma_\mu$ -- the reflection that reverses the subspace orthogonal to $x^\mu$. As discussed above, we can construct this group element from the infinitesimal generators by performing a left-handed or right-handed rotation by $\pi$ in a pair of totally orthogonal planes. In the higher-spin algebra, such rotations are generated by the bilinears $y^a_L y^b_L$ or $y^a_R y^b_R$, respectively. As before, we fix $x^\mu = e_0^\mu$, and use the representation \eqref{eq:gamma_timelike} for the gamma matrices. A twistor $Y^a$ can now be decomposed as:
\begin{align}
 Y^a = \begin{pmatrix} y_L^0 \\ y_L^1 \\ y_R^0 \\ y_R^1 \end{pmatrix} \ ,
\end{align}
where the top and bottom halves correspond to the left-handed and right-handed parts of $Y^a$ at $x^\mu$. Now, consider again a right-handed rotation along the bivector $e_1\wedge e_2 + e_3\wedge e_4$. In our chosen basis, we read off from eq. \eqref{eq:generators} that a rotation by an infinitesimal angle $\varepsilon$ in each of the two planes $e_1\wedge e_2,e_3\wedge e_4$ is represented in HS algebra by $1 + (\varepsilon/2)y_R^0 y_R^1$. Exponentiating with the star product, we obtain the operator for rotation by a finite angle $\theta$:
\begin{align}
 R_\theta(Y) = \exp_\star\left(\frac{\theta}{2}\,y_R^0\, y_R^1 \right) = \frac{1}{\cos(\theta/2)}\exp\left(\tan\frac{\theta}{2}\,y_R^0\, y_R^1\right) \ . \label{eq:RH_Gaussian}
\end{align}
One can verify this formula for the star-exponential by differentiating both sides with respect to $\theta$. Note the appearance of $\theta/2$ in eq. \eqref{eq:RH_Gaussian}, as opposed to $\theta$ in its Clifford-algebra analog \eqref{eq:RH_Clifford}. In particular, for $\theta = 2\pi$, we get:
\begin{align}
 R_{2\pi}(Y) = -1 \ . \label{eq:sign_inconsistency}
\end{align}
This signals a problem: even in a spinor representation, a $2\pi$ rotation in a \emph{pair} of planes must return the identity. To resolve this contradiction, we must recall from section \ref{sec:geometry:twistors:integrals} that the star product of Gaussians is only defined \emph{up to sign}. Now that we understand Gaussians as $Spin(1,4)$ elements, eq. \eqref{eq:sign_inconsistency} is teaching us that there is \emph{no globally consistent way} to fix this sign ambiguity. For instance, one attempt to fix the sign ambiguity may be to define ``the'' Gaussian representing a $Spin(1,4)$ element as the one obtained by the shortest direct route from the identity, i.e. by exponentiating a generator through the smallest possible angle. However, this definition would break down for precisely the case we're interested in: the reversal  $-(\delta^\mu_\nu + 2x^\mu x_\nu)$ of a 4d subspace in $\bbR^{1,4}$, which may be realized as a rotation \eqref{eq:RH_Gaussian} with $\theta=\pi$.

Consider, then, the rotation \eqref{eq:RH_Gaussian} with $\theta = \pi$. In this limit, the coefficients both outside and inside the exponent diverge, and the Gaussian becomes a delta function over the $y_R$ spinor space: $R_\pi(Y) \sim \delta^R_x(Y)$. To find the normalization, we integrate \eqref{eq:RH_Gaussian} over $y_R$ using eq. \eqref{eq:Gaussian_spinor}:
\begin{align}
 \int R_\theta(y_R)\,d^2y_R = \frac{\pm i}{\sin(\theta/2)} \ .
\end{align}
Thus, in the limit $\theta = \pi$, the integral is $\pm i$, and we identify the reflection operator as:
\begin{align}
 R_\pi(Y) = \pm i\delta^R_x(Y) \ . \label{eq:reflection_delta}
\end{align}
As we will shortly see, the sign ambiguity here cannot be fixed. If we were to construct the reflection $-(\delta^\mu_\nu + 2x^\mu x_\nu)$ via a left-handed rotation, we would instead get $\pm i\delta^L_x(Y)$ as the reflection operator. In other words, $\pm i\delta^L_x(Y)$ and $\pm i\delta^R_x(Y)$ in higher-spin algebra play the same geometric role as do $x$ and $-x$ in Clifford algebra. In fact, we've already seen in \eqref{eq:delta_Klein_bulk} that the adjoint action of $\pm i\delta^{L/R}_x(Y)$ directly realizes the reflection $Y\rightarrow \pm xY$:
\begin{align}
 \begin{split}
    \left(\pm i\delta^L_x(Y)\right)\star f(Y)\star \left(\pm i\delta^L_x(Y)\right)^{-1}_\star &= \delta^L_x(Y)\star f(Y)\star\delta^L_x(Y) = f(xY) \ ; \\
    \left(\pm i\delta^R_x(Y)\right)\star f(Y)\star \left(\pm i\delta^R_x(Y)\right)^{-1}_\star &= \delta^R_x(Y)\star f(Y)\star\delta^R_x(Y) = f(-xY) \ .
 \end{split} \label{eq:reflection_HS}
\end{align}
As with eq. \eqref{eq:reflection_Clifford} in Clifford algebra, we can now take the reflection \eqref{eq:reflection_HS} as the fundamental geometric operation in place of the infinitesimal generators $Y_a Y_b$. In particular, the product \eqref{eq:delta_x_x} of two reflections with respect to the unit vectors $x,x'$ gives a rotation in the corresponding plane by twice the angle between $x$ and $x'$:
\begin{align}
 \left(\pm i\delta^L_x(Y)\right)\star\left(\mp i\delta^L_{x'}(Y)\right) = \left(\pm i\delta^R_x(Y)\right)\star\left(\mp i\delta^R_{x'}(Y)\right) = \frac{2}{1 - x\cdot x'}\exp\frac{iYxx'Y}{2(x\cdot x' - 1)} \ .
\end{align}
Note that in order to recover the identity in the limit $x=x'$, we must choose opposite signs for the two reflection operators. This demonstrates that the sign ambiguity \eqref{eq:reflection_delta} cannot be fixed consistently.

As we can see, HS algebra is a kind of square root of Clifford algebra. In a sense, this is already clear from the definitions \eqref{eq:Clifford_HS}, since spinors are the ``square roots'' of vectors. However, the ``square root'' relationship between the algebras is more concrete than that. Indeed, we see e.g. in \eqref{eq:reflection_HS} that the \emph{adjoint} action of HS algebra realizes the \emph{fundamental} action of Clifford algebra on the twistor $Y$. We also saw the angle $\theta/2$ appearing in \eqref{eq:RH_Gaussian} as opposed to $\theta$ in \eqref{eq:RH_Clifford}, which led to a sign ambiguity \emph{on top of} the ordinary double cover $SO(1,4)\rightarrow Spin(1,4)$.

We are now ready to apply this section's geometric viewpoint to linearized higher-spin gravity. We recognize immediately that while the adjoint action \eqref{eq:reflection_HS} of $\pm i\delta^{L/R}(Y)$ directly realizes the reflection $-(\delta^\mu_\nu + 2x^\mu x_\nu)$ on $Y$, the fundamental action \eqref{eq:C_solution_R}-\eqref{eq:C_solution_L} realizes the Penrose transform. In this sense, the Penrose transform is the ``square root'' of a reflection. While the adjoint reflection \eqref{eq:reflection_HS} acts on the argument $Y$, the Penrose transform acts on functions $f(Y)$ as a whole. This must be the case, since the $SO(1,4)$ transformation of an individual twistor does not have a square root: twistors are \emph{already} a square root of $\bbR^{1,4}$ vectors. 

For the final touch to our interpretation of the Penrose transform, we should spell out the spacetime significance of the reflection $-(\delta^\mu_\nu + 2x^\mu x_\nu)$. So far, the vector $x^\mu$ has been timelike, representing a radius vector on the $EAdS_4$ hyperboloid. However, eventually, the more physical case is Lorentzian $dS_4$ spacetime, given in $\bbR^{1,4}$ by \emph{spacelike} unit vectors $x^\mu$. There, the subspace orthogonal to $x^\mu$ is the $dS_4$ tangent space at the point $x$, and the reflection $-(\delta^\mu_\nu + 2x^\mu x_\nu)$ is the de Sitter analog of CPT, with $x$ as the origin. Our statement \eqref{eq:sqrt_CPT} now follows: the Penrose transform is a square root of CPT.

\subsubsection{Rotations by $2\pi$ and the antipodal map} \label{sec:spacetime_subgroup:HS:antipodal}

So far in this section, we've been careful to distinguish $SO(1,4)$ from the full $O(1,4)$. The geometric transformations we've constructed up to now only cover $SO(1,4)$, i.e the even elements of $O(1,4)$. This includes the CPT reflections \eqref{eq:reflection_Clifford},\eqref{eq:reflection_HS}, since they reverse an even number of axes. To enlarge our scope to the full $O(1,4)$, we must add to our menu its central element: the antipodal map $x^\mu\rightarrow -x^\mu$, which reverses all 5 axes in $\bbR^{1,4}$. On a single twistor $Y^a$, the only way to represent this transformation non-trivially is by complex conjugation, which we will not consider here. With that option closed, we must resort, as with the Penrose transform, to acting on whole functions $f(Y)$. In fact, in section \ref{sec:linear_HS:antipodal}, we've already seen how this happens -- the antipodal map on bulk master fields $C(x;Y)$ is realized by multiplying either $C(x;Y)$ itself or its Penrose transform by $\delta(Y)$:
\begin{align}
 F_{L/R}(Y)\rightarrow F_{L/R}(Y)\star\delta(Y) \quad \Longleftrightarrow \quad C(x;Y) \rightarrow C(-x;Y) = C(x;Y)\star\delta(Y)  \ .
\end{align}
This follows from decomposing $\delta(Y) = \delta_x^L(Y)\star\delta_x^R(Y)$, which can be interpreted as the Penrose transform at $x$ followed by the inverse transform at $-x$. Thus, while $SO(1,4)$ is manifestly realized by the adjoint action of HS algebra, the antipodal map is realized by acting with $\delta(Y)$ in the fundamental.

To complete the picture, it remains to understand the geometric role of $\delta(Y)$ \emph{when acting in the adjoint}. As we can see from \eqref{eq:delta_Klein}, the answer is simply a $2\pi$ rotation:
\begin{align}
 \delta(Y)\star f(Y)\star\delta(Y) = f(-Y) \ .
\end{align}
This can again be understood in terms of the decomposition $\delta(Y) = \delta_x^L(Y)\star\delta_x^R(Y)$: the product of two $\pi$ rotations along e.g. the bivectors $e_1\wedge e_2 + e_3\wedge e_4$ and $e_1\wedge e_2 - e_3\wedge e_4$ is simply a $2\pi$ rotation along $e_1\wedge e_2$.

In light of the above two roles of $\delta(Y)$, one can rephrase the $\sqrt{\text{CPT}}$ nature of the Penrose transform as follows: \emph{the Penrose transform is to CPT as the antipodal map is to a $2\pi$ rotation}.

\subsection{The null limit} \label{sec:spacetime_subgroup:null}

So far, we've considered spacetime symmetries through the lens of reflections around timelike (or spacelike) vectors $x^\mu$. As we've seen, this geometry relates naturally to the bulk higher-spin theory. To discuss the boundary theory, we must take the limit \eqref{eq:limit}, where the reflection vector $x^\mu$ becomes null. The reflection matrix $-(\delta^\mu_\nu + 2x^\mu x_\nu)$ then becomes:
\begin{align}
 -(\delta^\mu_\nu + 2x^\mu x_\nu) \ \longrightarrow \ \frac{2}{z^2}\left(-\ell^\mu\ell_\nu + O(z^2) \right) \ . \label{eq:null_reflection}
\end{align}
The leading-order part of this matrix, renormalized so as to make it finite, is the degenerate ``reflection matrix'' $-\ell^\mu\ell_\nu$, which projects any vector onto $\ell^\mu$. Combining two such ``reflections'' with respect to a pair of null vectors $\ell,\ell'$, we get the matrix:
\begin{align}
 (-\ell^\mu\ell_\rho)(-\ell'^\rho\ell'_\nu) \sim -\ell^\mu\ell'_\nu \ . \label{eq:infinite_boost}
\end{align}
Treating $\ell,\ell'$ as the null limits of highly boosted timelike vectors $x,x'$, we recognize the matrix \eqref{eq:infinite_boost} as a boost by an infinite angle in the $\ell\wedge\ell'$ plane (again, renormalized for finiteness). This boost shrinks $\ell'^\mu$ to zero, stretches $\ell^\mu$ to infinity, and leaves untouched the subspace orthogonal to both. As a result, the renormalized matrix $-\ell^\mu\ell'_\nu$ leaves the $\ell^\mu$ component finite, while annihilating both the $\ell'^\mu$ and orthogonal components.

The degenerate ``reflections'' $-\ell^\mu\ell_\nu$ and ``infinite boosts'' $-\ell^\mu\ell'_\nu$ satisfy a ``forgetful property'': any linear operation sandwiched between two reflections is reduced to the corresponding boost \eqref{eq:infinite_boost}. Explicitly, for any matrix $M^\mu{}_\nu$, we trivially have:
\begin{align}
(-\ell^\mu\ell_\rho)M^\rho{}_\sigma(-\ell'^\sigma\ell'_\nu) \sim -\ell^\mu\ell'_\nu \ . \label{eq:forgetful_matrix}
\end{align}

In Clifford algebra, the analog of the matrix $-\ell^\mu\ell_\nu$ is the algebra element $\ell = \ell^\mu\gamma_\mu$; in higher-spin algebra, the corresponding element is the boundary spinor delta function $\delta_\ell(Y)$. These algebra elements and their products do not quite represent $SO(1,4)$ transformations, but renormalized limiting cases thereof. In fact, the renormalization is different in the different algebras: $-\ell^\mu\ell_\nu$, $\ell$ and $\delta_\ell(Y)$ all scale differently with $\ell^\mu$. Thus, it is tempting to apply the geometry of \eqref{eq:null_reflection}-\eqref{eq:forgetful_matrix} in the context of Clifford or HS algebra, but one must be mindful that not every property might carry over.

It turns out that the HS algebra element $\delta_\ell(Y)$ closely resembles in its properties the ``null reflection'' matrix $-\ell^\mu\ell_\nu$, while the Clifford algebra element $\ell$ does not. In Clifford algebra, the analog of the ``forgetful property'' \eqref{eq:forgetful_matrix} does not hold: the products $\ell\Gamma\ell'$ are not all proportional to each other, but span a 4d subspace, parameterized by varying $\Gamma$ over the Clifford algebra 
of the 3d hyperplane orthogonal to $\ell,\ell'$. In contrast, in HS algebra, the ``forgetful property'' \emph{does} hold, as we've seen in eq. \eqref{eq:forgetful}: all products of the form $\delta_\ell\star f\star\delta_{\ell'}$ are proportional to each other, since they must lie in the intersection of the subalgebras $\calA_+(\ell)\cap\calA_-(\ell')$. In particular, the product of three ``null reflections'' behaves similarly in spacetime and in HS algebra:
\begin{align}
 \begin{split}
   (-\ell^\mu\ell_\rho)(-\ell'^\rho\ell'_\sigma)(-\ell''^\sigma\ell''_\nu) &= -(\ell\cdot\ell')(\ell'\cdot\ell'')\ell^\mu\ell''_\nu \\
   &\text{vs.} \\
   \delta_\ell(Y)\star\delta_{\ell'}(Y)\star\delta_{\ell''}(Y) &= \pm i\sqrt{-\frac{\ell\cdot\ell''}{2(\ell\cdot\ell')(\ell'\cdot\ell'')}}\,\delta_\ell(Y)\star\delta_{\ell''}(Y) \ ,
 \end{split}
\end{align}
where the different proportionality coefficients arise from the different scaling properties of $-\ell^\mu\ell_\nu$ and $\delta_\ell(Y)$.

\subsection{Manifest $SO(1,4)$ in the higher-spin fundamental} \label{sec:spacetime_subgroup:manifest}

We are now ready to understand the geometric action \eqref{eq:K_sqrt_CPT} of CPT -- and thus of the entire $SO(1,4)$ -- on the boundary two-point product \eqref{eq:K_summary} in the higher-spin fundamental. In other words, we wish to calculate the action $\delta_\ell\star\delta_{\ell'}\star\delta^R_x$ of the CPT operator $\delta^R_x(Y)$ on the boundary two-point product $\delta_\ell(Y)\star\delta_{\ell'}(Y)$. The first step is to notice that the boundary-bulk product $\delta_{\ell'}\star\delta^R_x$ can be reduced to a boundary-boundary product. Specifically, we can read off from \eqref{eq:delta_ell_ell}-\eqref{eq:delta_ell_x} the identity (switching temporarily from $\ell'$ to $\ell$ to simplify notations):
\begin{align}
 \delta_\ell(Y)\star\delta^R_x(Y) = -2(\ell\cdot x)\,\delta_\ell(Y)\star\delta_{\tilde\ell}(Y) \ , \label{eq:x_ell_to_ell_ell}
\end{align}
where $\tilde\ell^\mu$ is the result of the CPT reflection $-(\delta^\mu_\nu + 2x^\mu x_\nu)$ acting on the null vector $\ell^\mu$:
\begin{align}
 \tilde\ell^\mu = -\ell^\mu - 2(\ell\cdot x)x^\mu \ . \label{eq:ell_tilde}
\end{align}
In the context of $EAdS_4$ geometry, $\tilde\ell$ is the second boundary endpoint of the geodesic that begins at $\ell$ and passes through $x$.

Within the geometric framework of this section, the equality \eqref{eq:x_ell_to_ell_ell} is not surprising. Up to renormalization, the product $\delta_\ell(Y)\star\delta^R_x(Y)$ represents an infinite boost in the timelike plane $\ell\wedge x$. The same boost can also be represented by $\delta_\ell(Y)\star\delta_{\tilde\ell}(Y)$, where $\tilde\ell$ is the second null vector in this plane. That is precisely the statement of eqs. \eqref{eq:x_ell_to_ell_ell}-\eqref{eq:ell_tilde}. Returning now to the task of calculating $\delta_\ell\star\delta_{\ell'}\star\delta^R_x$, we use \eqref{eq:x_ell_to_ell_ell} and then \eqref{eq:delta_ell_ell_ell} to find:
\begin{align}
 \begin{split}
   \delta_\ell(Y)\star\delta_{\ell'}(Y)\star\delta^R_x(Y) &= -2(\ell'\cdot x)\,\delta_\ell(Y)\star\delta_{\ell'}(Y)\star\delta_{\tilde\ell'}(Y) 
     = \pm i\sqrt{\frac{\ell\cdot\tilde\ell'}{\ell\cdot\ell'}}\,\delta_\ell(Y)\star\delta_{\tilde\ell'}(Y) \ .
 \end{split}
\end{align}
We have thus confirmed the first equation in \eqref{eq:K_sqrt_CPT}:
\begin{align}
 K(\ell,\ell';Y)\star i\delta^R_x(Y) = \pm K(\ell,\tilde\ell';Y) \ , \label{eq:K_sqrt_CPT_1}
\end{align}
where:
\begin{align}
K(\ell,\ell';Y) \sim \sqrt{-\ell\cdot\ell'}\,\delta_\ell(Y)\star\delta_{\ell'}(Y) \ . \label{eq:K_raw2}
\end{align}
For multiplication on the left, we similarly derive:
\begin{align}
 i\delta^R_x(Y)\star K(\ell,\ell';Y) = \pm K(\tilde\ell,\ell';Y) \ , \label{eq:K_sqrt_CPT_2}
\end{align}
and identical formulas hold for $\delta^L_x(Y)$ in place of $\delta^R_x(Y)$. Note that the sign ambiguities in \eqref{eq:K_sqrt_CPT_1},\eqref{eq:K_sqrt_CPT_2} cannot be consistently resolved. If we insisted on choosing a particular sign, then applying e.g. eq. \eqref{eq:K_sqrt_CPT_1} twice, we would find a contradiction with the identity $\delta^R_x\star\delta^R_x = +1$.

As promised, we see that the CPT reflection operators $\pm i\delta^{R/L}_x(Y)$, acting on $K(\ell,\ell';Y)$ in the higher-spin fundamental, have the effect of applying CPT to \emph{one} of the two boundary points $\ell,\ell'$. Thus, when acting on these bilocals, the Penrose transform is manifestly a square root of CPT. Furthermore, since all of $SO(1,4)$ can be constructed by combining reflections around different points $x$, we conclude that the same is true for a \emph{general} $SO(1,4)$ operator $g$: acting with $g$ on $K(\ell,\ell';Y)$ in the higher-spin fundamental will result in the corresponding $SO(1,4)$ transformation of one of the two points $\ell,\ell'$. We can verify this directly by applying the $SO(1,4)$ generators \eqref{eq:generators} to find the result quoted in \eqref{eq:K_sqrt_infinitesimal_summary}:
\begin{align}
 \begin{split} 
    M_{\mu\nu}\star K(\ell,\ell';Y) &= \ell_\mu\frac{\del K}{\del\ell^\nu} - \ell_\nu\frac{\del K}{\del\ell^\mu} \ ; \\
    -K(\ell,\ell';Y)\star M_{\mu\nu} &= \ell'_\mu\frac{\del K}{\del\ell'^\nu} - \ell'_\nu\frac{\del K}{\del\ell'^\mu} \ . 
 \end{split} \label{eq:K_sqrt_infinitesimal}
\end{align} 
Note that in this case, there are no sign ambiguities in the star products, since $M_{\mu\nu}\sim Y\gamma_{\mu\nu}Y$ is polynomial in $Y$. 

\section{The CFT in higher-spin-covariant twistor language} \label{sec:CFT}

\subsection{Overview}

In this section, we express the 3d free $U(N)$ vector model in twistor language, making its higher-spin conformal invariance manifest. We represent the conformal 3-sphere on which the CFT lives as the projective lightcone $(\ell_\mu\ell^\mu = 0,\ \ell^0 > 0,\ \ell^\mu\cong\lambda\ell^\mu)$ in $\bbR^{1,4}$. Thus, we are using the ``embedding-space formalism'' for CFT (see e.g. \cite{Weinberg:2010fx}). In section \ref{sec:CFT:local}, we express our CFT in the standard language of local operators and sources. In section \ref{sec:CFT:bilocal}, as a first step towards the twistor formalism, we express the theory and its correlators at separated points in a \emph{bilocal} language. For the very special case of a free vector model, this language is more natural than the standard local one, because all single-trace operators are quadratic in the fundamental fields. Our bilocal formalism is inspired by the one in \cite{Das:2003vw}. However, unlike the authors of \cite{Das:2003vw}, we do not treat the bilocal operators as a new ``fundamental'' field. Instead, we treat them straightforwardly as composite operators, coupled to bilocal sources. Of course, these quadratic CFT operators \emph{do} become fundamental fields once we switch to the bulk description. In this sense, the CFT is a ``square root'' of the bulk theory. The results of the present section can be viewed as a consequence of this ``square root'' relation, combined with the ``square root'' relation \eqref{eq:CPT}-\eqref{eq:sqrt_CPT} between the Penrose transform and CPT.

The local and bilocal languages for the CFT share some qualitative features. The bilocal sources, like the local gauge potentials, are gauge-redundant (in fact, their gauge redundancy is even larger). Conversely, the bilocal operators, like the local conserved currents, satisfy constraints. Finally, as discussed in section \ref{sec:summary:contact}, in a region with non-vanishing (either local or bilocal) sources, one would need contact terms to obtain finite and conserved expectation values for the local currents.

In section \ref{sec:CFT:twistor}, having established the bilocal language, we use it as a springboard towards a fully nonlocal, twistorial formulation of the CFT. In this formulation, the sources are no longer gauge-redundant, while the single-trace ``currents'' are constraint-free. At the same time, the theory's global higher-spin symmetry becomes manifest. Our transform between the twistor and bilocal formulations is a boundary version of the bulk Penrose transform. Since the CFT is free even when the bulk is interacting, this boundary/twistor transform allows us to express the full partition function in the twistor language. Finally, as we discuss in section \ref{sec:holography:general_currents}, the twistor language appears to automatically include all the necessary contact terms, so that we end up with conserved currents even at finite sources.

\subsection{Local language} \label{sec:CFT:local}

We begin with the action of $N$ free massless scalars in the fundamental representation of an internal $U(N)$ symmetry:
\begin{align}
S_{\text{CFT}} = -\int d^3\ell\,\bar\phi_I\Box\phi^I \ . \label{eq:S_free}
\end{align}
Here, $I=1,\dots,N$ is an internal index; $\phi^I$ and their complex conjugates $\bar\phi_I$ are dynamical fields with conformal weight $\Delta=1/2$. We consider only $U(N)$ singlets to be observable (for example, one might imagine that the $U(N)$ is gauged with a very weak coupling). The single-trace primaries of the theory \eqref{eq:S_free} consist of an infinite tower of conserved currents $j^{(s)}$, one for each spin $s$. To write these out explicitly, we can use a flat 3d frame as in \eqref{eq:flat}, with 3d spatial indices $(i,j,k,\dots)$. Then the spin-$s$ current $j^{(s)}$ reads \cite{Craigie:1983fb,Anselmi:1999bb}:
\begin{align}
j^{(s)}_{k_1\dots k_s} = \frac{1}{(2i)^s}\,\bar\phi_I
\left(\sum_{m=0}^s (-1)^m \binom{2s}{2m} \overset{\leftarrow}{\del}_{(k_1}\dots\overset{\leftarrow}{\del}_{k_m} \overset{\rightarrow}{\del}_{k_{m+1}}\dots\overset{\rightarrow}{\del}_{k_s)} - \text{traces}\right) \phi^I \ .
\label{eq:explicit_j}
\end{align}
Here, the $1/i^s$ prefactor ensures that $j^{(s)}$ is real, while the $1/2^s$ prefactor is chosen for agreement with the bulk asymptotics in section \ref{sec:holography:example_currents}. We include in \eqref{eq:explicit_j} also the case $s=0$, i.e. the scalar ``current'' $j^{(0)} = \bar\phi_I\phi^I$. The spin-1 current $j^{(1)}_i$ is $1/2$ times the ordinary charge current for the $U(1)$ component of $U(N)$:
\begin{align}
j^{(1)}_i = \frac{1}{2i}\,\bar\phi_I\overset{\leftrightarrow}\del_i\phi^I \ ,
\end{align}
while the spin-2 current $j^{(2)}_{ij}$ is $2$ times the theory's stress-energy tensor:
\begin{align}
j^{(2)}_{ij} = 2T_{ij} \ ; \quad T_{ij} = -\frac{1}{8}\left(\bar\phi_I\del_i\del_j\phi^I + \phi^I\del_i\del_j\bar\phi_I - 6\del_{(i}\bar\phi_I\del_{j)}\phi^I + 2g_{ij}\del_k\bar\phi_I\del^k\phi^I \right) \ . \label{eq:stress_tensor}
\end{align}
For the stress tensor and the conserved currents of spin $s>2$, there are various related definitions that all satisfy a conservation law. The definition \eqref{eq:explicit_j} is the unique one that is totally symmetric and traceless. In particular, the stress tensor \eqref{eq:stress_tensor} is the one derived by varying the metric in a theory of free massless scalars with conformal coupling. 

When we're not using the explicit formula \eqref{eq:explicit_j} with its flat 3d derivatives, we can use $\bbR^{1,4}$ indices $(\mu,\nu,\dots)$ for the currents $j^{(s)}$, as in section \ref{sec:geometry:spacetime:currents}. Introducing sources $A^{(s)}_{\mu_1\dots\mu_s}$ for the single-trace operators \eqref{eq:explicit_j}, the free action \eqref{eq:S_free} becomes:
\begin{align}
S_{\text{CFT}} = -\int d^3\ell\,\bar\phi_I\Box\phi^I - \int d^3\ell \sum_{s=0}^\infty A^{(s)}_{\mu_1\dots\mu_s}(\ell)\, j_{(s)}^{\mu_1\dots\mu_s}(\ell) \ . \label{eq:S_local}
\end{align}
The sources $A^{(s)}_{\mu_1\dots \mu_s}$ are spin-$s$ gauge potentials. In particular, $A^{(1)}_\mu$ is an ordinary $U(1)$ gauge potential (times $2$), while $A^{(2)}_{\mu\nu}$ is a metric perturbation (times $1/2$).

\subsection{Bilocal language} \label{sec:CFT:bilocal}

Having formulated our theory in the ordinary language of local operators and sources, let us now present its much simpler formulation in terms of bilocals. The idea is to notice that the local primaries $j^{(s)}_{\mu_1\dots\mu_s}(\ell)$ are just a complicated-looking Taylor expansion of the two-point inner product $\phi^I(\ell)\bar\phi_I(\ell')$. Note that before imposing the free field equations $\Box\phi^I = 0$, there is not enough information in the totally symmetric and traceless $j^{(s)}_{\mu_1\dots\mu_s}(\ell)$ to encode all possible configurations of $\phi^I(\ell)\bar\phi_I(\ell')$. However, \emph{after} imposing the field equations, there is \emph{too much} information. The currents $j^{(s)}_{\mu_1\dots\mu_s}(\ell)$ then satisfy constraints, i.e. conservation laws. In this situation, we might as well directly use the bilocal $\phi^I(\ell)\bar\phi_I(\ell')$ as our basic single-trace operator. The role of current conservation laws is then played by the field equations themselves. Coupling a bilocal source $\Pi(\ell',\ell)$ to the bilocal operator $\phi^I(\ell)\bar\phi_I(\ell')$, we write the CFT action in the form:
\begin{align}
 S_{\text{CFT}}[\Pi(\ell',\ell)] = -\int d^3\ell\,\bar\phi_I\Box\phi^I - \int d^3\ell' d^3\ell\,\bar\phi_I(\ell')\Pi(\ell',\ell)\phi^I(\ell) \ . \label{eq:S}
\end{align}
The ``conservation laws'' (actually, just field equations) on $\phi^I(\ell)\bar\phi_I(\ell')$ induce a gauge redundancy on $\Pi(\ell',\ell)$:
\begin{align}
 \Pi(\ell',\ell) \rightarrow \Pi(\ell',\ell) + \Box_\ell f(\ell',\ell) + \Box_{\ell'} g(\ell',\ell) \ . \label{eq:gauge_Pi}
\end{align}
In the large-$N$ limit, there are no constraints on $\phi^I(\ell)\bar\phi_I(\ell')$ other than the field equations, and thus \eqref{eq:gauge_Pi} captures the full gauge redundancy of $\Pi(\ell',\ell)$. For finite $N$, this is not the case: for example, for $N=1$, the product $\phi(\ell)\bar\phi(\ell')$ is determined by two functions $\phi(\ell),\bar\phi(\ell)$ of a single point $\ell$. Thus, for finite $N$, the redundancy in $\Pi(\ell',\ell)$ is greater. However, even then, this redundant parameterization of the single-trace sources remains legitimate.

The partition function of the theory \eqref{eq:S} is very easy to write down in the bilocal language. First, we write the action in a ``matrix'' notation:
\begin{align}
S_{\text{CFT}}[\Pi(\ell',\ell)] = -\bar\phi_I(\Box + \Pi)\phi^I \ , \label{eq:S_matrix}
\end{align}
where $\phi(\ell)$ is viewed as an infinite-dimensional vector, $\bar\phi(\ell)$ as a dual vector, and $\Box,\Pi$ as matrices/operators. The Gaussian path integral over $\phi$ and $\bar\phi$ immediately gives the partition function in the form:
\begin{align}
Z_{\text{CFT}}[\Pi(\ell',\ell)] \sim \left(\det{(\Box + \Pi)}\right)^{-N} \sim \left(\det{(1 + G\Pi)}\right)^{-N} = \exp\left(-N\tr\ln(1+G\Pi)\right) \ , \label{eq:Z}
\end{align}
where $G = \Box^{-1}$ is the boundary-to-boundary propagator:
\begin{align}
G(\ell,\ell') = -\frac{1}{4\pi\sqrt{-2\ell\cdot\ell'}} \ , \label{eq:G}
\end{align}
i.e. $G(\mathbf{r},\mathbf{r'}) = -1/(4\pi|\mathbf{r} - \mathbf{r'}|)$ in the flat frame \eqref{eq:flat}.

\begin{figure}%
	\centering%
	\includegraphics[scale=1]{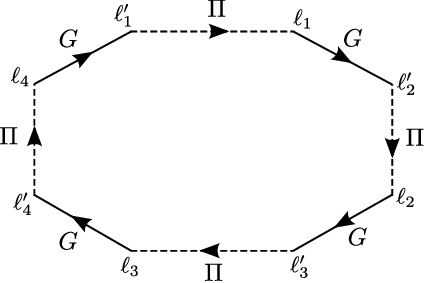} \\
	\caption{The Feynman diagram for a single-trace contribution $\tr(G\Pi)^n$ to the CFT partition function, drawn for $n=4$. Dashed lines represent the bilocal sources $\Pi(\ell',\ell)$. Solid lines represent propagators $G(\ell,\ell')$. Note that the diagram is in coordinate space rather than momentum space, and there is no loop integration involved.}
	\label{fig:polygon} 
\end{figure}

The partition function \eqref{eq:Z} is a combination of single-trace pieces of the form $\tr(G\Pi)^n$, which can be represented by ``1-loop'' Feynman diagrams as in figure \ref{fig:polygon}. The $U(N)$ ``color'' factor is taken into account by the $N$ in the exponent in eq. \eqref{eq:Z}.

Any UV divergences in the CFT's Feynman diagrams (such as the diagram in figure \ref{fig:polygon}) are associated with the short-distance divergence of the propagator \eqref{eq:G}, i.e. with the limit where some of the ``external legs'' of the $\Pi(\ell',\ell)$ factors coincide. As long as we are only interested in the bilocal source couplings \eqref{eq:S} and partition function \eqref{eq:Z}, these short-distance singularities don't seem to require any special treatment: the propagator \eqref{eq:G} behaves as $\sim 1/r$, which is integrable under the 3d volume measure $\sim r^2dr$. Thus, the partition function \eqref{eq:Z} is well-defined without any contact-term corrections. 

In contrast, the conserved local currents \eqref{eq:explicit_j}, the local source couplings \eqref{eq:S_local} and the bilocal gauge symmetry \eqref{eq:gauge_Pi} are all given up to contact terms, i.e. assuming separated points. We will not investigate these contact terms directly here. Instead, we will now switch to twistor language, where the need for contact terms, even for calculating local currents, seems to disappear entirely.

\subsection{Twistor language} \label{sec:CFT:twistor}

\subsubsection{From bilocals to twistor functions}  \label{sec:CFT:twistor:transform}

So far, we've made manifest the conformal $O(1,4)$ symmetry of the theory \eqref{eq:S}, but not its higher-spin extension. To this end, we will now employ the HS algebra of section \ref{sec:algebra}. The boundary two-point products of section \ref{sec:algebra:two_point} will play a central role. First, let us package the bilocal source $\Pi(\ell',\ell)$ into a twistor function $\Pi(Y)$:
\begin{align}
 F(Y) &= \int d^3\ell\,d^3\ell'K(\ell,\ell';Y)\,\Pi(\ell',\ell) \ , \label{eq:Pi_transform}
\end{align}
where the bilocal kernel $K(\ell,\ell';Y)$ is given by:
\begin{align}
  K(\ell,\ell';Y) = \frac{\sqrt{-2\ell\cdot\ell'}}{4\pi}\,\delta_\ell(Y)\star\delta_{\ell'}(Y) 
   = \frac{1}{\pi\sqrt{-2\ell\cdot\ell'}}\exp\frac{iY\ell\ell' Y}{2\ell\cdot\ell'} \ . \label{eq:K}
\end{align}
The kernel $K(\ell,\ell';Y)$ is an even function of $Y$, and has conformal weight $\Delta = 1/2$ with respect to each of the boundary points $\ell,\ell'$. 

The transform \eqref{eq:Pi_transform} involves a loss of information: the original bilocal $\Pi(\ell',\ell)$ is a function of 6 coordinates, while $F(Y)$ only depends on 4. Nevertheless, we will see that $F(Y)$ is sufficient to express the partition function, i.e. it is a \emph{complete} encoding of the ``physically relevant'' data in $\Pi(\ell',\ell)$. In fact, our transform can be viewed as stripping away the gauge redundancy in $\Pi(\ell',\ell)$. Indeed, we see from eq. \eqref{eq:ell_ell_wave} that $K(\ell,\ell';Y)$ satisfies, up to contact terms, the same field equations $\Box_\ell K = \Box_{\ell'} K = 0$ as the bilocal operator $\phi^I(\ell)\bar\phi_I(\ell')$. Therefore, $F(Y)$ is invariant under the gauge symmetry \eqref{eq:gauge_Pi}. Thus, at large $N$, $F(Y)$ constitutes a \emph{non-redundant} parameterization of the theory's sources (recall that at finite $N$, there is additional redundancy in $\Pi(\ell',\ell)$, which is not captured by eq. \eqref{eq:gauge_Pi}). What's more, while the gauge redundancy is lost, the true global HS symmetry can now be made manifest. Indeed, we will see below that the partition function in terms of $F(Y)$ is manifestly HS-invariant, with $F(Y)$ transforming in the adjoint.

The remaining question is whether $F(Y)$ constructed through \eqref{eq:Pi_transform} is an \emph{arbitrary} even function of $Y$, i.e. whether the functions $K(\ell,\ell';Y)$ form a spanning set for the HS algebra. Skipping slightly ahead in the narrative, the answer is essentially yes. Specifically, there's a one-to-one correspondence between $F(Y)$ and linearized bulk solutions (via the Penrose transform), and a one-to-one correspondence between linearized bulk solutions and allowed configurations of the linearized expectation values of the local currents $j^{(s)}_{\mu_1\dots\mu_s}(\ell)$. More precisely, the above statements are \emph{almost} true, due to a pair of related subtleties. First, the Penrose transform involves contour ambiguities. Second, the one-to-one mapping between expectation values $\langle j^{(s)}_{\mu_1\dots\mu_s}(\ell)\rangle$ and bulk solutions involves a requirement of regularity on $EAdS_4$, without which one loses the relationship between the boundary data corresponding to $\langle j^{(s)}_{\mu_1\dots\mu_s}(\ell)\rangle$ and the boundary data corresponding to the sources $A^{(s)}_{\mu_1\dots\mu_s}(\ell)$. This regularity on $EAdS_4$ can be enforced by an $i\epsilon$ prescription on boundary-to-bulk propagators, i.e. it is yet another contour issue. Again related to the above is the question of how $F(Y)$ behaves under the ``antipodal map'' $F(Y)\rightarrow F(Y)\star\delta(Y)$. As we can see from \eqref{eq:boundary_delta_odd}, $K(\ell,\ell';Y)$ is odd under this map, at least for $\ell\neq\ell'$:
\begin{align}
 K(\ell,\ell';Y)\star\delta(Y) = -K(\ell,\ell';Y) \quad \forall \ell\neq\ell' \ . \label{eq:K_odd}
\end{align} 
However, as we've seen in section \ref{sec:linear_HS:antipodal}, one cannot conclude the same for a linear superposition such as \eqref{eq:Pi_transform}, and in fact $F(Y)\star\delta(Y)$ cannot be defined consistently for generic functions $F(Y)$.

To summarize, the encoding \eqref{eq:Pi_transform} of the CFT sources into a twistor function $F(Y)$ is 1) complete, 2) free of the infinite-dimensional HS gauge redundancy, 3) capturing all the gauge-invariant information, 4) making global HS symmetry manifest, and 5) constraint-free, up to a set of closely related subtleties regarding contour choices, analiticity and discrete symmetries. 

\subsubsection{The partition function in twistor language}

In this section, we express the CFT partition function in terms of $F(Y)$. Remarkably, this can be done by rewriting each individual element in the CFT Feynman diagrams as an HS-covariant operation. The mechanism is captured by the following pair of identities:
\begin{align}
 \tr_\star K(\ell,\ell';Y) &= -4G(\ell,\ell') \ ; \label{eq:str_K} \\
 K(\ell_1,\ell'_1;Y)\star K(\ell_2,\ell'_2;Y) &= G(\ell_2,\ell_1')\,K(\ell_1,\ell'_2;Y) \ . \label{eq:K_K}
\end{align}
Here, the trace identity \eqref{eq:str_K} is just a restatement of eq. \eqref{eq:str_ell_ell}. The star-product identity \eqref{eq:K_K} follows from applying the three-point product formula \eqref{eq:delta_ell_ell_ell} twice:
\begin{align}
 \delta_{\ell_1}(Y)\star\delta_{\ell_1'}(Y)\star\delta_{\ell_2}(Y)\star\delta_{\ell_2'}(Y) 
   = -\frac{1}{2}\sqrt{\frac{(\ell_1\cdot\ell_2')}{(\ell_1\cdot\ell_1')(\ell_1'\cdot\ell_2)(\ell_2\cdot\ell_2')}}\,\delta_{\ell_1}(Y)\star\delta_{\ell_2'}(Y) \ . \label{eq:delta_ell_ell_ell_ell}
\end{align}
Here, the sign ambiguity in \eqref{eq:delta_ell_ell_ell} is removed by squaring $(\pm i)^2 = -1$. [NOTE: a previous version of this paper fixed the sign in an opposite manner, effectively choosing $+i$ in one application of \eqref{eq:delta_ell_ell_ell} and $-i$ in the other, which was justified by a reasonable regularization procedure. It appears the bilocal/twistor treatment in the present paper simply does not allow a clear resolution of the sign ambiguity, though the latter is in fact important for such basic things as the sign in the exponent in \eqref{eq:Z_HS} below. The present sign choice is informed by the \emph{local}/twistor analysis in the later work \cite{Local}.]

Having established the identity \eqref{eq:K_K}, we note that the star product there radically alters the spatial dependence of the $K$'s. Indeed, the $K$ factors on the LHS have essential singularities at $\ell_1=\ell_1'$ and $\ell_2=\ell_2'$, while the RHS has an essential singularity at $\ell_1'=\ell_2$ and a simple pole at $\ell_1' = \ell_2$. This is an example of how the star product's nonlocality in twistor space can get translated into nonlocality in spacetime.

We are now ready to employ eqs. \eqref{eq:str_K}-\eqref{eq:K_K} to compute the CFT partition function. The single-trace products that form the building blocks of $Z_{\text{CFT}}$, i.e. the one-loop Feynman diagrams from figure \ref{fig:polygon}, can be rewritten as:
\begin{align}
 \tr(G\Pi)^n = -\frac{1}{4}\tr_\star\Big(\underbrace{F(Y)\star F(Y)\star\ldots\star F(Y)}_{n\text{ times}}\Big) \ . \label{eq:loop}
\end{align}
Thus, the CFT Feynman diagrams make equally good sense in both the bilocal and higher-spin languages. In fact, the correspondence is at the level of individual diagram elements: via the identities \eqref{eq:str_K}-\eqref{eq:K_K}, every individual star product or higher-spin trace can be identified with a $G(\ell,\ell')$ propagator in the Feynman diagram. 

The entire partition function \eqref{eq:Z} can now be written in HS language, yielding the result \eqref{eq:Z_summary}:
\begin{align}
 Z_{\text{CFT}}[F(Y)] \sim \exp\left(\frac{N}{4}\tr_\star\ln_\star[1+F(Y)]\right) = \left(\textstyle\det_\star[1+F(Y)]\right)^{N/4} \ . \label{eq:Z_HS}
\end{align}
Here, $\ln_\star[1+F(Y)]$ is defined by substituting star products in the Taylor expansion of $\ln(1+x)$, and we introduce the ``star determinant''
$\det_\star f \equiv \exp(\tr_\star\ln_\star f)$.

The partition function \eqref{eq:Z_HS} is manifestly invariant under global HS symmetry, with the source $F(Y)$ transforming in the adjoint:
\begin{align}
 \delta F(Y) = \varepsilon(Y)\star F(Y) - F(Y)\star\varepsilon(Y) \ ; \quad \delta Z_{\text{CFT}} = 0 \ . \label{eq:Z_symmetry}
\end{align}
Conversely, the symmetry \eqref{eq:Z_symmetry} completely fixes the invariant traces \eqref{eq:loop} as the only possible ingredient in the partition function (up to a possible $\delta(Y)$ factor inside the trace; however, see eq. \eqref{eq:K_odd} and the surrounding discussion). For this reason, the traces \eqref{eq:loop} were introduced from a bulk perspective in \cite{Colombo:2012jx,Didenko:2012tv} as the unique expressions for the $n$-point functions, with only their coefficients left undetermined. Here, we derived the traces \eqref{eq:loop} directly from the boundary CFT, allowing us to fix their coefficients by writing down the full partition function \eqref{eq:Z_HS}.

\subsubsection{The single-trace currents in twistor space} \label{sec:CFT:twistor:currents}

We can construct  a ``current'' operator conjugate to the source $F(Y)$ as an HS-covariant variational derivative:
\begin{align}
\Phi(Y) = \frac{D}{DF(Y)} \ , \label{eq:Phi}
\end{align}
where the derivative $D/Df(Y)$ a functional $\Theta[f(Y)]$ is defined via:
\begin{align}
\delta\Theta = \tr_\star\left(\frac{D\Theta}{Df(Y)}\star\delta f(Y)\right) \ .
\end{align}
As we can see from \eqref{eq:str}, this implies that $D/Df(Y)$ is actually a Fourier transform of the ordinary variational derivative:
\begin{align}
\frac{D}{Df(Y)} = \int d^4U e^{iYU}\,\frac{\delta}{\delta f(U)} \ .
\end{align}

The expectation value $\left<\Phi(Y)\right>$ reads:
\begin{align}
 \left<\Phi(Y)\right> = \frac{D\ln Z_{\text{CFT}}}{DF(Y)} = \frac{N}{4}[1+F(Y)]_\star^{-1} = \frac{N}{4}\left(1 - F(Y) + O(F^2) \right) \ . \label{eq:Phi_expectation}
\end{align}
In particular, the linear piece of $\left<\Phi(Y)\right>$ is just a constant multiple of $F(Y)$. There is in fact no other possibility compatible with HS symmetry (again, with the subtle exception of multiplication by $\delta(Y)$, which we will touch on again in section \ref{sec:holography}).

For comparison, in the original bilocal language of eqs. \eqref{eq:S}-\eqref{eq:Z}, the current conjugate to $\Pi(\ell',\ell)$ is simply the bilocal operator $\phi^I(\ell)\bar\phi_I(\ell')$:
\begin{align}
 \phi^I(\ell)\bar\phi_I(\ell') &= \frac{\delta}{\delta\Pi(\ell',\ell)} \ ; \\
 \begin{split}
   \left<\phi^I(\ell)\bar\phi_I(\ell')\right> &= \frac{\delta\ln Z_{\text{CFT}}}{\delta\Pi(\ell',\ell)} = -N(1+G\Pi)^{-1}G \\
    &= N\left(-G(\ell,\ell') + \int d^3\ell_1 d^3\ell_2\,G(\ell,\ell_1)\,\Pi(\ell_1,\ell_2)\,G(\ell_2,\ell') + O(\Pi^2) \right) \ .
 \end{split} \label{eq:phi_phi_expectation}
\end{align}
The currents $\Phi(Y)$ and $\phi^I(\ell)\bar\phi_I(\ell')$ are related via the chain rule for variational derivatives:
\begin{align}
 \begin{split}
   \phi^I(\ell)\bar\phi_I(\ell') &= \int d^4Y \frac{\delta F(Y)}{\delta\Pi(\ell',\ell)}\frac{\delta}{\delta F(Y)}
    = \int d^4Y K(\ell,\ell';Y) \int d^4U e^{-iYU} \Phi(U) \\
    &= \tr_\star\left(\Phi(Y)\star K(\ell,\ell';Y)\right) \ .
 \end{split} \label{eq:phi_phi}
\end{align}
Equivalently, the perturbation of $F(Y)$ that couples to the operator $\phi^I(\ell)\bar\phi_I(\ell')$ is simply:
\begin{align}
 \frac{\delta F(Y)}{\delta\Pi(\ell',\ell)} = \frac{\delta{\left(\phi^I(\ell)\bar\phi_I(\ell')\right)}}{\delta\Phi(Y)} = K(\ell,\ell';Y) \ . \label{eq:phi_phi_source}
\end{align}

Since the source $F(Y)$ is gauge-invariant (at large $N$) and constraint-free, we conclude that the operator $\Phi(Y)$ is constraint-free (at large $N$) and gauge-invariant. In contrast, the bilocal $\phi^I(\ell)\bar\phi_I(\ell')$, which depends on 6 rather than 4 coordinates, is constrained by field equations. In regions where the source $\Pi(\ell',\ell)$ vanishes, $\phi^I(\ell)\bar\phi_I(\ell')$ inherits from $K(\ell,\ell';Y)$ the source-free wave equation \eqref{eq:ell_ell_wave} at $\ell\neq\ell'$. The same is \emph{not} true in regions with non-vanishing $\Pi(\ell',\ell)$, even though this is not evident from eq. \eqref{eq:phi_phi}. The important subtlety here is that the star product has the power to reshuffle the singularity structure in the $\ell,\ell'$ dependence, as in \eqref{eq:K_K}. The true spatial dependence of $\phi^I(\ell)\bar\phi_I(\ell')$ in regions where $\Pi(\ell',\ell)$ is non-vanishing can be seen from the expansion \eqref{eq:phi_phi_expectation}. 

\section{Holography} \label{sec:holography}

We are now ready to tie together our treatments of the bulk and boundary. Normally in AdS/CFT, one thinks in terms of local bulk fields, each of which is associated with asymptotic boundary data of two types, i.e. two complementary conformal weights. For massless fields in $AdS_4$, these conformal weights are integers, and the corresponding boundary data takes on an additional layer of meaning. For the conformally coupled massless scalar, the conformal weights are $\Delta = 1,2$; we refer to the corresponding boundary data as Dirichlet and Neumann, since that is their precise nature under a bulk conformal transformation that turns the asymptotic boundary into an ordinary hypersurface. For gauge fields with spin $s\geq 1$, the two different conformal weights arise for the gauge potential; in terms of the field strength, they correspond to its electric and magnetic parts, which have the same conformal weight but different parities. 

In different setups, HS gravity is dual to a variety of vector models. The free vector model \eqref{eq:S_local} is the simplest case, which, from the bulk point of view, relies on two choices. First, one must choose the bulk interactions to be those of the type-A (i.e. parity-even) model; this will not make an explicit appearance in the present paper, since we only consider the bulk interactions indirectly, through the CFT. Second, one must choose Neumann boundary conditions for the scalar, and magnetic boundary conditions for the gauge fields of spin $s\geq 1$, i.e. treat the Neumann \& magnetic boundary data as external sources, while the Dirichlet \& electric boundary data will correspond to the CFT operators. As discussed in \cite{Vasiliev:2012vf}, this is the choice of boundary data that preserves (global) HS symmetry, which is then reflected in the CFT.

In the present section, we will describe the HS/free-CFT holography from the twistorial perspective of sections \ref{sec:linear_HS},\ref{sec:CFT}. We will then make contact with the standard language of bulk fields vs. local CFT operators by comparing the bulk fields' Dirichlet/electric boundary data with the expectation values of the CFT currents.

\subsection{The basic dictionary}

The fundamental entry in our holographic dictionary is to identify the twistor function $F(Y)$ that encodes the CFT sources in \eqref{eq:Pi_transform} with either of the two functions $F_{R/L}(Y)$ that define the linearized bulk solution in \eqref{eq:C_solution_R}-\eqref{eq:C_solution_L}:
\begin{align}
 F(Y) = F_R(Y) \quad \text{or} \quad F(Y) = F_L(Y) \ . \label{eq:dictionary}
\end{align} 
In terms of the bulk master fields $C(x;Y)$, this implies:
\begin{align}
 F(Y) = -iC(x;Y)\star\delta^R_x(Y) \quad \text{or} \quad F(Y) = iC(x;Y)\star\delta^L_x(Y) \ . \label{eq:dictionary_C}
\end{align}
With the substitution \eqref{eq:dictionary_C}, the CFT partition function \eqref{eq:Z_HS} becomes a \emph{nonlinear functional of the linearized bulk solution}, as envisaged in \cite{Didenko:2012tv,Colombo:2012jx}:
\begin{align}
 Z \sim \left(\textstyle\det_\star[1-iC(x;Y)\star\delta^R_x(Y)]\right)^{N/4} \ \text{or} \quad Z \sim \left(\textstyle\det_\star[1+iC(x;Y)\star\delta^L_x(Y)]\right)^{N/4} \ . \label{eq:Z_bulk}
\end{align}
As in \cite{Didenko:2012tv,Colombo:2012jx}, the partition function \eqref{eq:Z_bulk} is given in terms of the master field at \emph{any single bulk point} $x$, which, by virtue of the unfolded formulation, encodes the entire linearized bulk solution. In this way, having placed the bulk and boundary on a common footing via twistor functions, we are able to express the full partition function, including the effects of bulk interactions, in bulk terms.

To establish the relation \eqref{eq:dictionary}, we will calculate on both sides the linearized expectation values of the local HS currents $j^{(s)}_{\mu_1\dots\mu_s}(\ell)$, as induced by the bilocal source $\Pi(\ell',\ell)$ at separated points. On the bulk side, this means calculating the electric field strengths at infinity. The distinction between $F_R(Y)$ and $F_L(Y)$ in \eqref{eq:dictionary} does not affect these expectation values. This distinction is instead related to the value of the boundary gauge fields $A^{(s)}_{\mu_1\dots\mu_s}(\ell)$, as well as to the contour issues discussed in section \ref{sec:CFT:twistor:transform}. This is because the two choices \eqref{eq:dictionary} are related by $F(Y)\rightarrow -F(Y)\star\delta(Y)$, or, in terms of the bulk fields, by $C(x;Y)\rightarrow -C(-x;Y)$. The electric field strengths at infinity and the CFT currents are unaffected by this transformation, since they are associated \cite{Vasiliev:2012vf,Neiman:2014npa} with the antipodally odd part of the bulk solution.

In a sense, an explicit calculation of $\left<j^{(s)}_{\mu_1\dots\mu_s}(\ell)\right>$ on both sides of the duality is actually unnecessary. The results are guaranteed to agree, simply because the higher-spin algebra contains every spin exactly once (actually twice for $s\geq 1$, corresponding to the two helicities; however, we can then use parity to distinguish their ``electric'' combination from the ``magnetic'' one). This is as it should be: since the higher-spin/free-CFT duality is in a sense the simplest of all holographic models, it \emph{should} appear trivial -- as trivial as eq. \eqref{eq:dictionary} -- once the correct language has been identified.  

Thus, in practice, our calculation of the boundary currents will serve two aims: to provide a consistency check for the formalism, and to fix the proportionality coefficients between the boundary currents and the bulk electric fields at infinity.
 
\subsection{Asymptotics of the bulk fields} \label{sec:holography:asymptotics}

In this section, we express the asymptotic boundary data of a bulk solution $C(x;Y)$ in terms of the twistor function $F_R(Y)$ (a similar analysis applies for $F_L(Y)$, with some sign changes). In accordance with the standard AdS/CFT prescription, we will focus on the asymptotics of the ``fundamental'' massless bulk fields \eqref{eq:HS_fields}, as opposed to the unfolded tower of derivatives \eqref{eq:unfolding}. As a result, our expressions will generally not be HS-covariant, i.e. they will contain spinor integrals that cannot be reduced to star products.

\subsubsection{Spin 0} \label{sec:holography:asymptotics:scalar}

The conformally-coupled massless scalar $C^{(0,0)}(x) = C(x;0)$ admits boundary data of two types: ``Dirichlet data'' $\varphi(\ell)$ with conformal weight $\Delta = 1$ and ``Neumann data'' $\pi(\ell)$ with weight $\Delta = 2$. At a bulk point $x$, the value of the scalar field can be found from the Penrose transform \eqref{eq:C_solution_R} or \eqref{eq:Penrose_R_to_scalar} as:
\begin{align}
 C(x;0) = i\tr_\star\left(F_R(Y)\star\delta^R_x(Y)\right) = i\int_{P_R(x)} d^2u_R\,F_R(u_R) \ . \label{eq:bulk_scalar}
\end{align}
The Dirichlet boundary data can be read off directly from the bulk-to-boundary limit \eqref{eq:limit},\eqref{eq:delta_limit}:
\begin{align}
 \varphi(\ell) = \lim_{x\rightarrow\ell/z} \frac{1}{z}\,C(x;0) = i\tr_\star\left(F_R(Y)\star\delta_\ell(Y)\right) = i\int_{P(\ell)} d^2u\,F_R(u) \ . \label{eq:Dirichlet}
\end{align}

The Neumann boundary data will be given by the second term in the Taylor series in $z$:
\begin{align}
C(x;0) = z\varphi(\ell) + z^2\pi(\ell) + O(z^3) \ .
\end{align}
To extract it, we must take the bulk-to-boundary limit more carefully, as in \eqref{eq:approach}, using a second boundary point $n$ to define the direction from which $x$ approaches $\ell$. Under  \eqref{eq:approach}, the chiral projector $P_R(x)$ takes the form:
\begin{align}
 P_R(x) = \frac{1}{2z}(\ell + z + z^2 n) = \frac{1}{z}P(\ell) + \frac{1}{2}(1 + zn) \ . \label{eq:approach_P}
\end{align}
We can now rewrite the $P_R(x)$ integral in \eqref{eq:bulk_scalar} as an integral over $P(\ell)$, via the change of variables:
\begin{align}
 u_R = 2P_R(x)u = (1+zn)u \ .
\end{align}
The measures $d^2u$ and $d^2u_R$ turn out to be related by a factor of $z$:
\begin{align}
 d^2u_R = \frac{P^R_{ab}(x)\,du_R^a du_R^b}{2(2\pi)} = \frac{2P^R_{ab}(x)\,du^a du^b}{2\pi} = \frac{zn_{ab}\,du^a du^b}{2\pi} = z d^2u \ .
\end{align}
Here, in the third equality, we used $du_a du^a = 0$ for $u\in P(\ell)$, while the fourth equality follows from contracting eq. \eqref{eq:measure} with $n_{ab}$. The bulk scalar \eqref{eq:bulk_scalar} thus becomes:
\begin{align}
 C(x;0) = iz\int_{P(\ell)} d^2u\,F_R\big((1+zn)u\big) = z\phi(\ell) + iz^2 n^{ab} \int_{P(\ell)}d^2u\,u_a\left.\frac{\del F_R(U)}{\del U^b}\right|_u + O(z^3) \ ,
\end{align}
from which we extract:
\begin{align}
 \pi(\ell) = in^\mu\gamma_\mu^{ab} \int_{P(\ell)}d^2u\,u_a\left.\frac{\del F_R(U)}{\del U^b}\right|_u \ . \label{eq:Neumann_n}
\end{align}
Now, recall that $n^\mu$ is an arbitrary null vector satisfying $\ell\cdot n = -1/2$. Since the result \eqref{eq:Neumann_n} should not depend on the choice of $n$, we can rewrite it as:
\begin{align}
 \pi(\ell)\ell_\mu = -\frac{i}{2}\gamma_\mu^{ab} \int_{P(\ell)}d^2u\,u_a\left.\frac{\del F_R(U)}{\del U^b}\right|_u \ . \label{eq:Neumann}
\end{align}
One can verify explicitly, using integration by parts, that the antisymmetric traceless part of the integral in \eqref{eq:Neumann} is indeed proportional to $\ell_{ab}$ (or, equivalently, that it vanishes upon contraction with $\ell^b{}_c$). 

Finally, let us point out the relation between the Dirichlet/Neumann boundary data and antipodal symmetry \cite{Vasiliev:2012vf,Ng:2012xp,Neiman:2014npa}. The Dirichlet data $\varphi(\ell)$ and the Neumann data $\pi(\ell)$ are associated with antipodally odd and even solutions respectively, in the sense that odd solutions have only $\varphi(\ell)$ non-vanishing, and even solutions have only $\pi(\ell)$ non-vanishing. As discussed in \cite{Halpern:2015zia}, this property can be deduced from the fact that the conformal weight of $\varphi$ ($\pi$) is an odd (even) positive integer. In our present language, the antipodal symmetry of $\varphi$ and $\pi$ can be seen in two ways. First, one can read off from \eqref{eq:Dirichlet},\eqref{eq:Neumann} the properties:
\begin{align}
 \varphi(-\ell) = -\varphi(\ell) \ ; \quad \pi(-\ell) = \pi(\ell) \ ,
\end{align}
where we used the fact that the measure \eqref{eq:measure} is odd under $\ell^\mu\rightarrow -\ell^\mu$. Second, we can apply the antipodal map to the bulk solution via $F_R(Y)\rightarrow F_R(Y)\star\delta(Y)$, which again results in:
\begin{align}
 \varphi(\ell) \rightarrow -\varphi(\ell) \ ; \quad \pi(\ell) \rightarrow \pi(\ell) \ .
\end{align}
In deriving this result, it is crucial to keep track of the sign factor in \eqref{eq:measure_product_boundary}.

\subsubsection{Spin $\geq 1$: chiral field strengths} 

The asymptotics for all the gauge fields with spin $s\geq 1$ can be described in a unified way using master fields. From the Penrose transform \eqref{eq:C_solution_R}, we extract two generating functions for the field strengths at a bulk point $x$:
\begin{align}
 C(x;y_L) = i\int_{P_R(x)} d^2u_R\,F_R(y_L + u_R) \ ; \quad C(x;y_R) = i\int_{P_R(x)} d^2u_R\,F_R(u_R)\,e^{iu_R y_R} \ . \label{eq:bulk_chiral_fields_raw}
\end{align}
Here, $y_L$ ($y_R$) is a left-handed (right-handed) spinor at $x$. The Taylor coefficients of $C(x;y_L)$ and $C(x;y_R)$ with respect to their spinor variables encode respectively the left-handed and right-handed field strengths $C^{(2s,0)}_{\alpha_1\dots\alpha_{2s}}(x),C^{(0,2s)}_{\dot\alpha_1\dots\dot\alpha_{2s}}(x)$ via eq. \eqref{eq:master_field}. The zeroth-order Taylor coefficient in both $C(x;y_L)$ and $C(x;y_R)$ is the spin-0 field $C(x;0)$. 

As a step towards taking the boundary limit, let us note that the integrals in \eqref{eq:bulk_chiral_fields_raw} do not change if we add to $y_L$ or $y_R$ a spinor of the opposite chirality. In other words, the chiral master fields \eqref{eq:bulk_chiral_fields_raw} can be extended trivially into functions of an entire twistor $Y$:
\begin{align}
 \begin{split}
   C_L(x;Y) &= C(x;P_L(x)Y) = i\int_{P_R(x)} d^2u_R\,F_R(Y + u_R) \ ; \\
   C_R(x;Y) &= C(x;P_R(x)Y) = i\int_{P_R(x)} d^2u_R\,F_R(u_R)\,e^{iu_R Y} \ .
 \end{split}  \label{eq:bulk_chiral_fields}
\end{align}
In the bulk-to-boundary limit \eqref{eq:limit}, the left-handed and right-handed fields \eqref{eq:bulk_chiral_fields} become:
\begin{align}
 C_L(x;Y) &= z\,\calC_L(\ell;Y) + O(z^2) \ ; & \calC_L(\ell;Y) &= i\int_{P(\ell)} d^2u\,F_R(Y + u) \ ; \label{eq:boundary_C_L} \\
 C_R(x;Y) &= z\,\calC_R(\ell;Y) + O(z^2) \ ; & \calC_R(\ell;Y) &= i\int_{P(\ell)} d^2u\,F_R(u)\,e^{iuY} \ , \label{eq:boundary_C_R}
\end{align}
where the factor of $z$ arises from the ratio of the measures $d^2u_R$ and $d^2u$. The boundary master fields $\calC_{L/R}(\ell;Y)$ have conformal weight $\Delta=1$, and depend only on the $P^*(\ell)$ spinor component $y^*$ of the twistor $Y$:
\begin{align}
 \calC_{L/R}(\ell,Y+u) = \calC_{L/R}(\ell;Y) \qquad \forall u\in P(\ell) \ . \label{eq:C_boundary_projective}
\end{align}
The Taylor expansion of $\calC_{L/R}(\ell;y^*)$ in powers of $y^*$ generates the individual left-handed/right-handed field strengths of various spins. The fields defined in this way have spinor indices in $P(\ell)$, as in the $j_\ell$ representation of boundary currents from section \ref{sec:geometry:spinors_boundary:currents}. 

The asymptotic field strengths $\calC_{L/R}(\ell;y^*)$ satisfy a Gauss law, i.e. each of the component fields with spin $s\geq 1$ has a vanishing divergence. To express and verify this fact explicitly, we must first use eq. \eqref{eq:u_u*} or \eqref{eq:j_P_j*} to convert the component fields into spinors with indices in $P^*(\ell)$. This is equivalent to converting $\calC_{L/R}(\ell;y^*)$ into a function of $y\in P(\ell)$:
\begin{align}
 \calC_{L/R}(\ell;Y) = \calC_{L/R}^*(\ell;P(\ell)Y) \ , 
\end{align}
where $\calC_{L/R}^*(\ell,y)$ can be given explicitly as:
\begin{align}
 \begin{split}
   \calC^*_L(\ell;y) &= -i\int d^4V\,F_R(V) \int_{P^*(\ell)}d^2u^*\,e^{iu^*(P(\ell)V - y)} \ ; \\
   \calC^*_R(\ell;y) &= i\int_{P^*(\ell)} d^2u^*\,F_R(P(\ell)u^*)\, e^{iu^*y} \ . 
 \end{split} \label{eq:C*}
\end{align}
The vanishing of the divergence \eqref{eq:spinor_div} for each of the component field strengths can now be expressed as:
\begin{align}
 \ell_\mu\gamma_{ab}^{\mu\nu}\, \frac{\del^3 \calC_{L/R}^*(\ell;y)}{\del\ell^\nu\del y_a\del y_b} = 0 \ , \label{eq:div_C}
\end{align}
and one can easily check that this constraint is in fact satisfied by the expressions \eqref{eq:C*}. Taking the $\del/\del\ell$ derivative of an integral over $P^*(\ell)$ requires some care, due to the $\ell$-dependence of the integration domain. The trick is to fix the integration range to some arbitrary 2d subspace of twistor space, which may then represent $P^*(\ell)$ for different values of $\ell$. One should keep track, however, of the $\ell$-dependence \eqref{eq:dual_measure} of the integration measure.

Finally, let us show how the boundary fields encoded in $\calC_{L/R}(\ell;Y)$ can be arrived at through tensor language. We approach the boundary as in section \ref{sec:geometry:spinors_boundary:limit}, moving the bulk point $x$ along towards the boundary point $\ell$ along the outwards-pointing tangent vector $t^\mu$. In the orthonormal tangent frame $t^\mu,e_i^\mu$, the components of the field strengths $C^{L/R}_{\mu_1\nu_1\dots\mu_s\nu_s}$ will scale as $z^{s+1}$ (this can be derived e.g. from the 4d conformal invariance of the free massless field equations). On the other hand, in a fixed frame in $\bbR^{1,4}$, the basis vector $t^\mu$ behaves asymptotically as $t^\mu\rightarrow\ell^\mu/z$, while the other basis vectors $e_i^\mu$ remain constant. Thus, in the fixed frame, $C^{L/R}_{\mu_1\nu_1\dots\mu_s\nu_s}$ will be dominated by components where the largest possible number of indices is pointing along $t^\mu$. This leaves us with the asymptotics:
\begin{align}
 \begin{split}
   C^{L/R}_{\mu_1\nu_1\dots\mu_s\nu_s}(x) &= z^{s+1}\left(2^s t_{[\mu_1}\delta_{\nu_1]}^{\rho_1}\dots t_{[\mu_s}\delta_{\nu_s]}^{\rho_s}\,\calC^{L/R}_{\rho_1\dots\rho_s}(\ell) + O(z) \right) \\
     &= z\left(2^s \ell_{[\mu_1}\delta_{\nu_1]}^{\rho_1}\dots\ell_{[\mu_s}\delta_{\nu_s]}^{\rho_s}\,\calC^{L/R}_{\rho_1\dots\rho_s}(\ell) + O(z) \right) \ .
 \end{split} \label{eq:C_tensor_asymptotics}
\end{align}
Here, in the first line, we insist on keeping the leading-order term within the tangent space of $EAdS_4$ at $x$; in the second line, we drop this requirement and substitute $t^\mu \rightarrow \ell^\mu/z$.
The tensors $\calC^{L/R}_{\mu_1\dots\mu_s}(\ell)$ are totally symmetric and traceless, with indices along $e_i^\mu$. More covariantly, these are boundary tensors in the sense of \eqref{eq:current_tensor_constraint}-\eqref{eq:current_tensor_equivalence}, with conformal weight $\Delta = s+1$. The equivalence \eqref{eq:current_tensor_equivalence} is associated with the different directions from which we could approach the boundary point $\ell$. Converting \eqref{eq:C_tensor_asymptotics} into spinor form as in \eqref{eq:C_tensor}, we get:
\begin{align}
 C^{(2s,0)}_{a_1\dots a_{2s}}(x) = z\,\calC^L_{a_1\dots a_{2s}}(\ell) + O(z^2) \ ; \quad C^{(0,2s)}_{a_1\dots a_{2s}}(x) = z\,\calC^R_{a_1\dots a_{2s}}(\ell) + O(z^2) \ , \label{eq:C_spinor_asymptotics}
\end{align}
where $\calC^L_{a_1\dots a_{2s}}(\ell)$ are totally symmetric boundary spinors with conformal weight $\Delta = 1$ and with indices in $P(\ell)$:
\begin{align}
 \calC^{L/R}_{a_1\dots a_{2s}}(\ell) = \gamma_{a_1 a_2}^{\mu_1\nu_1}\dots\gamma_{a_{2s-1}a_{2s}}^{\mu_s\nu_s} \ell_{\mu_1}\dots\ell_{\mu_s} \calC^{L/R}_{\nu_1\dots\nu_s}(\ell) \ . \label{eq:C_boundary_spinor_from_tensor}
\end{align}
We can now pack these into master fields, in analogy with \eqref{eq:master_field}:
\begin{align}
 \calC_{L/R}(\ell;Y) = \sum_{s=0}^\infty \frac{1}{(2s)!}\, Y^{a_1}\dots Y^{a_{2s}}\,\calC^{L/R}_{a_1\dots a_{2s}}(\ell) \ . \label{eq:C_master_asymptotics}
\end{align}
It is clear from eqs. \eqref{eq:master_field} and \eqref{eq:C_spinor_asymptotics} that the boundary master fields constructed in this way coincide with the ones in \eqref{eq:boundary_C_L}-\eqref{eq:boundary_C_R}.

\subsubsection{Spin $\geq 1$: electric and magnetic field strengths}

A more standard decomposition of the asymptotic field strengths is into their electric and magnetic parts. These are given by the sum and difference of the chiral field strengths \eqref{eq:boundary_C_L}-\eqref{eq:boundary_C_R}:
\begin{align}
 \calE(\ell;Y) = \calC_R(\ell;Y) + \calC_L(\ell;Y) \ ; \quad \calB(\ell;Y) = \calC_R(\ell;Y) - \calC_L(\ell;Y) \ . \label{eq:E_B}
\end{align}
$\calE(\ell;Y)$ and $\calB(\ell;Y)$ again have conformal weight $\Delta = 1$, and depend only on the $P^*(\ell)$ spinor component of $Y$. Thus, their Taylor coefficients in $Y$ are totally symmetric spinors with indices in $P(\ell)$, which encode the electric and magnetic field tensors for the various spins. The (higher-spin) electric and magnetic Gauss laws follow directly from those for $\calC_{L/R}(\ell;Y)$.

Let us unpack the definitions \eqref{eq:E_B} by working out their implications in tensor language. First, we identify the spin-0 components of $\calE(x;Y)$ and $\calB(x;Y)$:
\begin{align}
\calE(\ell;0) = 2\varphi(\ell) \ ; \quad \calB(\ell;0) = 0 \ . \label{eq:E_B_spin_0}
\end{align}
Thus, the spin-0 component of $\calE$ is proportional to the Dirichlet data for the bulk scalar, while the spin-0 component of $\calB$ vanishes. Next, we turn to the nonzero-spin components. Consider the bulk spin-$s$ field strength tensor \eqref{eq:C_tensor}. On a ``time slice'' (in quotes, since our bulk is Euclidean) with outward-pointing normal $t^\mu$, the field strength decomposes into electric and magnetic parts:
\begin{align}
 E_{\nu_1\nu_2\dots\nu_s}(x) &= t^{\mu_1}t^{\mu_2}\dots t^{\mu_s} C_{\mu_1\nu_1\mu_2\nu_2\dots\mu_s\nu_s}(x) \ ; \\
 B_{\nu_1\nu_2\dots\nu_s}(x) &= t^{\mu_1}t^{\mu_2}\dots t^{\mu_s}\left(-\frac{1}{2}\epsilon_{\mu_1\nu_1}{}^{\lambda\rho\sigma}x_\lambda\right) C_{\rho\sigma\mu_2\nu_2\dots\mu_s\nu_s}(x) \ . \label{eq:magnetic_tensor}
\end{align}
Thanks to the (anti)-self-duality \eqref{eq:self_dual} of the field strength's right-handed and left-handed components, this can be expressed equivalently as:
\begin{align}
 \begin{split} 
   E_{\nu_1\dots\nu_s}(x) &= t^{\mu_1}\dots t^{\mu_s} \left(C^R_{\mu_1\nu_1\dots\mu_s\nu_s}(x) + C^L_{\mu_1\nu_1\dots\mu_s\nu_s}(x) \right) \ ; \\
   B_{\nu_1\dots\nu_s}(x) &= t^{\mu_1}\dots t^{\mu_s} \left(C^R_{\mu_1\nu_1\dots\mu_s\nu_s}(x) - C^L_{\mu_1\nu_1\dots\mu_s\nu_s}(x) \right) \ .
 \end{split} \label{eq:E_B_bulk}
\end{align}
Here, we can already see the origin of eqs. \eqref{eq:E_B}. To make the relation explicit, let us work out the asymptotics of $E_{\mu_1\dots\mu_s}(x)$ and $B_{\mu_1\dots\mu_s}(x)$ as our ``time slice'' approaches the boundary. From eq. \eqref{eq:C_tensor_asymptotics}, we can read off immediately:
\begin{align}
 E_{\mu_1\dots\mu_s}(x) = z^{s+1}\calE_{\mu_1\dots\mu_s}(\ell) + O(z^2) \ ; \quad B_{\mu_1\dots\mu_s}(x) = z^{s+1}\calB_{\mu_1\dots\mu_s}(\ell) + O(z^2) \ ,
\end{align}
where:
\begin{align}
 \calE_{\mu_1\dots\mu_s}(\ell) = \calC^R_{\mu_1\dots\mu_s}(\ell) + \calC^L_{\mu_1\dots\mu_s}(\ell) \ ; \quad \calB_{\mu_1\dots\mu_s}(\ell) = \calC^R_{\mu_1\dots\mu_s}(\ell) - \calC^L_{\mu_1\dots\mu_s}(\ell) \ .
\end{align}
To arrive at eqs. \eqref{eq:E_B}, all that remains is to convert the boundary tensors $\calE_{\mu_1\dots\mu_s}(\ell)$ and $\calB_{\mu_1\dots\mu_s}(\ell)$ into spinor form as in \eqref{eq:C_boundary_spinor_from_tensor}:
\begin{align}
 \begin{split}
   \calE_{a_1\dots a_{2s}}(\ell) &= \gamma_{a_1 a_2}^{\mu_1\nu_1}\dots\gamma_{a_{2s-1}a_{2s}}^{\mu_s\nu_s} \ell_{\mu_1}\dots\ell_{\mu_s} \calE_{\nu_1\dots\nu_s}(\ell) \ ; \\
   \calB_{a_1\dots a_{2s}}(\ell) &= \gamma_{a_1 a_2}^{\mu_1\nu_1}\dots\gamma_{a_{2s-1}a_{2s}}^{\mu_s\nu_s} \ell_{\mu_1}\dots\ell_{\mu_s} \calB_{\nu_1\dots\nu_s}(\ell) \ ,
 \end{split} \label{eq:E_B_spinor_from_tensor}
\end{align}
and then package them into master fields as in \eqref{eq:C_master_asymptotics}:
\begin{align}
 \begin{split}
   \calE(\ell;Y) &= \sum_{s=0}^\infty \frac{1}{(2s)!}\, Y^{a_1}\dots Y^{a_{2s}}\,\calE_{a_1\dots a_{2s}}(\ell) \ ; \\ 
   \calB(\ell;Y) &= \sum_{s=0}^\infty \frac{1}{(2s)!}\, Y^{a_1}\dots Y^{a_{2s}}\,\calB_{a_1\dots a_{2s}}(\ell) \ .
 \end{split} \label{eq:E_B_master}
\end{align}

Finally, we should address the antipodal symmetry of $\calE(\ell;Y)$ and $\calB(\ell;Y)$. The antipodal map $F_R(Y)\rightarrow F_R(Y)\star\delta(Y)$ sends each of the integrals \eqref{eq:boundary_C_L}-\eqref{eq:boundary_C_R} into $-1$ times the other:
\begin{align}
 C_L(\ell;Y) \rightarrow -C_R(\ell;Y) \ ; \quad C_R(\ell;Y) \rightarrow -C_L(\ell;Y)
\end{align}
We can therefore read off from \eqref{eq:E_B} that the electric fields $\calE(\ell;Y)$ are antipodally odd, while the magnetic fields $\calB(x;Y)$ are antipodally even \cite{Neiman:2014npa}:
\begin{align}
 \calE(\ell;Y) \rightarrow -\calE(\ell;Y) \ ; \quad \calB(\ell;Y) \rightarrow \calB(\ell;Y) \ .
\end{align}
The same conclusion can be reached by the alternative methods that we've used for the spin-0 boundary data, i.e. by sending $\ell^\mu\rightarrow-\ell^\mu$ or examining the parity of the conformal weights of $\calE(\ell;Y)$ and $\calB(\ell;Y)$. From this point of view, the different antipodal parities of $\calE(\ell;Y)$ and $\calB(\ell;Y)$ arise from the antipodally odd $\epsilon_{\mu_1\nu_1}{}^{\lambda\rho\sigma}x_\lambda$ factor in the definition \eqref{eq:magnetic_tensor} of the magnetic fields.

\subsection{Electric fields at infinity from a bilocal boundary source} \label{sec:holography:example_asymptotics}

Now that we've defined the electric fields at infinity, let us evaluate them for the particular case of a bilocal source concentrated at a pair of points $\ell_0,\ell'_0$. Thus, in the language of eq. \eqref{eq:Pi_transform}, we choose the CFT sources as:
\begin{align}
  \Pi(\ell',\ell) = \delta^{5/2,1/2}(\ell,\ell_0)\,\delta^{5/2,1/2}(\ell',\ell'_0) \quad \Longrightarrow \quad F(Y) = K(\ell_0,\ell'_0;Y) \ , \label{eq:example_source}
\end{align}
where the superscripts on the delta functions indicate their conformal weight with respect to each argument. Now, according to our holographic dictionary \eqref{eq:dictionary}, we should construct the linearized bulk solution as the (right-handed or left-handed) Penrose transform of the twistor function $F(Y)$:
\begin{align}
 C(x;Y) = iK(\ell_0,\ell'_0;Y)\star\delta^R_x(Y) \quad \text{or} \quad C(x;Y) = -iK(\ell_0,\ell'_0;Y)\star\delta^L_x(Y) \ . \label{eq:ell_ell_x_raw}
\end{align}
As we've seen in section \ref{sec:spacetime_subgroup:manifest}, the result in both cases reads:
\begin{align}
 \begin{split}
   C(x;Y) &= \pm K(\ell_0,\ -\ell'_0-2(\ell'_0\cdot x)x;\ Y) \\
     &= \frac{\pm 1}{\pi\sqrt{2[\ell_0\cdot\ell'_0 + 2(\ell_0\cdot x)(\ell'_0\cdot x)]}}\exp\frac{iY[\ell_0\ell'_0 + 2(\ell'_0\cdot x)\ell_0 x] Y}{2[\ell_0\cdot\ell'_0 + 2(\ell_0\cdot x)(\ell'_0\cdot x)]} \ ,
 \end{split} \label{eq:ell_ell_x}
\end{align}
where the overall sign is ambiguous due to an intrinsic ambiguity in the star product. The bulk solution \eqref{eq:ell_ell_x} can be termed a ``boundary-boundary-bulk'' propagator. Note that one shouldn't conclude from the expression \eqref{eq:ell_ell_x} that this propagator is even under the antipodal map $x\rightarrow -x$: the sign ambiguity in \eqref{eq:ell_ell_x} can be resolved in opposite ways for future-pointing vs. past-pointing $x^\mu$. In fact, we should conclude from \eqref{eq:K_odd} that the propagator \eqref{eq:ell_ell_x} satisfies $C(x;Y)\star\delta(Y) = -C(x;Y)$, i.e. that it's antipodally \emph{odd}. This will be substantiated by our analysis of the solution's asymptotic behavior.

The next step is to extract the left-handed and right-handed field strengths, as in eq. \eqref{eq:bulk_chiral_fields}. Substituting $Y\rightarrow P_{L/R}(x)Y$ into the propagator \eqref{eq:ell_ell_x}, we get:
\begin{align}
 \begin{split}
   C_L(x;Y) &= \frac{\pm 1}{\pi\sqrt{2[\ell_0\cdot\ell'_0 + 2(\ell_0\cdot x)(\ell'_0\cdot x)]}} 
     \exp\frac{iY[\ell_0\ell'_0 + (\ell'_0\cdot x)\ell_0 x - (\ell_0\cdot x)\ell'_0 x - \ell_0\ell'_0 x] Y}{4[\ell_0\cdot\ell'_0 + 2(\ell_0\cdot x)(\ell'_0\cdot x)]} \ ; \\
   C_R(x;Y) &= \frac{\pm 1}{\pi\sqrt{2[\ell_0\cdot\ell'_0 + 2(\ell_0\cdot x)(\ell'_0\cdot x)]}}
     \exp\frac{iY[\ell_0\ell'_0 + (\ell'_0\cdot x)\ell_0 x - (\ell_0\cdot x)\ell'_0 x + \ell_0\ell'_0 x] Y}{4[\ell_0\cdot\ell'_0 + 2(\ell_0\cdot x)(\ell'_0\cdot x)]} \ ,
 \end{split}
\end{align}
where the only difference between the two expressions is in the sign of the last term in the exponent's numerator. We can now take the bulk-to-boundary limit \eqref{eq:limit} as in \eqref{eq:boundary_C_L}-\eqref{eq:boundary_C_R}, to get the asymptotic chiral field strengths:
\begin{align}
 \calC_L(\ell;Y) = \calC_R(\ell;Y) = \frac{\pm 1}{2\pi\sqrt{(\ell_0\cdot\ell)(\ell'_0\cdot\ell)}} \exp\frac{iY[(\ell'_0\cdot\ell)\ell_0\ell - (\ell_0\cdot\ell)\ell'_0\ell] Y}{8(\ell_0\cdot\ell)(\ell'_0\cdot\ell)} \ .
\end{align}
From these, we find the electric and magnetic boundary data as in \eqref{eq:E_B}:
\begin{align}
 \calE(\ell;Y) &= \frac{\pm 1}{\pi\sqrt{(\ell_0\cdot\ell)(\ell'_0\cdot\ell)}} \exp\frac{iY[(\ell'_0\cdot\ell)\ell_0\ell - (\ell_0\cdot\ell)\ell'_0\ell] Y}{8(\ell_0\cdot\ell)(\ell'_0\cdot\ell)} \ ; \label{eq:example_E_master} \\
 \calB(\ell;Y) &= 0 \quad \forall\ell\neq\ell_0,\ell'_0 \ ,
\end{align}
where we're careful to note that the magnetic field strengths vanish \emph{away from the source points $\ell_0,\ell'_0$}. Our analysis here doesn't capture the behavior at the source points themselves, and in fact we expect nonzero delta-function-like magnetic fields with support on $\ell_0,\ell'_0$. Since $\calB(\ell;Y)$ is associated with antipodally even solutions, its vanishing substantiates our identification of the propagator \eqref{eq:ell_ell_x_raw}-\eqref{eq:ell_ell_x} as antipodally odd. The possible \emph{non}-vanishing of $\calB(\ell;Y)$ at the source points themselves is related to the subtle interplay between antipodal symmetry and analyticity, which we discussed in section \ref{sec:linear_HS:antipodal}.

Let us now extract the various tensor components of the electric master field \eqref{eq:example_E_master}. We begin with the spin-0 Dirichlet data \eqref{eq:E_B_spin_0}:
\begin{align}
 \varphi(\ell) = \frac{1}{2}\calE(\ell;0) = \frac{\pm 1}{2\pi\sqrt{(\ell_0\cdot\ell)(\ell'_0\cdot\ell)}} \ . \label{eq:example_Dirichlet_covariant}
\end{align}
To extract the components with spin $s>0$, we expand \eqref{eq:example_E_master} into a Taylor series in $Y$ and compare with \eqref{eq:E_B_master}:
\begin{align}
 \begin{split}
   \calE^{a_1\dots a_{2s}}(\ell) ={}& \frac{\pm i^s (2s)!\,\gamma_{\mu_1\nu_1}^{(a_1 a_2}\dots\gamma_{\mu_s\nu_s}^{a_{2s-1}a_{2s})}}{8^s s!\,\pi \sqrt{(\ell_0\cdot\ell)(\ell'_0\cdot\ell)}}
     \left(\frac{\ell^{\mu_1}\ell_0^{\nu_1}}{\ell_0\cdot\ell} - \frac{\ell^{\mu_1}\ell'^{\nu_1}_0}{\ell'_0\cdot\ell} \right) \ldots
     \left(\frac{\ell^{\mu_s}\ell_0^{\nu_s}}{\ell_0\cdot\ell} - \frac{\ell^{\mu_s}\ell'^{\nu_s}_0}{\ell'_0\cdot\ell} \right)
 \end{split}
\end{align}
Next, we use \eqref{eq:E_B_spinor_from_tensor} to convert from spinors to tensors:
\begin{align}
 \calE^{\mu_1\dots\mu_s}(\ell) = \frac{\pm i^s (2s)!}{8^s s!\,\pi \sqrt{(\ell_0\cdot\ell)(\ell'_0\cdot\ell)}} 
   \left(\frac{\ell_0^{\mu_1}}{\ell_0\cdot\ell} - \frac{\ell'^{\mu_1}_0}{\ell'_0\cdot\ell} \right) \ldots \left(\frac{\ell_0^{\mu_s}}{\ell_0\cdot\ell} - \frac{\ell'^{\mu_s}_0}{\ell'_0\cdot\ell} \right) 
   - \text{traces} \ . \label{eq:example_E_tensor_covariant}
\end{align}
The trace pieces that are subtracted in \eqref{eq:example_E_tensor_covariant} can be represented using any 3d metric of the form $\eta_{\mu\nu} + 4\ell_{(\mu} n_{\nu)}$, where $n^\mu\in\bbR^{1,4}$ is a null vector satisfying $\ell\cdot n = -1/2$. Different choices of this 3d metric lead to tensors $\calE^{\mu_1\dots\mu_s}(\ell)$ that are equivalent under \eqref{eq:current_tensor_equivalence}.

Finally, let us make the boundary tensors \eqref{eq:example_E_tensor_covariant} more concrete by translating them into flat 3d boundary coordinates. To do this, we express $\ell_0$, $\ell'_0$ and $\ell$ in the flat conformal frame \eqref{eq:flat}. As it stands, the tensor \eqref{eq:example_E_tensor_covariant} is not tangential to the flat section \eqref{eq:flat_hyperplane} of the $\bbR^{1,4}$ lightcone. However, this can be fixed by adding a suitable multiple of $\ell^\mu$ to each tensor factor in \eqref{eq:example_E_tensor_covariant}. The $\ell_0^\mu/(\ell_0\cdot\ell)$ factors then become:
\begin{align}
 \frac{\ell_0^\mu}{\ell_0\cdot\ell} \cong \frac{\ell_0^\mu - \ell^\mu}{\ell_0\cdot\ell} = \frac{1}{|\mathbf{r} - \mathbf{r_0}|^2} \left(r^2 - r_0^2\, , \, 2(\mathbf{r} - \mathbf{r_0})\, , \, -(r^2 - r_0^2) \right) \ ,
\end{align}
and likewise for the $\ell'^\mu_0/(\ell'_0\cdot\ell)$ factors. Plugging these back into \eqref{eq:example_E_tensor_covariant} and keeping only the values $\mu = 1,2,3$ for each index, we end up with the 3d tensor:
\begin{align}
 \begin{split}
   \calE_{k_1\dots k_s}(\mathbf{r}) ={}& \frac{\pm i^s (2s)!}{2^{2s-1} s! \pi |\mathbf{r} - \mathbf{r_0}| |\mathbf{r} - \mathbf{r'_0}|} \\
     &\times \left(\frac{(\mathbf{r} - \mathbf{r_0})_{k_1}}{|\mathbf{r} - \mathbf{r_0}|^2} - \frac{(\mathbf{r} - \mathbf{r'_0})_{k_1}}{|\mathbf{r} - \mathbf{r'_0}|^2} \right) \ldots 
                   \left(\frac{(\mathbf{r} - \mathbf{r_0})_{k_s}}{|\mathbf{r} - \mathbf{r_0}|^2} - \frac{(\mathbf{r} - \mathbf{r'_0})_{k_s}}{|\mathbf{r} - \mathbf{r'_0}|^2} \right) - \text{traces} \ .
 \end{split} \label{eq:example_E_tensor}
\end{align}
This time, the subtracted trace pieces can be written out unambiguously, using the flat 3d metric $\delta_{ij}$. For completeness, we translate into the flat frame also the scalar boundary data \eqref{eq:example_Dirichlet_covariant}:
\begin{align}
 \varphi(\mathbf{r}) = \frac{\pm 1}{\pi |\mathbf{r} - \mathbf{r_0}| |\mathbf{r} - \mathbf{r'_0}|} \ . \label{eq:example_Dirichlet}
\end{align}
Note that eqs. \eqref{eq:example_E_tensor}-\eqref{eq:example_Dirichlet} have the same geometric structure as a 3-point function between two spin-0, $\Delta=1/2$ operators at $\mathbf{r_0},\mathbf{r'_0}$ and a spin-$s$, $\Delta=s+1$ operator at $\mathbf{r}$. This ``coincidence'' is of course predetermined by the boundary conformal symmetry.

\subsection{Boundary currents from a bilocal source} \label{sec:holography:example_currents}

In this section, we calculate the linearized expectation values of the CFT currents induced by the bilocal source \eqref{eq:example_source}. First, we write the linearized expectation value of the bilocal operator \eqref{eq:phi_phi_expectation}:
\begin{align}
 \left<\phi^I(\ell)\bar\phi_I(\ell')\right>_{\text{linear}} = NG(\ell,\ell'_0)G(\ell_0,\ell') = \frac{N}{32\pi^2\sqrt{(\ell'_0\cdot\ell)(\ell_0\cdot\ell')}} \ , \label{eq:example_phi_phi_covariant}
\end{align}
where we ignore both the zeroth order and all orders higher than 1 in the source dependence. 

Translating \eqref{eq:example_phi_phi_covariant} into the flat boundary coordinates \eqref{eq:flat}, we get:
\begin{align}
 \left<\phi^I(\mathbf{r})\bar\phi_I(\mathbf{r'})\right>_{\text{linear}} = \frac{N}{16\pi^2 |\mathbf{r} - \mathbf{r'_0}| |\mathbf{r'} - \mathbf{r_0}|} \ . \label{eq:example_phi_phi}
\end{align} 
The Taylor expansion of this around $\mathbf{r} = \mathbf{r'}$ reads:
\begin{align}
\begin{split}
  &\left<\phi^I(\mathbf{r})\overset{\leftarrow}{\del}_{i_1}\dots\overset{\leftarrow}{\del}_{i_m} \overset{\rightarrow}{\del}_{j_1}\dots\overset{\rightarrow}{\del}_{j_n} \bar\phi_I(\mathbf{r})\right>_{\text{linear}} 
    = \frac{(-1)^{m+n}(2m)!(2n)! N}{2^{m+n+4} \pi^2 m!n!} \\
  &\qquad \times 
    \frac{(\mathbf{r} - \mathbf{r'_0})_{i_1}\dots(\mathbf{r} - \mathbf{r'_0})_{i_m}(\mathbf{r} - \mathbf{r_0})_{j_1}\dots(\mathbf{r} - \mathbf{r_0})_{j_n}}{|\mathbf{r} - \mathbf{r'_0}|^{2m+1} |\mathbf{r} - \mathbf{r_0}|^{2n+1}}
   \ + \,  \text{trace terms} \ ,  
\end{split} \label{eq:example_phi_phi_taylor}
\end{align}
where by ``trace terms'' we mean terms proportional to the flat 3d metric $\delta_{ij}$. We can now combine the derivatives \eqref{eq:example_phi_phi_taylor} to obtain the spin-$s$ currents \eqref{eq:explicit_j}:
\begin{align}
 \begin{split}
   &\left<j^{(s)}_{k_1\dots k_s}(\mathbf{r})\right>_{\text{linear}} = \frac{(2s)!N}{4^{s+2}i^s\pi^2} \sum_{m=0}^s \frac{(-1)^{s-m}}{m!(s-m)!} \\
    &\qquad \times \frac{(\mathbf{r} - \mathbf{r'_0})_{(k_1}\dots(\mathbf{r} - \mathbf{r'_0})_{k_m}(\mathbf{r} - \mathbf{r_0})_{k_{m+1}}\dots(\mathbf{r} - \mathbf{r_0})_{k_s)}}
                                       {|\mathbf{r} - \mathbf{r'_0}|^{2m+1} |\mathbf{r} - \mathbf{r_0}|^{2(s-m)+1}} - \text{traces} \ .
 \end{split}
\end{align}
The above sum evaluates neatly into:
\begin{align}
 \begin{split}
   &\left<j^{(s)}_{k_1\dots k_s}(\mathbf{r})\right>_{\text{linear}} = \frac{i^s(2s)! N}{4^{s+2}s!\pi^2 |\mathbf{r} - \mathbf{r_0}| |\mathbf{r} - \mathbf{r'_0}|} \\
     &\qquad \times \left(\frac{(\mathbf{r} - \mathbf{r_0})_{k_1}}{|\mathbf{r} - \mathbf{r_0}|^2} - \frac{(\mathbf{r} - \mathbf{r'_0})_{k_1}}{|\mathbf{r} - \mathbf{r'_0}|^2} \right) \ldots 
                                \left(\frac{(\mathbf{r} - \mathbf{r_0})_{k_s}}{|\mathbf{r} - \mathbf{r_0}|^2} - \frac{(\mathbf{r} - \mathbf{r'_0})_{k_s}}{|\mathbf{r} - \mathbf{r'_0}|^2} \right) - \text{traces} \ .
 \end{split} \label{eq:example_j}
\end{align}
If we had taken into account the source-independent term in the bilocal $\left<\phi^I(\ell)\bar\phi_I(\ell')\right>$, we would have gotten an additional divergent contribution to the spin-0 ``current'' $\langle j^{(0)}(\mathbf{r})\rangle = \langle\phi^I(\mathbf{r})\bar\phi_I(\mathbf{r})\rangle$, with no change to the currents of spin $s>0$.

Comparing now with the results \eqref{eq:example_E_tensor}-\eqref{eq:example_Dirichlet} for the asymptotics of the linearized bulk fields, we find:
\begin{align}
 \left<\phi^I(\mathbf{r})\bar\phi_I(\mathbf{r})\right>_{\text{linear}} = \pm \frac{N}{16\pi}\,\varphi(\mathbf{r}) \ ; \quad 
 \left<j^{(s)}_{k_1\dots k_s}(\mathbf{r})\right>_{\text{linear}} = \pm \frac{N}{32\pi}\,\calE_{k_1\dots k_s}(\mathbf{r}) \ .  \label{eq:currents_holography}
\end{align}
where, in the second equality, we take $s>0$. We've thus demonstrated the proportionality, and found the proportionality coefficients, between the Dirichlet/electric boundary data and the linearized expectations values of the corresponding single-trace operators. The $s$-independence of the coefficients in \eqref{eq:currents_holography} results from the particular normalization choice in our definition \eqref{eq:explicit_j} of the CFT currents.

The sign ambiguity in \eqref{eq:currents_holography}, which we've been carrying from eq. \eqref{eq:ell_ell_x}, can be fixed by hand by comparing with the standard dictionary in the spin-2 case, i.e. the correspondence between the bulk graviton and the CFT stress tensor. To do this, we'd have to fix a sign convention for the relation between the Weyl tensor $C_{\mu\nu\rho\sigma}$ and the corresponding bulk metric perturbation. In any case, this fixing of the signs doesn't seem essential: in GR, the sign of the metric perturbation only becomes meaningful at the interacting level, where detailed analogies with HS gravity are not very useful.

\subsection{General boundary currents and the extent of the holographic dictionary} \label{sec:holography:general_currents}

Let us now extract the general lessons from our calculation in sections \ref{sec:holography:example_asymptotics}-\ref{sec:holography:example_currents}. In analogy with the bulk asymptotics $\calE(\ell;Y)$ and $\calB(\ell;Y)$, let us define local master-field operators on the CFT side via: 
\begin{align}
 J(\ell;Y) &= \frac{1}{8\pi i}\int_{P(\ell)} d^2u \left(\Phi(u)\,e^{iuY} + \Phi(Y+u)\right) \ ; \label{eq:J} \\
 H(\ell;Y) &= \frac{i}{2\pi N}\int_{P(\ell)} d^2u \left(\Phi(u)\,e^{iuY} - \Phi(Y+u)\right) \ . \label{eq:H}
\end{align}
From eq. \eqref{eq:Phi_expectation}, we see that at first order in the sources, these operators have the expectation values:
\begin{align}
 \langle J(\ell;Y)\rangle_{\text{linear}} &= \frac{iN}{32\pi}\int_{P(\ell)} d^2u \left(F(u)\,e^{iuY} + F(Y+u)\right) \ ; \\
 \langle H(\ell;Y)\rangle_{\text{linear}} &= \frac{1}{8\pi i}\int_{P(\ell)} d^2u \left(F(u)\,e^{iuY} - F(Y+u)\right) \ .
\end{align}
The $Y$ dependence of these master fields is only through the spinor component $y^*\in P^*(\ell)$.

The implication of the result \eqref{eq:currents_holography} is that, in regions where the source $\Pi(\ell',\ell)$ vanishes, $\langle J(\ell;Y)\rangle_{\text{linear}}$ encodes the linearized currents $\left<j^{(s)}_{\mu_1\dots \mu_s}(\ell)\right>_{\text{linear}}$ in the same sense that $\calE(\ell;Y)$ encodes the electric boundary data $E_{\mu_1\dots\mu_s}(\ell)$, up to the sign ambiguity in the Penrose transform \eqref{eq:ell_ell_x}. At the same time, in complete analogy with $\calE(\ell;Y)$ and $\calB(\ell;Y)$, the tensor components of the operators \eqref{eq:J}-\eqref{eq:H} are automatically divergence-free. Putting everything together, we see that $J(\ell;Y)$ encodes a tower of spin-$s$ conformal primaries $j^{(s)}_{\mu_1\dots \mu_s}(\ell)$, conserved to all orders in the source $\Pi(\ell',\ell)$, which at linear order correctly reproduce the expectation values of the CFT currents in regions where $\Pi(\ell',\ell)$ vanishes. The most natural conclusion, then, is that $J(\ell;Y)$ encodes the conserved CFT currents to all orders in the source, \emph{with all the necessary contact terms automatically included}.

As for the master field $H(\ell;Y)$, the result of section \ref{sec:holography:example_asymptotics} implies that its linearized expectation value vanishes in regions with $\Pi(\ell',\ell)=0$. By construction, $\langle H(\ell;Y)\rangle_{\text{linear}}$ is proportional to the magnetic boundary data $\calB(\ell;Y)$. Thus, we expect that in regions with $\Pi(\ell',\ell)\neq 0$, it will encode the linearized magnetic field strengths associated with the source. This is the reason for our choice of coefficient in \eqref{eq:H}: we wanted to emphasize that $H(\ell;Y)$ is more closely related to the source $F(Y)$ than to the ``current'' $\Phi(Y)$. The interpretation of the full non-linear expectation value of $H(\ell;Y)$ is not entirely clear to us. Perhaps the most natural possibility is that it still encodes the source's magnetic field strength, but with the non-abelian structure of higher-spin symmetry taken into account.

\section{Discussion} \label{sec:discuss}

In this paper, we've shown how a twistorial description underlies both bulk and boundary pictures in the higher-spin/free-CFT holography. In particular, our boundary/twistor transform \eqref{eq:Pi_transform},\eqref{eq:phi_phi} does the same for single-trace bilocals in the free $U(N)$ vector model as the Penrose transform has done for free massless fields in 4d.

Our main evidence that the bulk and boundary pictures as derived from twistor space are indeed holographically equivalent is the calculation of linearized boundary currents away from sources. Beyond this, much of the relationship between our twistor language and the standard local descriptions was left implicit. It should be worthwhile to explore this relationship further. In particular, one would like to express the boundary sources' local field strengths in the bilocal language, and compare to the twistorial expression \eqref{eq:H}. One should also understand explicitly the local currents in regions where the source doesn't vanish, and then check how (or whether) eq. \eqref{eq:J} contains the necessary information about contact terms.

On the bulk side, the main missing component in our approach, as in \cite{Didenko:2012tv}, is the relation to the nonlinear Vasiliev equations. The unbroken global HS symmetry has allowed us to ``cheat'' by encoding the interactions as functionals of the linearized master fields. However, to make contact with the broader realm of higher-spin theory, one should understand how to go back and forth between this approach and Vasiliev's picture of nonlinear bulk master fields.

From a fundamental perspective, the picture we laid out in this paper is very appealing: all the three geometric frameworks of bulk, boundary and twistor space are manifestly unified. Furthermore, the twistor function $F(Y)$ provides a clean diff-invariant \& gauge-invariant encoding of the physical data on both bulk and boundary. Ideally, one would like to apply this kind of picture to more realistic holographic models, which contain General Relativity in the bulk. However, at the moment, it is unclear to us how that might happen. In our construction, we relied heavily on the fact that the boundary CFT is a free vector model -- that is what allowed the bilocal formulation of the single-trace operators. Similarly, in the bulk, we made crucial use of the unfolded formulation of HS theory. It is what enabled us to cleanly encode a linearized bulk solution in terms of a master field at a single point $x$, which we could then use as an input for the ``bulk'' partition function \eqref{eq:Z_bulk}.

While there are many reasons to study HS theory, the author's personal motivation is that it provides the only known working model of \dSCFT{4}{3} \cite{Anninos:2011ui}. In that context, I am pursuing a program \cite{Halpern:2015zia} to extract the physics inside observers' cosmological horizons. A key component in this program is the idea \cite{Parikh:2002py} to replace $dS_4$ with its ``folded-in-half'' version $dS_4/Z_2$, where the $Z_2$ refers to the antipodal map $x^\mu\rightarrow -x^\mu$. In \cite{Halpern:2015zia}, we managed to derive the physics inside the horizon in this picture for the linearized limit of HS gravity, i.e. for free massless fields in the bulk. The motivation of the present work was to develop tools in order to translate those preliminary results into the twistor language of HS theory, and then extend them beyond the linearized limit. It is our hope that the language we developed here will prove powerful enough for the task.

\section*{Acknowledgements}

I am grateful to Eugene Skvortsov, Slava Didenko, Mikhail Vasiliev, Antal Jevicki, Maxim Grigoriev, Vasudev Shyam and Adrian David for discussions. This work was supported by the Quantum Gravity Unit of the Okinawa Institute of Science and Technology Graduate University (OIST). During early stages of the work, YN was employed at Perimeter Institute, where he was supported by the Government of Canada through Industry Canada, by the Province of Ontario through the Ministry of Research \& Innovation, as well as by NSERC Discovery grants. The initial calculations were performed while attending the MIAPP Programme ``Higher Spin Theory and Duality''.

\end{document}